\documentclass[aip,jcp,reprint,amsmath,amssymb]{revtex4-2}

\newcommand{\dd}{\mathop{}\!\mathrm{d}}
\newcommand{\vv}{\mathbf}
\newcommand{\vvg}{\boldsymbol}

\usepackage{graphicx}
\usepackage[suffix=]{epstopdf}
\usepackage{natmove}

\usepackage{xcolor}

\begin{document}
\title{Wertheim's thermodynamic perturbation theory with double-bond association and its application to colloid--linker mixtures}
\author{Michael P. Howard}
\affiliation{McKetta Department of Chemical Engineering, University of Texas at Austin, Austin, Texas 78712, USA}

\author{Zachary M. Sherman}
\affiliation{McKetta Department of Chemical Engineering, University of Texas at Austin, Austin, Texas 78712, USA}

\author{Delia J. Milliron}
\affiliation{McKetta Department of Chemical Engineering, University of Texas at Austin, Austin, Texas 78712, USA}

\author{Thomas M. Truskett}
\email{truskett@che.utexas.edu}
\affiliation{McKetta Department of Chemical Engineering, University of Texas at Austin, Austin, Texas 78712, USA}
\affiliation{Department of Physics, University of Texas at Austin, Austin, Texas 78712, USA}

\begin{abstract}
We extend Wertheim's thermodynamic perturbation theory to derive the association free energy of a multicomponent mixture for which double bonds can form between any two pairs of the molecules' arbitrary number of bonding sites. This generalization reduces in limiting cases to prior theories that restrict double bonding to at most one pair of sites per molecule. We apply the new theory to an associating mixture of colloidal particles (``colloids'') and flexible chain molecules (``linkers''). The linkers have two functional end groups, each of which may bond to one of several sites on the colloids. Due to their flexibility, a significant fraction of linkers can ``loop'' with both ends bonding to sites on the same colloid instead of bridging sites on different colloids. We use the theory to show that the fraction of linkers in loops depends sensitively on the linker end-to-end distance relative to the colloid bonding-site distance, which suggests strategies for mitigating the loop formation that may otherwise hinder linker-mediated colloidal assembly.
\end{abstract}

\maketitle

\section{Introduction}
It has been over 35 years since Wertheim developed a thermodynamic perturbation theory (TPT) for fluids
of molecules that associate through strong directional attractions \cite{Wertheim:1984a,Wertheim:1984b,Wertheim:1986a,Wertheim:1986b}. Wertheim's TPT has proven to be powerful and broadly applicable, capturing the thermodynamic behavior of fluids comprising, e.g., small molecules with chemically reactive sites, associating macromolecules such as DNA nanostars \cite{Rovigatti:2014,Locatelli:2017}, and patchy colloidal particles (``colloids'') \cite{Bianchi2006,Bianchi2007,Russo:2009,Russo:2011}. Indeed, Wertheim's TPT is a foundation for statistical associating fluid theory (SAFT) \cite{Chapman:1986,Joslin:1987,Jackson:1988,Chapman:1988}, which has been widely used to develop equations of state for complex fluid mixtures \cite{Muller:2001}.

\begin{figure}
    \centering
    \includegraphics{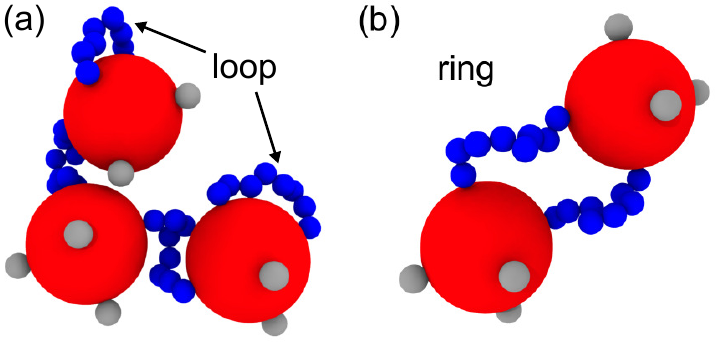}
    \caption{Snapshots of associating colloids (red) and linkers (blue), rendered using OVITO 2.9.0 \cite{Stukowski:2010}. Free bonding sites on the colloids are shown in gray. (a) Two linkers ``loop'' to form a double bond with a colloid, while the other two linkers bridge between different colloids. (b) Two colloids and two linkers form a ``ring.'' Loop and ring bonding structures are both neglected in TPT1.}
    \label{fig:snap}
\end{figure}

In practice, many applications of Wertheim's TPT, including SAFT, adopt a first-order approximation that treats association as pairwise bonding into tree-like networks and is often referred to as TPT1 \cite{Wertheim:1984b,Wertheim:1986b}. We previously used TPT1 to predict the phase behavior of mixtures of colloids and difunctional chain molecules (``linkers'') in which linker ends associate with sites on the colloids through strong reversible bonds (Figure~\ref{fig:snap}) \cite{Howard:2019}. Homogeneous colloid--linker mixtures of this type lose thermodynamic stability for a range of colloid or linker concentrations \cite{Lindquist:2016,Howard:2019}, phase separating into colloid-rich and colloid-lean phases and potentially gelling \cite{SaezCabezas:2018,Dominguez:2020}. TPT1 qualitatively predicts this transition, but we have found that it underestimates the amount of linker required for phase separation \cite{Howard:2019,Dominguez:2020}.
We hypothesized that this underestimation was due in part to TPT1's neglect of other prevalent bonding motifs such as linkers that ``loop'' so that both ends associate to the same colloid (Figure~\ref{fig:snap}a), i.e., form a double bond between a linker and a colloid.

TPT1 has been extended to include double bonds for restricted numbers of components, bonding sites, and/or pairs of sites that can participate in the bonds \cite{Sear:1994a,Sear:1994b,Galindo:2002,Avlund:2011,Tavares:2012,Rovigatti:2013,Marshall:2013a,Marshall:2013b}. The most general result to date is due to Marshall \cite{Marshall:2018}, which includes multiple components with multiple bonding sites but permits only one pair of sites per component to double bond. This restriction is well justified when the geometry of the sites limits potential double bonding pairs, e.g., direct association of colloids with small numbers of attractive patches. However, it can break down for common linking schemes for self-assembly. For example, nanoparticles are often coated with flexible ligands that are functionalized to bond with a complementary functional group on another nanoparticle \cite{Hoppe:2002,Maneeprakorn:2010,Macfarlane:2011,Borsley:2016,Wang:2019,Marro:2020} or a linking molecule \cite{Mirkin:1996,Alivisatos:1996,Xiong:2009,Dominguez:2020}. Control over the specific number of functionalized ligands on each nanoparticle is infeasible or produces low yields \cite{Zanchet:2001}, and due to their flexibility and surface coverage, multiple ligands on the same nanoparticle might readily double bond with another nanoparticle or with a linker \cite{Dominguez:2020}. Given the success of TPT1 for other associating fluids, we aim to develop the theory to handle such cases.

Building on previous TPTs, we derive a general expression for the association free energy of a mixture when a double bond can form between any two pairs of sites on any of the components; this expression reduces to TPT1 \cite{Chapman:1988} and Marshall's result \cite{Marshall:2018} as special cases. We apply the theory to the colloid--linker mixture we previously studied, where the linker readily double bonded with nearly any pair of sites on the colloids \cite{Howard:2019}. We compute the fraction of linkers forming double-bond loops. In the strong association limit, this fraction is strongly dependent on the distribution of linker end-to-end distances. Modifications to linker length or flexibility that make the linker end-to-end distance incompatible with the distance between colloid bonding sites inhibit loop formation; this consideration may prove helpful in the design of molecular linkers for self-assembly. We also show that the theory readily incorporates other prevalent bonding motifs with two-site association, such as ``rings'' made from multiple linkages between colloids (Figure~\ref{fig:snap}b), and that incorporating these motifs improves its accuracy.

The rest of the article is organized as follows. In Section~\ref{sec:wt}, we review the fundamentals of Wertheim's theory, which we hope will serve as a pedagogical reference complementary to and expanding on excellent recent reviews \cite{Marshall:2016,Zmpitas:2016}. In Section~\ref{sec:tptdb}, we derive the new theory for double bonding between multiple components with multiple sites. We apply the theory to the colloid--linker mixture in Section~\ref{sec:colllink} before summarizing in Section~\ref{sec:conclude}.

\section{Wertheim's theory\label{sec:wt}}
We begin by outlining Wertheim's theory for associating fluids \cite{Wertheim:1984a,Wertheim:1984b,Wertheim:1986a,Wertheim:1986b}, which is based on a graphical expansion in the grand canonical ensemble using the densities of various bonding states. Our aim is to provide a useful entry point for those seeking to understand and build upon this work, as well as for those who wish to similarly apply or extend Wertheim's theory. We will follow Wertheim's derivation for molecules with multiple association sites \cite{Wertheim:1986a} with small modifications to include multiple components; any subscripts or superscripts denoting component type in this article should be neglected when comparing to notation from his work.

\subsection{Mixture model}
Consider a mixture of $k$ components that are able to associate with each other through short-ranged attractions. We will adopt a physical model for the mixture that refers to the constituents of a component as ``molecules'' that are covered with bonding ``sites,'' but here a molecule should be interpreted liberally to include not only small molecules but also macromolecules, colloids, etc. A molecule of component $i$ has a set of sites $\Gamma_i = \{A_1,..., A_{n_i}\}$, which we will assume for now are rigidly fixed within the molecule so that its conformation can be specified by its position $\vv{r}$ and orientation $\vvg{\Omega}$. The total potential energy of the mixture $U$ is taken to be the sum of a one-body potential $u_1^{(i)}$ acting on each molecule of component $i$ and a pairwise interaction $u_2^{(ij)}$ between two molecules of components $i$ and $j$. The pairwise interaction is split into a repulsive core $u_{\rm R}^{(ij)}$ and a short-ranged attraction $u_{AB}^{(ij)}$ between sites,
\begin{equation}
u_2^{(ij)}(\vv{1},\vv{2}) = u_{\rm R}^{(ij)}(\vv{1},\vv{2}) +
    \sum_{A \in \Gamma_i} \sum_{B \in \Gamma_j} u_{AB}^{(ij)}(\vv{1},\vv{2}),
\label{eq:pot}
\end{equation}
where $\vv{1} = (\vv{r}_1,\vvg{\Omega}_1)$ is a shorthand notation for the position and orientation of molecule 1. The site attractions are functions of $\vv{1}$ and $\vv{2}$ through the positions of sites $A$ and $B$, $\vv{r}_A^{(i)}(\vv{1})$ and $\vv{r}_B^{(j)}(\vv{2})$. The summation is taken over all pairs of sites in the two molecules, but some sites may be noninteracting.

In the grand canonical ensemble, the temperature $T$, volume $V$, and all chemical potentials $\mu_i$ are held constant, and the numbers of molecules $\{N_i\}$ fluctuate. The grand partition function is
\begin{equation}
\Xi = \sum_{N_1=0}^\infty \cdots \sum_{N_k=0}^\infty \int \Bigg(\prod_{i=1}^k  \dd{\vv{N}_i} \frac{\Lambda_i^{N_i} e^{\beta\mu_i N_i}}{N_i!}\Bigg) e^{-\beta U(\{\vv{N}_i\})},
\label{eq:xi}
\end{equation}
where $\vv{N}_i$ indicates all the positions and orientations of the $N_i$ molecules of component $i$, $\Lambda_i$ accounts for integrals over translational and rotational momenta (i.e., is related to the thermal wavelength), $N_i!$ is a factor for permutating labels of the molecules, and $\beta=1/(k_{\rm B}T)$ with $k_{\rm B}$ being Boltzmann's constant. The one-body contributions $u_1^{(i)}$ to the potential energy $U$ can be grouped with the chemical potential $\mu_i$ to define a modified fugacity
\begin{equation}
z_i(\vv{1}) = \Lambda_i \exp\left[\beta(\mu_i-u_1^{(i)}(\vv{1}))\right]
\end{equation}
so that for our model
\begin{align}
\Xi = \sum_{N_1=0}^\infty \cdots \sum_{N_k=0}^\infty &\int \Bigg(\prod_{i=1}^k \dd{\vv{N}_i} \frac{1}{N_i!} \prod_{\vv{1}} z_i(\vv{1})\Bigg) \nonumber \\
&\times \exp\Bigg(-\beta \sum_{\vv{1},\vv{2}} u_2^{(ij)}(\vv{1},\vv{2}) \Bigg).
\label{eq:xi2}
\end{align}
The product of $z_i$ is taken over all molecules of component $i$ (indexed $\vv{1}$), and the sum of $u_2^{(ij)}$ is taken over all unique pairs of molecules (indexed $\vv{1}$ and $\vv{2}$).

\subsection{Graphical expansion}
\begin{figure}
    \centering
    \includegraphics{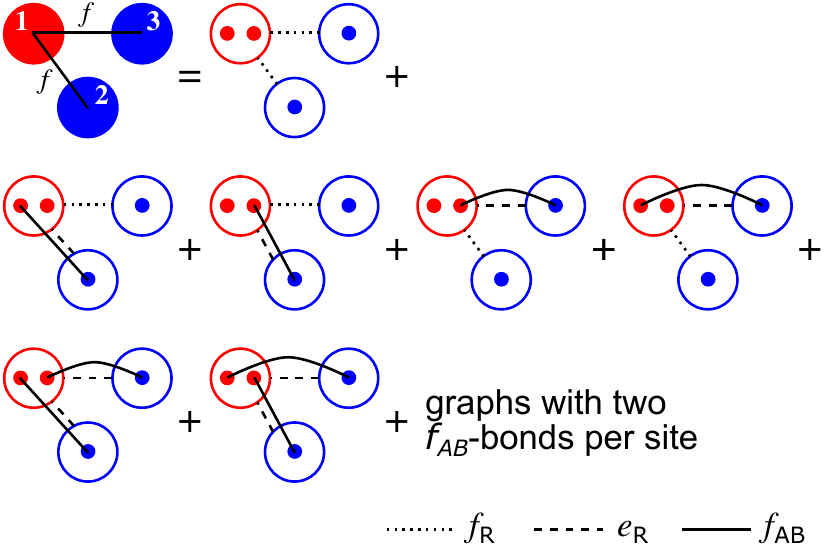}
    \caption{Graphical representations of the integral $\int \dd\vv{1}\dd\vv{2}\dd\vv{3} z_1(\vv{1}) z_2(\vv{2}) z_2(\vv{3}) f^{(12)}(\vv{1},\vv{2}) f^{(12)}(\vv{1},\vv{3})$. The first graph is a connected $z$-graph with $f$-bonds between molecule 1 of component 1 (red) and molecules 2 and 3 of component 2 (blue). The other graphs show the expansion of $f$ using Eq.~\eqref{eq:f} into hyperpoints (larger open circles) and bonding sites (smaller filled circles). Hyperpoints connected by an $f_{AB}$-bond also have an $e_{\rm R}$-bond (dashed line) between them, while those without an $f_{AB}$-bond have an $f_{\rm R}$-bond (dotted line). In this example, a molecule of component 1 has two bonding sites, and a molecule of component 2 has one bonding site. We omit the graphs having two $f_{AB}$-bonds at a site because these can usually be neglected due to steric hindrance, and we do not show the label-permutation prefactor ($1/2$) for the $z$-graph from Eq.~\eqref{eq:xi2}.}
    \label{fig:graphs}
\end{figure}

We now define $e^{(ij)}(\vv{1},\vv{2}) = \exp[-\beta u_2^{(ij)}(\vv{1},\vv{2})]$ as the $e$-function (Boltzmann weight) associated with an interaction between two molecules of components $i$ and $j$, respectively, and $f^{(ij)}(\vv{1},\vv{2}) = e^{(ij)}(\vv{1},\vv{2})-1$ as the corresponding Mayer $f$-function. The integrand in Eq.~\eqref{eq:xi2} can be written as a product of $f$-functions using
\begin{equation}
\exp\bigg(-\beta \sum_{\vv{1},\vv{2}} u_2^{(ij)}(\vv{1},\vv{2}) \bigg) = \prod_{\vv{1},\vv{2}} \bigg(1+f^{(ij)}(\vv{1},\vv{2})\bigg).
\end{equation}
The integrals in Eq.~\eqref{eq:xi2} can then be expanded into multiple integrals over different products of $f$-functions, and it is useful to express these integrals as graphs (Figure~\ref{fig:graphs}) \cite{Andersen:1977}. Each integral is represented by a ``$z$-graph'' whose nodes, called field points, are the integration variables and are each assigned a factor of the fugacity $z_i$. Two field points in a graph may be connected by at most one $f$-bond, whose presence is represented by an edge between the points that carries a weight $f^{(ij)}$. There is an infinite number of $z$-graphs in $\Xi$ of which only a subset are connected, i.e., have all points joined to each other by $f$-bonds. It can be shown that $\ln \Xi$ is the sum of only these connected graphs, and the mixture thermodynamics can be computed using the grand potential $F = -\beta^{-1} \ln \Xi$. However, using only $f$-bonds to represent association can give a poorly convergent summation for $\ln \Xi$, which motivated Wertheim to further decompose the graphs \cite{Wertheim:1984a}.

For the model interactions defined by Eq.~\eqref{eq:pot}, an $f$-bond is separable into repulsive and attractive parts,
\begin{align}
f^{(ij)}&(\vv{1},\vv{2}) = f_{\rm R}^{(ij)}(\vv{1},\vv{2}) + e_{\rm R}^{(ij)}(\vv{1},\vv{2}) \nonumber \\
&\times\Bigg[\prod_{A \in \Gamma_i} \prod_{B \in \Gamma_j} \left(1+f_{AB}^{(ij)}(\vv{1},\vv{2})\right)-1\Bigg]
\label{eq:f},
\end{align}
where the subscripts denote the part of the potential. Graphically, each point in a $z$-graph is replaced by a hyperpoint encompassing the molecule's sites, and the $f$-bonds are replaced by repulsive bonds between hyperpoints and attractive bonds between sites (Figure~\ref{fig:graphs}). Each hyperpoint still carries a factor $z_i$, and repulsive $f_{\rm R}$-bonds are edges between two hyperpoints. Attractive $f_{AB}$-bonds are one or more edges between sites $A$ and $B$ in different hyperpoints, which have in parallel a repulsive $e_{\rm R}$-bond between the hyperpoints.

Two hyperpoints are ``directly connected'' if they have a repulsive $f_{\rm R}$-bond or any attractive $f_{AB}$-bond between their sites. Further, two sites are ``bond-connected'' if there is a path of attractive bonds between them, while two sites are ``constraint-connected'' if they are in the same hyperpoint. Now consider the set of $z$-graphs where all hyperpoints are connected by networks of attractive bonds; Wertheim called such a graph having $s$ hyperpoints an ``$s$-mer'' analogous to the association of monomers into a polymer. The connected $z$-graphs in $\ln \Xi$ can all be expressed as $s$-mers with $f_{\rm R}$-bonds between hyperpoints in distinct $s$-mers, including the $s=1$-mer or single hyperpoint without any attractive bonds.

Within an $s$-mer graph, there can be multiple networks of bond-connected sites; these networks must be constraint-connected. Consider two hyperpoints $\vv{1}$ and $\vv{2}$ in an $s$-mer that are not directly connected by an attractive bond. For such a hyperpoint pair, Wertheim proposed that if any site $A$ on $\vv{1}$ is bond-connected to any site $B$ on $\vv{2}$, two $s$-mer graphs with and without an $f_{\rm R}$-bond between $\vv{1}$ and $\vv{2}$ can be usefully combined into one graph with an $e_{\rm R}$-bond between $\vv{1}$ and $\vv{2}$. Hence, we will form $e_{\rm R}$-bonds between all pairs of hyperpoints in an $s$-mer with bond-connected sites, but pairs of hyperpoints in an $s$-mer that are connected only through a constraint connection will not have an $e_{\rm R}$-bond. By this procedure, all hyperpoints in a bond-connected network become irreducibly connected.%
\footnote{An articulation point of a connected graph is a point that, if removed, will disconnect the graph into two or more graphs \cite{Andersen:1977}. An irreducible graph is free of articulation points. For example, a pair of points or any closed cycle of points is irreducible, but three points bonded colinearly are not irreducible because the middle point is an articulation point.}

Define $\rho^{(i)}(\vv{1})$ to be the sum of $z$-graphs obtained by turning a field hyperpoint of component $i$, which is an integration variable, into a ``root'' hyperpoint labeled $\vv{1}$, which is not an integration variable. This is equivalent to the functional differentiation
\begin{equation}
\rho^{(i)}(\vv{1}) = z_i(\vv{1}) \frac{\delta \ln \Xi}{\delta z_i(\vv{1})},
\end{equation}
so $\rho^{(i)}(\vv{1})$ is the average ``singlet'' density of component $i$ at $\vv{1}$ in the grand canonical ensemble. The singlet density is related to the familiar number density $\rho_i(\vv{r}_1)$ by integration over orientations,
\begin{equation}
\rho_i(\vv{r}_1) = \int \dd{\vvg{\Omega}_1} \rho^{(i)}(\vv{1}).
\label{eq:rhon}
\end{equation}
The various $z$-graphs in the sum $\rho^{(i)}(\vv{1})$ can be further analyzed in terms of their bonding at the root hyperpoint. A $z$-graph is assigned to the sum $\rho_\alpha^{(i)}(\vv{1})$ if $\alpha$ is the set of bonded sites at $\vv{1}$ ($\alpha \subseteq \Gamma_i$),%
\footnote{$A \subseteq B$ denotes that $A$ is a subset of $B$, including the improper subset $A = B$.}
and as a result,
\begin{equation}
\rho^{(i)}(\vv{1}) = \sum_{\alpha \subseteq \Gamma_i} \rho_\alpha^{(i)}(\vv{1}).
\end{equation}
Now let $c_\varnothing^{(i)}(\vv{1})$ be the sum of the subset of graphs in $\rho_\varnothing^{(i)}(\vv{1})/z_i(\vv{1})$ for which $\vv{1}$ is not an articulation point. It can be shown that \cite{Zmpitas:2016}
\begin{equation}
\ln\left(\frac{\rho_\varnothing^{(i)}(\vv{1})}{z_i(\vv{1})} \right) = c_\varnothing^{(i)}(\vv{1}).
\label{eq:c0}
\end{equation}
Further let $c_\alpha^{(i)}(\vv{1})$ be the sum of the subset of graphs in $\rho_\alpha^{(i)}(\vv{1})/\rho_\varnothing^{(i)}(\vv{1})$ for which $\vv{1}$ is not a ``constraint'' articulation point, i.e., removal of the hyperpoint and any incident repulsive bonds at $\vv{1}$ leaves the graph connected. $\rho_\alpha^{(i)}(\vv{1})$ includes all graphs bonded at the sites $\alpha$, which can be formed by composing at the root point various $c_a^{(i)}(\vv{1})$ for $a \subseteq \alpha$, such that their union gives $\alpha$, along with the graphs that are not bonded,
\begin{equation}
\rho_\alpha^{(i)}(\vv{1}) = \rho_\varnothing^{(i)}(\vv{1}) \sum_{\gamma \in P(\alpha)} \prod_{a \in \gamma} c_a^{(i)}(\vv{1}).
\label{eq:rhoca}
\end{equation}
The sum is over all set partitions of $\alpha$, denoted $P(\alpha)$,%
\footnote{A partition of a set $A$ is a grouping of the elements of $A$ into one or more non-empty sets using every element exactly once. For example, if $A = \{a,b,c\}$, then $\{\{a\},\{b\},\{c\}\}$, $\{\{a\},\{b,c\}\}$, $\{\{a,b\},\{c\}\}$, $\{\{a,c\},\{b\}\}$, and $\{\{a,b,c\}\}$ are all partitions of $A$. $P(A)$ denotes the set of all possible partitions of $A$. The last partition, into only a single subset $\{A\}$, is called an improper partition.}
and the product is over all sets in each partition.

\subsection{Topological reduction}
As currently defined, $c_\alpha^{(i)}$ is a sum of both reducible and irreducible $z$-graphs, and it is computationally desirable to eliminate the reducible ones. To do so, Wertheim considered the procedure of adding at each field point of an irreducible graph either nothing or another graph with one root point. By the $e_{\rm R}$-bond replacement procedure for an $s$-mer, the root point of the added graph must have bonding complementary to the bonding at the field point of the irreducible graph, so the added graphs can be any $\rho_\gamma^{(i)}$ for $\gamma \subseteq \Gamma_i\setminus\alpha$.%
\footnote{$A \setminus B = \{a \in A | a \notin B \}$ denotes the set difference, i.e., all elements that are in $A$ but not in $B$.}
This naturally leads to definition of the auxiliary singlet densities
\begin{equation}
\sigma_\alpha^{(i)}(\vv{1}) = \sum_{\gamma \subseteq \alpha} \rho_\gamma^{(i)}(\vv{1}),
\label{eq:sigmarho}
\end{equation}
which are the sums of all graphs whose bonded sites at the root point $\vv{1}$ are a subset of $\alpha$. For example, $\sigma_A^{(i)}$ contains all graphs bonded at $A$ (and not bonded elsewhere) or having no bonded sites.%
\footnote{Forgiving some abuse of notation, the label of a single site $A$ should be replaced by a set $\{A\}$ when it represents a set of bonded sites.}
Two special cases are $\sigma_\varnothing^{(i)} = \rho_\varnothing^{(i)}$, the sum of graphs not bonded at any site, and $\sigma_{\Gamma_i}^{(i)} = \rho^{(i)}$, including all possible combinations of sites.
The bonded singlet densities $\rho_\alpha^{(i)}$ can be expressed in terms of $\sigma_\alpha^{(i)}$,%
\footnote{$|A|$ denotes the number of elements in a set $A$.}
\begin{equation}
\rho_\alpha^{(i)}(\vv{1}) = \sum_{\gamma\subseteq\alpha} (-1)^{|\alpha\setminus\gamma|} \sigma_\gamma^{(i)}(\vv{1}),
\label{eq:rhosigma}
\end{equation}
with details in Appendix~\ref{sec:alg}.

Each field point of an irreducible graph that has a set of bonded sites $\alpha$ is now reassigned a factor $\sigma_{\Gamma_i\setminus\alpha}^{(i)}$, which are precisely the graphs \textit{not} bonded at $\alpha$, and $c_\alpha^{(i)}$ can be expressed as a sum of only irreducible graphs. We define $c$ to be the sum of all irreducible graphs consisting of $s$-mer graphs and $f_{\rm R}$-bonds between $s$-mers where all hyperpoints are field points and carry these factors. (Wertheim called this sum $c^{(0)}$, but we will use $c$ to avoid confusion with the type superscripts.) All $c_\alpha^{(i)}$ can be obtained from $c$ by turning a field point bonded at $\alpha$ into a root point labeled $\vv{1}$ and removing its factor $\sigma_{\Gamma_i\setminus\alpha}^{(i)}$. This procedure is given by the functional differentiation,
\begin{equation}
c_\alpha^{(i)}(\vv{1}) = \frac{\delta c}{\delta \sigma_{\Gamma_i\setminus\alpha}^{(i)}(\vv{1})}.
\label{eq:ca}
\end{equation}
The $c_\alpha^{(i)}$ can also be expressed as functions of $\sigma_\alpha^{(i)}$ directly,
\begin{equation}
c_\alpha^{(i)} = -\delta_{|\alpha|,1} - \sum_{\gamma \in P(\alpha)} (-1)^{|\gamma|} (|\gamma|-1)! \prod_{a \in \gamma} \frac{\sigma_a^{(i)}}{\sigma_\varnothing^{(i)}},
\label{eq:casigma}
\end{equation}
for $\alpha \ne \varnothing$ and with $\delta_{i,j}$ being the Kronecker delta. (Note that this corrects a typographical error in Eq.~(27) of Ref.~\citenum{Wertheim:1986a}.) The inverse relationship is
\begin{equation}
\sigma_\alpha^{(i)} = \sigma_\varnothing^{(i)} \sum_{\gamma \in P(\alpha)} \prod_{a \in \gamma} \left(\delta_{|a|,1} + c_a^{(i)}\right).
\label{eq:sigmaca}
\end{equation}
See Appendix~\ref{sec:alg} for details required to arrive at these results.

\subsection{Thermodynamics}
Given the $\sigma_\alpha^{(i)}$ and $c_\alpha^{(i)}$, it can be shown using a Legendre transform of the grand potential that the Helmholtz free energy $A$ of the associating mixture is \cite{Zmpitas:2016}
\begin{align}
\beta A &= \sum_{i=1}^k \int \dd{\vv{1}}\Bigg[ \sigma_{\Gamma_i}^{(i)}(\vv{1}) \ln\left(\Lambda_i^{-1} \sigma_\varnothing^{(i)}(\vv{1})\right) - \sigma_{\Gamma_i}^{(i)}(\vv{1}) \nonumber \\
    &+ \beta u_1^{(i)}(\vv{1}) \sigma_{\Gamma_i}^{(i)}(\vv{1}) + \sum_{\substack{\alpha\subseteq\Gamma_i\\\alpha\ne\varnothing}}
    \sigma_{\Gamma_i\setminus\alpha}^{(i)}(\vv{1})c_\alpha^{(i)}(\vv{1}) \Bigg] - c.
\end{align}
The integrand can be expressed in terms of only $\sigma_\alpha^{(i)}$ using Eq.~\eqref{eq:casigma} to replace $c_\alpha^{(i)}$ (see Appendix~\ref{sec:alg}),
\begin{equation}
\sum_{\substack{\alpha\subseteq\Gamma_i\\\alpha\ne\varnothing}}
    \sigma_{\Gamma_i\setminus\alpha}^{(i)}(\vv{1}) c_\alpha^{(i)}(\vv{1})
= \sigma_{\Gamma_i}^{(i)}(\vv{1}) + Q^{(i)}(\vv{1})
\label{eq:Qca}
\end{equation}
with
\begin{equation}
Q^{(i)}= -\sum_{A \in \Gamma_i} \sigma_{\Gamma_i \setminus A}^{(i)}
    + \sigma_\varnothing^{(i)}
    \sum_{\substack{\gamma \in P(\Gamma_i)\\ \gamma \ne \{\Gamma_i\}}} (-1)^{|\gamma|}(|\gamma|-2)!
    \prod_{a \in \gamma} \frac{\sigma_a^{(i)}}{\sigma_\varnothing^{(i)}},
\label{eq:Q}
\end{equation}
to give
\begin{align}
\beta A = \sum_{i=1}^k &\int \dd{\vv{1}}\Bigg[ \sigma_{\Gamma_i}^{(i)}(\vv{1}) \ln\left(\Lambda_i^{-1} \sigma_\varnothing^{(i)}(\vv{1})\right) \nonumber \\
&+ \beta u_1^{(i)}(\vv{1}) \sigma_{\Gamma_i}^{(i)}(\vv{1}) + Q^{(i)}(\vv{1})\Bigg] - c.
\label{eq:A}
\end{align}
At equilibrium, $A$ should be minimized with respect to variations in $\sigma_\alpha^{(i)}$, subject to a constraint on $\sigma_{\Gamma_i}^{(i)}$ to give a fixed number of molecules. Taking the functional derivative of $A$ with respect to $\sigma_\alpha^{(i)}$ for $\alpha \subset \Gamma_i$ gives%
\footnote{$A \subset B$ denotes $A$ is a proper subset of $B$, i.e., there is at least one element of $B$ that is not in $A$ so $A \ne B$.}
\begin{equation}
0 = \frac{\delta (\beta A)}{\delta \sigma_\alpha^{(i)}(\vv{1})} =
    \delta_{|\alpha|,0} \frac{\sigma_{\Gamma_i}^{(i)}}{\sigma_\varnothing^{(i)}}
    + \frac{\partial Q^{(i)}}{\partial \sigma_\alpha^{(i)}} - c_{\Gamma_i\setminus\alpha}^{(i)}.
\end{equation}
That this relationship holds can be verified by taking derivatives of Eq.~\eqref{eq:Q} and comparing them to Eq.~\eqref{eq:casigma}. Functional differentiation of $A$ with respect to the total singlet density $\sigma_{\Gamma_i}^{(i)}$ gives the chemical potential $\mu_i$ after some manipulation, which is the correct thermodynamic relationship.

The results so far are formally exact, but Wertheim's theory has been most successfully applied as an approximate TPT \cite{Wertheim:1984b,Wertheim:1986b}. In TPT, the free energy is computed relative to a reference system with molecules whose pairwise interactions are given by only $u_{\rm R}^{(ij)}$ and whose free energy $A_{\rm R}$ is more readily computed. The reference system has no attractive bonds and is assumed to have the same total densities $\sigma_{\Gamma_i}^{(i)}$,
\begin{align}
\beta A_{\rm R} = \sum_{i=1}^k &\int \dd{\vv{1}}\Bigg[ \sigma_{\Gamma_i}^{(i)}(\vv{1})
    \ln\left(\Lambda_i^{-1} \sigma_{\Gamma_i}^{(i)}(\vv{1})\right) - \sigma_{\Gamma_i}^{(i)}(\vv{1}) \nonumber \\
    &+ \beta u_1^{(i)}(\vv{1}) \sigma_{\Gamma_i}^{(i)}(\vv{1}) \Bigg] - c_{\rm R},
\label{eq:AR}
\end{align}
where $c_{\rm R}$ is the sum of irreducible graphs for the reference fluid without any bonding, i.e., it contains only $f_{\rm R}$-bonds and should only be a functional of the total density $\sigma_{\Gamma_i}^{(i)}$. By identifying the first term as the ideal free energy, $-c_{\rm R}$ is the excess contribution to $\beta A_{\rm R}$. In practice, $A_{\rm R}$ is usually evaluated with a known equation of state for the reference system; for example, $A_{\rm R}$ for a spatially homogeneous hard-sphere mixture is well approximated by Boubl\'{i}k's equations of state \cite{Boublik:1970}.

The association free energy, $\Delta A = A - A_{\rm R}$, is then
\begin{align}
\beta \Delta A = \sum_{i=1}^k &\int \dd{\vv{1}} \Bigg[ \sigma_{\Gamma_i}^{(i)}(\vv{1})
    \ln\left(\frac{\sigma_\varnothing^{(i)}(\vv{1})}{\sigma_{\Gamma_i}^{(i)}(\vv{1})}\right) \nonumber \\
    &+ \sigma_{\Gamma_i}^{(i)}(\vv{1}) + Q^{(i)}(\vv{1}) \Bigg] - \Delta c.
\label{eq:dA}
\end{align}
The last term $\Delta c = c-c_{\rm R}$ is the key to Wertheim's theory, as it is the contributions to the sum of irreducible graphs from association. Although $\Delta c$ formally contains all irreducible graphs with at least one attractive bond, Wertheim discussed a few cases of ``steric hindrance'' that can simplify the graphs that need to be included. Two particularly convenient ones are:
\begin{enumerate}
\item The formation of an $f_{AB}$-bond between site $A$ on molecule 1 and site $B$ on molecule 2 prevents any site in a third molecule 3 from bonding with either $A$ or $B$. This hindrance usually occurs because of repulsion between the cores of the molecules.
\item Two sites $B$ and $C$ on a molecule 2 cannot bond to the same site $A$ on molecule 1. This can also be enforced by the site geometry or by restricting the model to form exclusive bonds (e.g., covalent chemistry).
\end{enumerate}
Together, these hindrances remove all graphs with multiple bonding of a single site, which greatly reduces the combinations that must be considered. Further simplification can be made by including only a limited subset of irreducible graphs, which we do next.

\section{Double-bond association\label{sec:tptdb}}
TPT1 usually includes the lowest-order irreducible graphs from pair association in $\Delta c$ \cite{Wertheim:1984b,Wertheim:1986b}. This admits at most one attractive bond between two molecules and enforces a tree-like bonded network. To extend TPT1 to include double bonding, we also include the irreducible graphs with two attractive bonds between two molecules,
\begin{align}
\Delta c \approx \frac{1}{2}&\sum_{i=1}^k \sum_{A \in \Gamma_i} \sum_{j=1}^k \sum_{B \in \Gamma_j}
    \int \dd{\vv{1}}\dd{\vv{2}} \bigg[\sigma_{\Gamma_i \setminus A}^{(i)}(\vv{1}) \nonumber \\
    &\times \sigma_{\Gamma_j \setminus B}^{(j)}(\vv{2}) g_{\rm R}^{(ij)}(\vv{1},\vv{2}) f_{AB}^{(ij)}(\vv{1},\vv{2})\bigg] \nonumber \\
+
\frac{1}{2} &\sum_{i=1}^k \sum_{\substack{AB \subseteq \Gamma_i \\ |AB| = 2}} \sum_{j=1}^k \sum_{\substack{CD \subseteq \Gamma_j\\|CD| = 2}}
    \int \dd{\vv{1}}\dd{\vv{2}} \bigg[ \sigma_{\Gamma_i \setminus AB}^{(i)}(\vv{1})  \nonumber \\
    &\times  \sigma_{\Gamma_j \setminus CD}^{(j)}(\vv{2}) g_{\rm R}^{(ij)}(\vv{1},\vv{2}) f_{ABCD}^{(ij)}(\vv{1},\vv{2}) \bigg],
\label{eq:c}
\end{align}
where $g_{\rm R}^{(ij)}$ is the pair correlation function in the reference fluid. The sums over $AB$ and $CD$ are taken over the two-site pairs on $i$ and $j$ (e.g., $AB = \{A,B\}$ is a two-element subset of $\Gamma_i$), and
\begin{equation}
f_{ABCD}^{(ij)} = f_{AC}^{(ij)} f_{BD}^{(ij)} + f_{AD}^{(ij)} f_{BC}^{(ij)}
\end{equation}
accounts for the two ways sites in $AB$ and $CD$ can associate with each other. The first term in Eq.~\eqref{eq:c} is standard in TPT1 and accounts for all ways a single bond can form between two molecules, while the second term accounts for all ways a double bond can form. In practice, some or all of these graphs may be zero or negligible for a given model due to the bonding chemistry and/or site geometry but we do not exclude any of these at this stage.

In expressing the graph sum in this way, we have neglected any rings (cycles) in the graph sum beyond the double bond, although all such graphs are irreducible \cite{Sear:1994b,Sear:1996}. Hence, the overall bonding network is still a tree of single and double bonds between molecules. Rings of molecules bonded at two sites can be included within the framework we develop here using the approximation of Sear and Jackson \cite{Sear:1994b}; we will revisit this point in Section~\ref{sec:colllink}. We also neglect effects of steric hindrance that prevent bonding at one site when a bond forms at another; these effects might be treated by including additional graphs or with a renormalization approach \cite{Wertheim:1987,Marshall:2013a,Marshall:2013b}.

We now differentiate Eq.~\eqref{eq:c} with respect to $\sigma_\alpha^{(i)}$, giving
\begin{align}
\Delta c_A^{(i)}(\vv{1})
    = \sum_{j=1}^k \sum_{B \in \Gamma_j} &\int \dd{\vv{2}} \bigg[\sigma_{\Gamma_j \setminus B}^{(j)}(\vv{2}) g_{\rm R}^{(ij)}(\vv{1},\vv{2})
    f_{AB}^{(ij)}(\vv{1},\vv{2})\bigg]
\label{eq:dcA}
\end{align}
and
\begin{equation}
\Delta c_{AB}^{(i)} = \sum_{j=1}^k \sum_{\substack{CD \subseteq \Gamma_j\\|CD| = 2}}
\int \dd{\vv{2}} \Bigg[\sigma_{\Gamma_j \setminus CD}^{(j)}(\vv{2})
    g_{\rm R}^{(ij)}(\vv{1},\vv{2}) f_{ABCD}^{(ij)}(\vv{1},\vv{2}) \Bigg],
\label{eq:dcAB}
\end{equation}
so
\begin{align}
\Delta c \approx \frac{1}{2}&\sum_{i=1}^k \sum_{A \in \Gamma_i}
    \int \dd{\vv{1}} \bigg[\sigma_{\Gamma_i\setminus A}^{(i)}(\vv{1}) \Delta c_A^{(i)}(\vv{1})\bigg] \nonumber \\
+ \frac{1}{2} &\sum_{i=1}^k \sum_{\substack{AB \subseteq \Gamma_i \\ |AB| = 2}}
    \int \dd{\vv{1}} \bigg[\sigma_{\Gamma_i\setminus AB}^{(i)}(\vv{1}) \Delta c_{AB}^{(i)}(\vv{1})\bigg].
\end{align}
The remaining $\Delta c_\alpha^{(i)}$ for $\alpha \subset \Gamma_i$ are zero. Because $c_{\rm R}$ is only a functional of $\sigma_{\Gamma_i}^{(i)}$, we can replace $c_\alpha^{(i)}$ by $\Delta c_\alpha^{(i)}$ in Eq.~\eqref{eq:Qca},
\begin{equation}
\sigma_{\Gamma_i}^{(i)} + Q^{(i)} = \sum_{A \in \Gamma_i} \sigma_{\Gamma_i\setminus A}^{(i)} \Delta c_A^{(i)}
+ \sum_{\substack{AB \subseteq \Gamma_i \\ |AB| = 2}} \sigma_{\Gamma_i\setminus AB}^{(i)} \Delta c_{AB}^{(i)}.
\end{equation}
We may further express $\Delta c_A^{(i)}$ and $\Delta c_{AB}^{(i)}$ in terms of $\sigma_\alpha^{(i)}$ using Eq.~\eqref{eq:casigma},
\begin{equation}
\Delta c_A^{(i)} = -1 + \frac{\sigma_{A}^{(i)}}{\sigma_\varnothing^{(i)}}
\label{eq:dcAsigma}
\end{equation}
and
\begin{equation}
\Delta c_{AB}^{(i)} = \frac{\sigma_{AB}^{(i)}}{\sigma_\varnothing^{(i)}} - \frac{\sigma_A^{(i)}\sigma_B^{(i)}}{(\sigma_\varnothing^{(i)})^2}.
\label{eq:dcABsigma}
\end{equation}
These results can be directly substituted in Eq.~\eqref{eq:dA} to define $\Delta A$ using only $\sigma_\alpha^{(i)}$; however, this system of equations is still incomplete for specifying all $\sigma_\alpha^{(i)}$ (and hence $\Delta A$).

We use Eq.~\eqref{eq:sigmaca} to find a relationship between $\sigma_\alpha^{(i)}$ in terms of $\Delta c_A^{(i)}$ and $\Delta c_{AB}^{(i)}$ given that all other $\Delta c_\alpha^{(i)}$ are zero for $\alpha \subset \Gamma_i$:
\begin{align}
\sigma_\alpha^{(i)}
&= \sigma_\varnothing^{(i)}\sum_{\substack{\gamma\in P(\alpha)\\|a| \le 2 \forall a \in \gamma}} \prod_{\substack{A \in \gamma\\|A| = 1}} \left(1+\Delta c_A^{(i)}\right)
   \prod_{\substack{AB \in \gamma\\|AB| = 2}} \Delta c_{AB}^{(i)}.
\label{eq:sigmacadb}
\end{align}
The sum is taken over all partitions of $\alpha$ having only one-element or two-element subsets, and the products are taken over each type of subset; all other partitions do not contribute. For $|\alpha| \ge 3$, Eq.~\eqref{eq:sigmacadb} can be restated as a recursive relationship,
\begin{equation}
\sigma_\alpha^{(i)} = \frac{\sigma_A^{(i)} \sigma_{\alpha\setminus A}^{(i)}}{\sigma_\varnothing^{(i)}}
    + \sum_{B \in \alpha\setminus A} \Bigg[\frac{\sigma_{AB}^{(i)}}{\sigma_\varnothing^{(i)}} - \frac{\sigma_A^{(i)}\sigma_B^{(i)}}{(\sigma_\varnothing^{(i)})^2}\Bigg] \sigma_{\alpha\setminus AB}^{(i)},
\label{eq:sigmacadbrecur}
\end{equation}
after substitution of Eqs.~\eqref{eq:dcAsigma} and \eqref{eq:dcABsigma}. The first term accounts for $A$ being a one-element subset in the first product of Eq.~\eqref{eq:sigmacadb}, which contributes a factor of $1 + \Delta c_A^{(i)}$; this factor then multiplies $\sigma_{\alpha\setminus A}^{(i)}$, which includes all partitions of $\alpha\setminus A$. The second term accounts for all two-element subsets containing $A$, which contribute a factor of $\Delta c_{AB}^{(i)}$ in the second product of Eq.~\eqref{eq:sigmacadb} for all $B \in \alpha\setminus A$. This must multiply $\sigma_{\alpha\setminus AB}^{(i)}$, which includes all remaining partitions of $\alpha\setminus AB$. For example, if $\alpha = ABC = \{A,B,C\}$ and we remove $A$, we obtain
\begin{align}
\sigma_{ABC}^{(i)} &= \frac{\sigma_A^{(i)} \sigma_{BC}^{(i)}}{\sigma_\varnothing^{(i)}}
    + \Bigg[\frac{\sigma_{AB}^{(i)}}{\sigma_\varnothing^{(i)}} - \frac{\sigma_A^{(i)}\sigma_B^{(i)}}{(\sigma_\varnothing^{(i)})^2}\Bigg] \sigma_C^{(i)} \nonumber \\
    &+ \Bigg[\frac{\sigma_{AC}^{(i)}}{\sigma_\varnothing^{(i)}} - \frac{\sigma_A^{(i)}\sigma_C^{(i)}}{(\sigma_\varnothing^{(i)})^2}\Bigg] \sigma_B^{(i)} \nonumber \\
&= \frac{\sigma_A^{(i)} \sigma_{BC}^{(i)}}{\sigma_\varnothing^{(i)}} + \frac{\sigma_B^{(i)} \sigma_{AC}^{(i)}}{\sigma_\varnothing^{(i)}} + \frac{\sigma_C^{(i)} \sigma_{AB}^{(i)}}{\sigma_\varnothing^{(i)}} - 2 \frac{\sigma_A^{(i)} \sigma_B^{(i)} \sigma_C^{(i)}}{(\sigma_\varnothing^{(i)})^2}.
\end{align}
As expected, the same expression is obtained if site $B$ or $C$ is initially remove from $\alpha$ rather than $A$. The partitions of $\alpha$ having only one-element or two-element subsets are $\{\{A\},\{B,C\}\}$, $\{\{B\},\{A,C\}\}$, $\{\{C\},\{A,B\}\}$, and $\{\{A\},\{B\},\{C\}\}$, so Eq.~\eqref{eq:sigmacadb} is also equivalent after substitution and some simplification.

We now recast these equations using the notation of Chapman and coworkers \cite{Chapman:1986,Joslin:1987,Jackson:1988,Chapman:1988}, and define $X_\alpha^{(i)}(\vv{1}) = \sigma_{\Gamma_i\setminus\alpha}^{(i)}(\vv{1}) / \sigma_{\Gamma_i}^{(i)}(\vv{1})$ as the local fraction of molecules of type $i$ not bonded at all sites in $\alpha$. (Note the special case $X_\varnothing^{(i)}(\vv{1}) = 1$.) The association free energy is
\begin{align}
\beta \Delta A = & \sum_{i=1}^k \int \dd{\vv{1}} \sigma_{\Gamma_i}^{(i)}(\vv{1}) \Bigg[\ln X_{\Gamma_i}^{(i)}(\vv{1}) \nonumber \\
    +\frac{1}{2} &\sum_{A\in\Gamma_i} X_A^{(i)}(\vv{1}) \Bigg(\frac{X_{\Gamma_i\setminus A}^{(i)}(\vv{1})}{X_{\Gamma_i}^{(i)}(\vv{1})} - 1\Bigg) \nonumber \\
    + \frac{1}{2} &\sum_{\substack{AB \subseteq \Gamma_i \\ |AB| = 2}} X_{AB}^{(i)}(\vv{1}) \Bigg(\frac{X_{\Gamma_i\setminus AB}^{(i)}(\vv{1})}{X_{\Gamma_i}^{(i)}(\vv{1})} - \frac{X_{\Gamma_i\setminus A}^{(i)}(\vv{1}) X_{\Gamma_i\setminus B}^{(i)}(\vv{1})}{(X_{\Gamma_i}^{(i)}(\vv{1}))^2}\Bigg)\Bigg],
\label{eq:dAXint}
\end{align}
while the ``mass action'' or ``chemical equilibrium'' conditions from $\Delta c_\alpha^{(i)}$ are
\begin{align}
&-1 + \frac{X_{\Gamma_i\setminus A}^{(i)}(\vv{1})}{X_{\Gamma_i}^{(i)}(\vv{1})} = \sum_{j=1}^k \sum_{B\in\Gamma_j} \nonumber \\
& \int \dd{\vv{2}} \bigg[\sigma_{\Gamma_j}^{(j)}(\vv{2}) X_B^{(j)}(\vv{2}) g_{\rm R}^{(ij)}(\vv{1},\vv{2})
    f_{AB}^{(ij)}(\vv{1},\vv{2})\bigg]
\label{eq:XAint}
\end{align}
and
\begin{align}
&\frac{X_{\Gamma_i\setminus AB}^{(i)}(\vv{1})}{X_{\Gamma_i}^{(i)}(\vv{1})} - \frac{X_{\Gamma_i\setminus A}^{(i)}(\vv{1}) X_{\Gamma_i\setminus B}^{(i)}(\vv{1})}{(X_{\Gamma_i}^{(i)}(\vv{1}))^2} = \sum_{j=1}^k \sum_{\substack{CD \subseteq \Gamma_j\\|CD| = 2}} \nonumber \\
& \int \dd{\vv{2}} \bigg[\sigma_{\Gamma_j}^{(j)}(\vv{2}) X_{CD}^{(j)}(\vv{2})
    g_{\rm R}^{(ij)}(\vv{1},\vv{2}) f_{ABCD}^{(ij)}(\vv{1},\vv{2}) \bigg].
\label{eq:XABint}
\end{align}

If the mixture is spatially homogeneous, none of the singlet densities (or $X_\alpha^{(i)}$) depends on position or orientation. Carrying out the integral in Eq.~\eqref{eq:dAXint} gives the association free-energy density $\Delta a = \Delta A / V$ in terms of the number densities $\rho_i$ and $X_\alpha^{(i)}$
\begin{align}
\beta \Delta a = &\sum_{i=1}^k \rho_i\Bigg[\ln X_{\Gamma_i}^{(i)} + \frac{1}{2} \sum_{A\in\Gamma_i} X_A^{(i)} \Bigg(\frac{X_{\Gamma_i \setminus A}^{(i)}}{X_{\Gamma_i}^{(i)}} - 1\Bigg) \nonumber \\
&+ \frac{1}{2} \sum_{\substack{AB \subseteq \Gamma_i \\ |AB| = 2}} X_{AB}^{(i)} \Bigg(\frac{X_{\Gamma_i \setminus AB}^{(i)}}{X_{\Gamma_i}^{(i)}} - \frac{X_{\Gamma_i \setminus A}^{(i)} X_{\Gamma_i \setminus B}^{(i)}}{(X_{\Gamma_i}^{(i)})^2}\Bigg)\Bigg],
\label{eq:dAX}
\end{align}
integrating Eq.~\eqref{eq:XAint} over $\vv{1}$ gives
\begin{equation}
-1 + \frac{X_{\Gamma_i \setminus A}^{(i)}}{X_{\Gamma_i}^{(i)}} = \sum_{j=1}^k \rho_j \sum_{B\in\Gamma_j} X_B^{(j)} \Delta_{AB}^{(ij)}
\label{eq:XA}
\end{equation}
with
\begin{align}
\Delta_{AB}^{(ij)} = (\Omega_i \Omega_j)^{-1} &\int\dd\vv{r}_{12} \dd\vvg{\Omega}_1 \dd\vvg{\Omega}_2 \bigg[g_{\rm R}^{(ij)}(\vv{r}_{12},\vvg{\Omega}_1,\vv{\Omega}_2) \nonumber \\
&\times f_{AB}^{(ij)}(\vv{r}_{12},\vvg{\Omega}_1,\vv{\Omega}_2)\bigg],
\label{eq:deltaAB}
\end{align}
and integrating Eq.~\eqref{eq:XABint} over $\vv{1}$ gives
\begin{equation}
\frac{X_{\Gamma_i\setminus AB}^{(i)}}{X_{\Gamma_i}^{(i)}} - \frac{X_{\Gamma_i\setminus A}^{(i)} X_{\Gamma_i\setminus B}^{(i)}}{(X_{\Gamma_i}^{(i)})^2} = \sum_{j=1}^k \rho_j \sum_{\substack{CD \subseteq \Gamma_j \\ |CD| = 2}} X_{CD}^{(j)} \Delta_{ABCD}^{(ij)}
\label{eq:XAB}
\end{equation}
with
\begin{align}
\Delta_{ABCD}^{(ij)} = (\Omega_i \Omega_j)^{-1} &\int\dd\vv{r}_{12} \dd\vvg{\Omega}_1 \dd\vvg{\Omega}_2 \bigg[g_{\rm R}^{(ij)}(\vv{r}_{12},\vvg{\Omega}_1,\vv{\Omega}_2) \nonumber \\
&\times f_{ABCD}^{(ij)}(\vv{r}_{12},\vvg{\Omega}_1,\vv{\Omega}_2)\bigg],
\label{eq:deltaABCD}
\end{align}
where $\vv{r}_{12} = \vv{r}_2 - \vv{r}_1$ is the separation between molecules and $\Omega_i$ is the integral over orientations of a molecule of component $i$. Here, we have used that $g_{\rm R}^{(ij)}$ should only depend on the relative separation between molecules in a spatially homogeneous mixture. $\Delta_{AB}^{(ij)}$ and $\Delta_{ABCD}^{(ij)}$ are then ``bond volumes'' integrated over separations and averaged over all orientations of the two molecules \cite{Jackson:1988,Chapman:1988}.

Equation~\eqref{eq:sigmacadb} can also be restated as
\begin{align}
X_{\Gamma_i\setminus\alpha}^{(i)} &= X_{\Gamma_i}^{(i)} \sum_{\substack{\gamma\in P(\alpha)\\|a| \le 2 \forall a \in \gamma}} \prod_{\substack{A \in \gamma\\|A| = 1}} \left(\frac{X_{\Gamma_i\setminus A}^{(i)}}{X_{\Gamma_i}^{(i)}}\right) \nonumber \\
&\times \prod_{\substack{AB \in \gamma\\|AB| = 2}} \Bigg[\frac{X_{\Gamma_i\setminus AB}^{(i)}}{X_{\Gamma_i}^{(i)}} - \frac{X_{\Gamma_i\setminus A}^{(i)} X_{\Gamma_i\setminus B}^{(i)}}{(X_{\Gamma_i}^{(i)})^2}\Bigg],
\label{eq:Xdb}
\end{align}
and equivalently, as a recursive relationship,
\begin{align}
X_{\Gamma_i\setminus\alpha}^{(i)}
&= \frac{X_{\Gamma_i\setminus A}^{(i)} X_{\Gamma_i\setminus(\alpha\setminus A)}^{(i)}}{X_{\Gamma_i}^{(i)}} \nonumber \\
&+ \sum_{B\in\alpha\setminus A} \Bigg[\frac{X_{\Gamma_i\setminus AB}^{(i)}}{X_{\Gamma_i}^{(i)}} - \frac{X_{\Gamma_i\setminus A}^{(i)} X_{\Gamma_i\setminus B}^{(i)}}{(X_{\Gamma_i}^{(i)})^2}\Bigg] X_{\Gamma_i\setminus(\alpha\setminus AB)}^{(i)}.
\label{eq:Xdbrecur}
\end{align}
These relationships allow $X_A^{(i)}$ and $X_{AB}^{(i)}$ to be computed in terms of $X_{\Gamma_i}^{(i)}$ and all the $X_{\Gamma_i\setminus A}^{(i)}$ and $X_{\Gamma_i\setminus AB}^{(i)}$, and they also self-consistently define $X_{\Gamma_i}^{(i)}$. Taken together, Eqs.~\eqref{eq:XA}, \eqref{eq:XAB}, and \eqref{eq:Xdb} comprise a system of nonlinear equations that can be numerically solved to obtain all $X_\alpha^{(i)}$, with Eq.~\eqref{eq:dAX} giving the free energy; in practice, this may be challenging for certain parameter values. There are, however, two cases that simplify considerably: (1) no double bonds can form and (2) a double bond can form at only one pair of sites on each molecule. We now consider these in turn before applying the general theory to a colloid--linker mixture in Section~\ref{sec:colllink}.

\subsection{No double bonds\label{sec:tptdb:nodb}}
In TPT1, at most one bond can form between two molecules \cite{Wertheim:1984b,Wertheim:1986b}, which is a reasonable approximation when the site geometry on the molecules inhibits double bonding, e.g., patchy colloids with two bonding sites on opposite hemispheres, and the formation of bonded rings of molecules is unlikely, e.g., because the bonding sites are rigidly fixed. (This amounts to setting $\Delta_{ABCD}^{(ij)} = 0$ for all pairs of sites on all molecules.) TPT1 is the foundation of SAFT \cite{Chapman:1986,Joslin:1987,Jackson:1988,Chapman:1988,Muller:2001}, so we will show that we recover the SAFT equations originally derived by Chapman for multicomponent mixtures with multiple bonding sites \cite{Chapman:1988} if we neglect double bonding between molecules.

Making the simplification that $\Delta_{ABCD}^{(ij)} = 0$, Eq.~\eqref{eq:Xdbrecur} becomes
\begin{equation}
X_{\Gamma_i\setminus\alpha}^{(i)}
= \frac{X_{\Gamma_i\setminus A}^{(i)} X_{\Gamma_i\setminus(\alpha\setminus A)}^{(i)}}{X_{\Gamma_i}^{(i)}},
\end{equation}
for any $A \in \alpha$, which when evaluated for $\alpha = \Gamma_i$ gives
\begin{equation}
1 = \frac{X_A^{(i)} X_{\Gamma_i\setminus A}^{(i)}}{X_{\Gamma_i}^{(i)}}
\end{equation}
and it follows that
\begin{equation}
X_{\Gamma_i}^{(i)} = \prod_{A \in \Gamma_i} X_A^{(i)}.
\end{equation}
This relationship captures the statistical independence of bonding in TPT1: the fraction of molecules of component $i$ that are not bonded at \textit{any} site is the product of the fraction not bonded at \textit{each} site. The free energy
\begin{equation}
\beta \Delta a = \sum_{i=1}^k \rho_i \sum_{A \in \Gamma_i} \left[ \ln X_A^{(i)}
  + \frac{1}{2} \left( 1 - X_A^{(i)}\right) \right]
\end{equation}
and chemical equilibrium equations
\begin{equation}
X_A^{(i)} = \left[1 + \sum_{j=1}^k \rho_j \sum_{B \in \Gamma_j} X_B^{(j)} \Delta_{AB}^{(ij)} \right]^{-1}
\label{eq:XAnodb}
\end{equation}
follow by substitution and are equivalent to Eqs.~(3) and (4) of Ref.~\citenum{Chapman:1988} after a conversion to express $\Delta A$ per molecule.

\subsection{One double bond\label{sec:tptdb:onedb}}
As another special case, suppose that exactly two sites $A_{12} = \{A_1,A_2\}$ on a molecule of component $i$ can form a double bond with a similar set of two sites on a molecule of component $j$. (We will label all of these sites $A_1$ and $A_2$, but $A_1$ on component $i$ and $A_1$ on component $j$ need not be the same, and similarly for $A_2$.) This restriction may physically correspond to a molecular geometry where only two of the sites are close enough to double bond with another molecule, which was studied by Marshall and Chapman for a one-component fluid \cite{Marshall:2013b} and by Marshall for a multicomponent mixture \cite{Marshall:2018}, or a restrictive bonding chemistry.

The immediate consequence is that only $\Delta_{A_{12} A_{12}}^{(ij)} \ne 0$, which drastically simplifies the expressions for $\sigma_\alpha^{(i)}$ because a maximum of two partitions of $\alpha$ contribute to Eq.~\eqref{eq:Xdb}---the partition into all singleton sets and the partition containing one two-element subset $A_{12}$---so
\begin{equation}
X_A^{(i)} = \begin{cases}
\dfrac{X_{\Gamma_i}^{(i)}}{X_{\Gamma_i\setminus A}^{(i)}},& A \notin A_{12} \\
\dfrac{X_{A_{12}}^{(i)} X_{\Gamma_i\setminus (A_{12}\setminus A)}^{(i)}}{X_{\Gamma_i}^{(i)}},& A \in A_{12}
\end{cases}.
\label{eq:XdbA12}
\end{equation}
The first case, when $A$ is not one of the double-bond sites, is obtained from Eq.~\eqref{eq:Xdbrecur} with $\alpha = \Gamma_i$ and is the same as in TPT1. The second case, when $A$ is one of the double-bond sites, is obtained from Eq.~\eqref{eq:Xdbrecur} with $\alpha = \Gamma_i \setminus A_1$ and $\Gamma_i \setminus A_2$ and removing $A_2$ and $A_1$, respectively. Combining Eq.~\eqref{eq:XdbA12} with Eq.~\eqref{eq:Xdbrecur} for $\alpha = \Gamma_i$ and either $A \in A_{12}$ removed gives the additional relationship
\begin{equation}
X_{A_{12}}^{(i)} \Bigg(\frac{X_{\Gamma_i\setminus A_{12}}^{(i)}}{X_{\Gamma_i}^{(i)}} - \frac{X_{\Gamma_i\setminus A_1}^{(i)} X_{\Gamma_i\setminus A_2}^{(i)}}{(X_{\Gamma_i}^{(i)})^2}\Bigg) = 1 - \frac{X_{A_1}^{(i)} X_{A_2}^{(i)}}{X_{A_{12}}^{(i)}}.
\end{equation}
Finally, it follows from applying Eq.~\eqref{eq:XdbA12} recursively that
\begin{equation}
X_{\Gamma_i}^{(i)} = X_{A_{12}}^{(i)} \prod_{A \in \Gamma_i\setminus A_{12}} X_A^{(i)}.
\end{equation}

With these results, the free energy is
\begin{align}
\beta \Delta a &= \sum_{i=1}^k \rho_i \Bigg[\sum_{A \in \Gamma_i} \left( \ln X_A^{(i)} + \frac{1}{2} \left( 1 - X_A^{(i)}\right) \right) \nonumber \\
&+ \ln\left(\frac{X_{A_{12}}^{(i)}}{X_{A_1}^{(i)} X_{A_2}^{(i)}}\right) - \frac{1}{2}\left(1-\frac{X_{A_1}^{(i)} X_{A_2}^{(i)}}{X_{A_{12}}^{(i)}}\right) \Bigg].
\label{eq:Adb1X}
\end{align}
The free energy has been written so that the sum over sites $A$ is the same as TPT1, while the additional terms give the free energy due to double-bond formation. This expression is equivalent to Eq.~(19) of Ref.~\citenum{Marshall:2018} after some manipulations. The corresponding chemical equilibrium equations for $A \notin A_{12}$ are the same as Eq.~\eqref{eq:XAnodb}, while for $\alpha \subseteq A_{12}$, the three $X_\alpha^{(i)}$ are given by the system of equations
\begin{equation}
-1 + \frac{X_{A_2}^{(i)}}{X_{A_{12}}^{(i)}} = \sum_{j=1}^k \rho_j \sum_{B \in \Gamma_j} X_B^{(j)} \Delta_{A_1 B}^{(ij)},
\end{equation}
a similar equation with the labels $A_1$ and $A_2$ reversed, and
\begin{equation}
\frac{1}{X_{A_{12}}^{(i)}} - \frac{X_{A_1}^{(i)} X_{A_2}^{(i)}}{(X_{A_{12}}^{(i)})^2} = \sum_{j=1}^k \rho_j X_{A_{12}}^{(j)} \Delta_{A_{12} A_{12}}^{(ij)}.
\end{equation}

\section{Colloid--Linker Mixture\label{sec:colllink}}
Having extended Wertheim's TPT to include general double-bond association, we now apply it to a model mixture of colloids and flexible linker molecules that we previously studied with molecular simulations and TPT1 \cite{Howard:2019}. In that work, we aimed to predict the phase behavior of the homogeneous mixture; however, TPT also gives valuable information about the prevalence of different bonding motifs. We measured a significant fraction of linkers forming loops (a double bond with both linker ends attached to the same colloid) in our simulations. Linker loops are neglected in TPT1, so here we investigate the linker loop fraction under different conditions as a useful test of the new theory.

\subsection{Model}
The linkers we studied were linear chains of $M = 8$ tangentially bonded hard-sphere segments with diameter $d_{\rm l}$, which we also called ``polymers'' in Ref.~\citenum{Howard:2019}, while the colloids were hard spheres with diameter $d_{\rm c} = 5\,d_{\rm l}$. The mixture composition was defined using the colloid volume fraction $\eta_{\rm c} = \rho_{\rm c} \pi d_{\rm c}^3/6$, which we varied from 0.01 to 0.15, and the linker-to-colloid number ratio $\rho_{\rm l}/\rho_{\rm c}$, which we fixed at 1.5. Each linker had $n_{\rm l} = 2$ bonding sites, one at the center of each end segment. Each colloid had $n_{\rm c} = 6$ bonding sites arranged at the vertices of an octahedron at a distance $d_{\rm cl}^* = d_{\rm cl} + (2^{1/6}-1)d_{\rm l} \approx 3.12\,d_{\rm l}$ from the center of the colloid, where $d_{\rm cl} = (d_{\rm c}+d_{\rm l})/2 = 3\,d_{\rm l}$ is the hard-sphere contact distance for a colloid and a linker segment. (Placement of the colloid bonding sites at $d_{\rm cl}^*$ rather than $d_{\rm cl}$ was due to our original work using nearly hard potentials for the spheres \cite{Weeks:1971}.) A linker could reversibly bond to a colloid through a short-ranged site attraction,
\begin{equation}
u_{AB}^{({\rm cl})}(\vv{1},\vv{2}) = \begin{cases}
-\varepsilon e^{-(r_{AB}/\ell)^2},& r_{AB} < 2.5\,\ell \\
0,& r_{AB} \ge 2.5\,\ell
\end{cases},
\label{eq:ugauss}
\end{equation}
where $r_{AB} = |\vv{r}_B^{({\rm l})}(\vv{2}) - \vv{r}_A^{({\rm c})}(\vv{1})|$ is the distance between the linker and colloid sites, $\varepsilon$ is the strength of the attraction, and $\ell = 0.2\,d_{\rm l}$ sets the range of the attraction. The potential was truncated for $r_{AB} \ge 2.5\,\ell = 0.5\,d_{\rm l}$. There was at most one bond at each site, and there were no bonds between two colloid sites or two linker sites.

\subsection{Perturbation theory with molecule flexibility}
We developed the theory in Sections~\ref{sec:wt} and \ref{sec:tptdb} with all sites fixed on rigid molecules, but in our colloid--linker model, the linker bonding sites fluctuate with the end-to-end vector $\vv{R}$. This intramolecular flexibility can be incorporated in the theory with a few modifications and approximations \cite{Wertheim:1987,Sear:1994a,Sear:1994b}. To specify the linker's conformation, we redefined $\vv{1} = (\vv{r}_1,\vv{R}_1)$ where $\vv{r}_1$ is the position of one end of the chain and $\vv{R}_1$ is its end-to-end vector, effectively coarse-graining the internal linker segments. To appropriately weight the conformations, the linkers were given an intramolecular potential
\begin{equation}
\beta u_1^{({\rm l})}(\vv{R}) \approx -\ln p(\vv{R}),
\end{equation}
where $p(\vv{R})$ was the probability distribution of end-to-end vectors in the reference mixture without bonding. The linker singlet densities now depend on $\vv{R}$, and states where both ends of a chain are bonded may have a different dependence on $\vv{R}$ than the unbonded state. As a result, $\sigma_{\Gamma_{\rm l}}^{({\rm l})}$ may have a different $\vv{R}$-dependence in the associating mixture than the reference mixture, and computing $A_{\rm R}$ at the same $\sigma_{\Gamma_{\rm l}}^{({\rm l})}$ may be inconvenient for equations of state that are functionals of the number density. Wertheim showed how this can be circumvented in certain cases by choosing a reference mixture having the same number density after integration over internal degrees of freedom and making a small modification to Eq.~\eqref{eq:dA} \cite{Wertheim:1987}. We accordingly take the reference state to be the hard-chain mixture at the same number density and without bonding ($\beta\varepsilon = 0$).

With these modifications, we specialized the graph sum given by Eq.~\eqref{eq:c} to our model. We first reduced the number of $\sigma_\alpha^{({\rm c})}$ that needed to be explicitly included. For a linker to double bond, the arc-length distance between colloid bonding sites must be less than the linker's contour length $L = (M-1)d_{\rm l} = 7\,d_{\rm l}$. In our model, the arc-length distance between nearest sites on the colloid was $d_{\rm cl}^* \pi/2 \approx 4.9\,d_{\rm l}$, so the linker was able to form double bonds with any of these pairs of sites. However, the arc-length distance between sites on opposite hemispheres was $d_{\rm cl}^* \pi \approx 9.8\,d_{\rm l}$, so a linker could not form a double bond with these sites. We accordingly excluded double bonds between the three pairs of sites on opposite sides of the colloid; the remaining $\nu_{\rm c} = n_{\rm c}(n_{\rm c}-2)/2$ double-bond pairs were all equivalent. We then defined the graph sum using representative sets of colloid bonding sites, $A_1$ for a single bond and $A_{12}=\{A_1,A_2\}$ for a double bond between nearest sites. The linker had $n_{\rm l} = 2$ equivalent bonding sites $B_{12} = \{B_1,B_2\}$.

We next simplified the chemical equilibrium equations for a spatially homogeneous mixture. The linker singlet densities were constant with respect to $\vv{r}$ but not $\vv{R}$, so we redefined $X_\alpha^{({\rm l})}(\vv{r}_1) = \sigma_{\Gamma_{\rm l}\setminus\alpha}^{({\rm l})}(\vv{r}_1)/\sigma_{\Gamma_{\rm l}}^{({\rm l})}(\vv{r}_1)$ for the linkers using singlet densities integrated over $\vv{R}_1$. We further assumed that the distribution of end-to-end vectors was the same for linkers having neither or one end bonded and with both given by $p(\vv{R})$, i.e., $\rho_\varnothing^{({\rm l})}(\vv{1}) = \rho_\varnothing^{({\rm l})}(\vv{r}_1)p(\vv{R}_1)$ and $\rho_{B_1}^{({\rm l})}(\vv{1}) = \rho_{B_1}^{({\rm l})}(\vv{r}_1)p(\vv{R}_1)$. Carrying through the integration and explicitly specializing for the difunctional linker gave
\begin{align}
-1 + \frac{X_{\Gamma_{\rm c}\setminus A_1}^{({\rm c})}}{X_{\Gamma_{\rm c}}^{({\rm c})}} &= 2 \rho_{\rm l} X_{B_1}^{({\rm l})} \Delta_1
\label{eq:eq1c} \\
-1 + \frac{X_{B_1}^{({\rm l})}}{X_{B_{12}}^{({\rm l})}} &= n_{\rm c} \rho_{\rm c} X_{A_1}^{({\rm c})} \Delta_1
\label{eq:eq1p}
\end{align}
and
\begin{align}
\frac{X_{\Gamma_{\rm c}\setminus A_{12}}^{({\rm c})}}{X_{\Gamma_{\rm c}}^{({\rm c})}} - \left(\frac{X_{\Gamma_{\rm c}\setminus A_1}^{({\rm c})}}{X_{\Gamma_{\rm c}}^{({\rm c})}}\right)^2 &= 2 \rho_{\rm l} X_{B_{12}}^{({\rm l})} \Delta_2
\label{eq:eq2c} \\
\frac{1}{X_{B_{12}}^{({\rm l})}} - \left(\frac{X_{B_1}^{({\rm l})}}{X_{B_{12}}^{({\rm l})}}\right)^2 &= 2 \nu_{\rm c} \rho_{\rm c} X_{A_{12}}^{({\rm c})} \Delta_2
\label{eq:eq2p}
\end{align}
with single-bond volume
\begin{align}
\Delta_1 \approx \Omega_{\rm c}^{-1} &\int \dd\vv{r}_{12} \dd\vvg{\Omega}_1 \dd\vv{R}_2  \bigg[p(\vv{R}_2) g_{\rm R}^{({\rm cl})}(\vv{r}_{12},\vv{R}_2) \nonumber \\
&\times f_{A_1 B_1}^{({\rm cl})}(\vv{r}_{12},\vvg{\Omega}_1) \bigg]
\label{eq:delta2}
\end{align}
and double-bond volume
\begin{align}
\Delta_2 \approx \Omega_{\rm c}^{-1} &\int \dd\vv{r}_{12} \dd\vvg{\Omega}_1 \dd\vv{R}_2 \bigg[p(\vv{R}_2) g_{\rm R}^{({\rm cl})}(\vv{r}_{12},\vv{R}_2) \nonumber \\
&\times f_{A_1 B_1}^{({\rm cl})}(\vv{r}_{12},\vvg{\Omega}_1) f_{A_2 B_2}^{({\rm cl})}(\vv{r}_{12},\vvg{\Omega}_1,\vv{R}_2) \bigg].
\label{eq:delta4}
\end{align}
The factors of 2 in Eqs.~\eqref{eq:eq2c} and \eqref{eq:eq2p} account for equivalent permutations of the site labels. Additionally, Eq.~\eqref{eq:Xdb} gives $X_{A_1}^{({\rm c})}$ and $X_{A_{12}}^{({\rm c})}$ as functions of $X_{\Gamma_{\rm c}\setminus A_{12}}^{({\rm c})}$, $X_{\Gamma_{\rm c}\setminus A_1}^{({\rm c})}$, and $X_{\Gamma_{\rm c}}^{({\rm c})}$, along with an implicit equation for $X_{\Gamma_{\rm c}}^{({\rm c})}$ in these variables. Together, these make a complete set of nonlinear equations in five variables that can be solved for a given mixture composition and attraction strength $\varepsilon$ after evaluation of $\Delta_1$ and $\Delta_2$. Complete details of how we evaluated these integrals and solved these equations are given in Appendix~\ref{sec:bvint}.

\subsection{Pair correlation function}
The pair correlation function between the colloids and linkers in the reference system $g_{\rm R}^{({\rm cl})}(\vv{r}_{12},\vv{R}_2)$ depends on not only the distance between the colloid and the linker $\vv{r}_{12}$ but also the linker's conformation $\vv{R}_2$. We considered approximating $g_{\rm R}^{({\rm cl})}$ using a superposition of the hard-sphere correlations $g_{\rm hs}^{({\rm cl})}$ in the mixture obtained by dissolving the linker's internal bonds, but we suspected this approach might overestimate $g_{\rm R}^{({\rm cl})}$ because segments of polymer chains are known to be depleted near the surface of colloids \cite{Fuchs:2002} whereas hard spheres are enriched \cite{Boublik:1970}.

We accordingly measured $g_{\rm R}^{({\rm cl})}$ in molecular dynamics simulations of the reference hard-chain mixture using the model and simulation methods of Ref.~\citenum{Howard:2019}. We simulated 1000 colloids and 1500 linkers in a cubic box with periodic boundary conditions using \textsc{lammps} (22 Aug 2018) \cite{Plimpton:1995,Weeks:1971,Grest:1986}. Starting at $\eta_{\rm c} = 0.01$, we first equilibrated the mixture for $1.5 \times 10^4\,\tau$, where $\tau$ is the unit of time in the simulations. We then sampled configurations for analysis every $10\,\tau$ over a $10^4\,\tau$ period (1000 configurations). Last, we linearly compressed the edge length of the simulation box over $5 \times 10^3\,\tau$ to reach $\eta_{\rm c} = 0.02$. We repeated this procedure up to and including $\eta_{\rm c} = 0.15$. We computed $g_{\rm R}^{({\rm cl})}$ as a histogram of three variables: the distance from the center of the colloid to one end of the linker $r_{B_1} = |\vv{r}_{12}|$, the end-to-end distance $R = |\vv{R}_2|$, and the polar angle $\phi$ between $\vv{R}_2$ and $\vv{r}_{12}$, i.e., $\cos\phi = (\vv{r}_{12}\cdot\vv{R}_2)/(r_{B_1}R_2)$. The histogram bin ranges and widths were $3\,d_{\rm l} \le r_{B_1} \le 4\,d_{\rm l}$ with width $0.25\,d_{\rm l}$, $0\,d_{\rm l} \le R \le 7\,d_{\rm l}$ with width $0.5\,d_{\rm l}$, and $0 \le \phi \le \pi$ with width $\pi/9$ (20 degrees). We have labeled $r_{B_1}$ using the end $B_1$, but we also took the other linker end $B_2$ as $\vv{r}_{12}$ to improve sampling.

\begin{figure}
    \centering
    \includegraphics{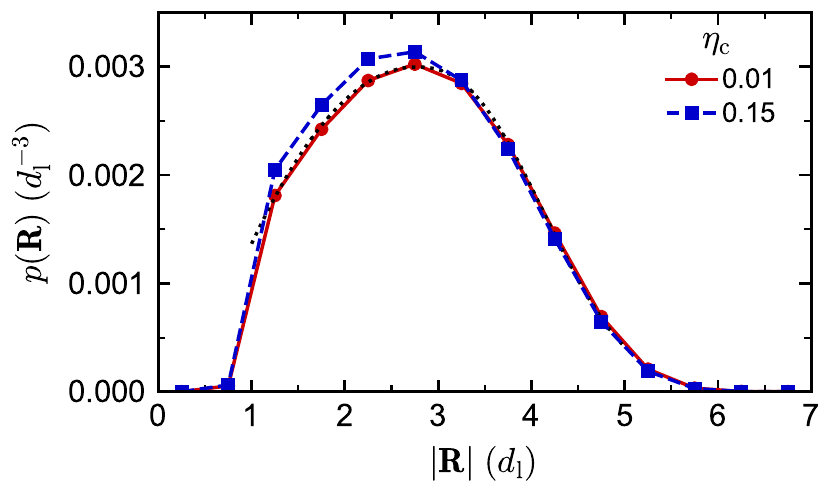}
    \caption{Probability density function $p$ for the linker end-to-end vector $\vv{R}$ as a function of the end-to-end distance $|\vv{R}|$ at $\eta_{\rm c} = 0.01$ (red circles) and $\eta_{\rm c} = 0.15$ (blue squares). The black dotted line shows an empirical fit to the $\eta_{\rm c} = 0.01$ distribution for $|\vv{R}| \ge \,d_{\rm l}$, which we used for numerical convenience when evaluating the bond-volume integrals (Appendix~\ref{sec:bvint}).}
    \label{fig:endtoend}
\end{figure}

We also computed the end-to-end vector distribution $p(\vv{R})$ as a histogram of $R$ using the same $R$-bins as for $g_{\rm R}^{({\rm cl})}$ (Figure~\ref{fig:endtoend}). This distribution was largely independent of $\eta_{\rm c}$ at the compositions we simulated, so we used $p(\vv{R})$ at $\eta_{\rm c} = 0.01$ to calculate $\Delta_1$ and $\Delta_2$. (We used the $p(\vv{R})$ measured at each composition, however, to normalize $g_{\rm R}^{({\rm cl})}$.) We noted that the probability of having $R$ commensurate with the distance between two colloid bonding sites ($\sqrt{2}d_{\rm cl}^* \approx 4.4\,d_{\rm l}$) was nonnegligible, meaning that it was likely that double bonds would form.

The integrand of $\Delta_1$ is nonzero only when one end of the linker interacts with a bonding site. Anticipating the hard-sphere exclusion between the colloid and linker segment, this requires $d_{\rm cl} \le r_{B_1} \le r^*$ where $r^* = d_{\rm cl}^* + 2.5\ell$ based on the range of Eq.~\eqref{eq:ugauss}. We found that $g_{\rm R}^{({\rm cl})}$ did not change significantly in this range of $r_{B_1}$, tending to increase slightly for larger $r_{B_1}$. In the strong association limit of large $\beta\varepsilon$, the dominant contribution to the integral occurs when the bonding sites overlap at $r_{B_1} \approx d_{\rm cl}^*$. We accordingly approximated $g_{\rm R}^{({\rm cl})}(r_{B_1},R,\phi) \approx g_{\rm R}^{({\rm cl})}(d_{\rm cl}^*,R,\phi)$, which still depended on both $R$ and $\phi$, e.g., because the other linker end cannot penetrate the colloid. However, further inspection of the data suggested a function of only one other variable, namely the distance $r_{B_2}$ from the center of the colloid to the other end of the linker,
\begin{equation}
r_{B_2} = |\vv{r}_{12}+\vv{R}_2| = (r_{B_1}^2 + R^2 + 2 r_{B_1} R \cos\phi)^{1/2}.
\label{eq:rB2}
\end{equation}
Conformations having $r_{B_2} < d_{\rm cl}$ had $g_{\rm R}^{({\rm cl})}(d_{\rm cl}^*,r_{B_2}) = 0$ because this would cause the linker end to penetrate the colloid. For larger values of $r_{B_2}$, we typically found $g_{\rm R}^{({\rm cl})}(d_{\rm cl}^*,r_{B_2}) < 1$ for most conformations, indicating depletion of the linker near the surface of the colloid \cite{Fuchs:2002}.

\begin{figure}
    \centering
    \includegraphics{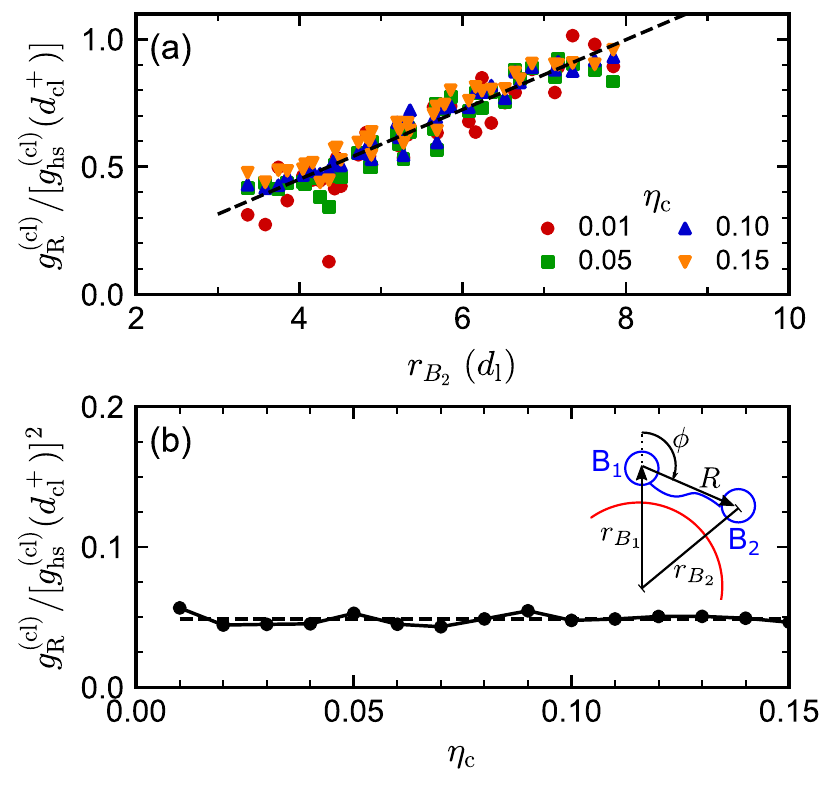}
    \caption{Simulated colloid--linker pair distribution function $g_{\rm R}^{({\rm cl})}$ in reference mixtures having varied $\eta_{\rm c}$ and fixed $\rho_{\rm l}/\rho_{\rm c} = 1.5$. $g_{\rm R}^{({\rm cl})}$ is normalized by the contact value of the pair distribution function $g_{\rm hs}^{({\rm cl})}(d_{\rm cl}^+)$ in the hard-sphere mixture that would be obtained by removing all bonds from the linkers \cite{Boublik:1970}. (a) When one linker end is fixed at $r_{B_1} = d_{\rm cl}^*$, $g_{\rm R}^{({\rm cl})}/[g_{\rm hs}^{({\rm cl})}(d_{\rm cl}^+)]$ collapses as a linear function (dashed line) of the distance from the colloid center to the other chain end $r_{B_2}$. Here, we show data only for $d_{\rm l} \le R \le 5\,d_{\rm l}$ and $0 \le \phi \le \pi/2$, for which we have reliable sampling. (b) When both ends of the linker are fixed at colloid bonding sites so that $r_{B_1} = d_{\rm cl}^*$, $R = \sqrt{2} d_{\rm cl}^*$ and $\phi = 3\pi/4$, $g_{\rm R}^{({\rm cl})}/[g_{\rm hs}^{({\rm cl})}(d_{\rm cl}^+)]^2$ is a constant (dashed line). The inset of (b) illustrates the definitions of $r_{B_1}$, $r_{B_2}$, $R$, and $\phi$,}
    \label{fig:g}
\end{figure}

We noted that $g_{\rm R}^{({\rm cl})}(d_{\rm cl}^*,r_{B_2})$ also had a volume fraction dependence. Conveniently, scaling $g_{\rm R}^{({\rm cl})}$ by $g_{\rm hs}^{({\rm cl})}(d_{\rm cl}^+)$, which we obtained from Boubl\'{i}k's equation of state \cite{Boublik:1970}, effectively accounted for this composition dependence and collapsed the data (Figure~\ref{fig:g}a). (We included in Figure~\ref{fig:g}a only data for $d_{\rm l} \le R \le 5\,d_{\rm l}$ and $0 \le \phi \le \pi/2$, for which we had sufficient sampling. These conformations are expected to contribute most significantly to $\Delta_1$, as others are less probable or wholly excluded.) Empirically, the data was well fit by a linear function of $r_{B_2}$ for $r_{B_2} > d_{\rm cl}$, which we used to evaluate $\Delta_1$.

The integrand of $\Delta_2$ is nonzero only when both ends of the linker interact with bonding sites. Using similar reasoning as for $\Delta_1$, we approximated $g_{\rm R}^{({\rm cl})}$ by its value when all the bonding sites overlap, $g_{\rm R}^{({\rm cl})}(r_{B_1},R,\phi) \approx g_{\rm R}^{({\rm cl})}(d_{\rm cl}^*,\sqrt{2}d_{\rm cl}^*,3\pi/4)$. The value of $g_{\rm R}^{({\rm cl})}$ again depended on $\eta_{\rm c}$, but we found that it was essentially a constant across volume fractions when scaled by $[g_{\rm hs}^{({\rm cl})}(d_{\rm cl}^+)]^2$ (Figure~\ref{fig:g}b). Qualitatively, the additional factor of $g_{\rm hs}^{({\rm cl})}(d_{\rm cl}^+)$ here accounts for the presence of the other chain end near the surface \cite{Sear:1994a,Sear:1994b,Sear:1996}. The value of $g_{\rm R}^{({\rm cl})}$ is significantly less than the superposition of the hard-sphere correlations, presumably due to the large entropic penalty to confine the linker near the surface of the colloid.

\subsection{Loop fractions\label{sec:colllink:loop}}
\begin{figure}
    \centering
    \includegraphics{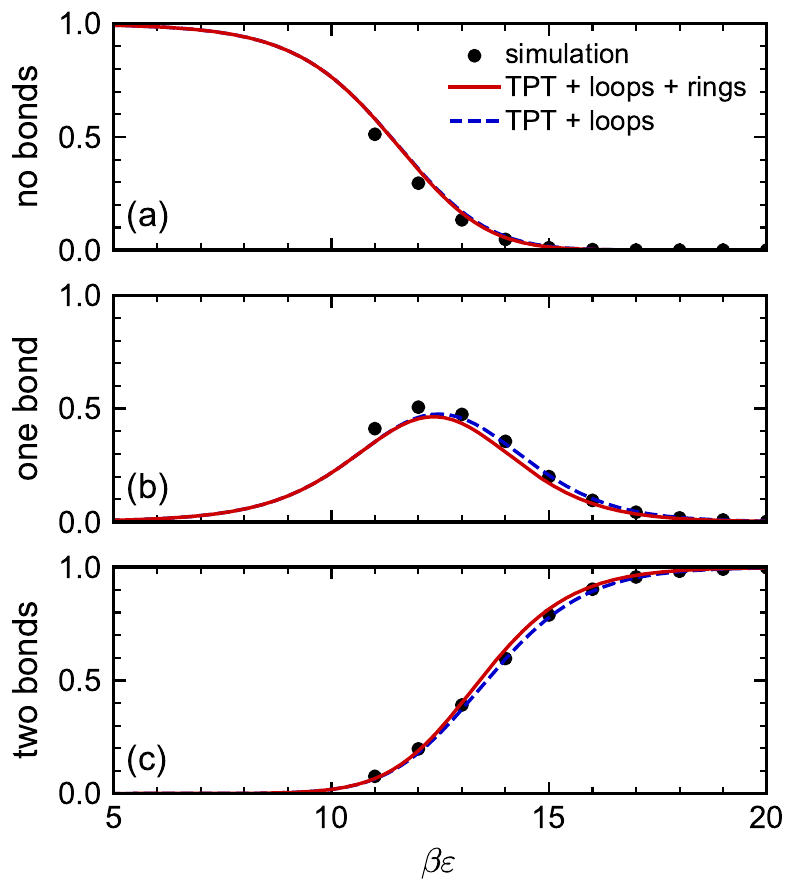}
    \caption{Fractions of linkers with (a) no bonded ends, (b) one bonded end, and (c) two bonded ends at $\eta_{\rm c} = 0.10$ and $\rho_{\rm l}/\rho_{\rm c} = 1.5$ as a function of attraction strength $\varepsilon$. The lines are TPT predictions accounting for only loops (dashed blue lines) and for both loops and two-colloid rings (solid red lines), and the black circles are simulation data \cite{Howard:2019}.}
    \label{fig:bonds}
\end{figure}

With all inputs determined, we proceeded to compute the fraction of colloids and linkers in different bonding states at various attraction strengths $\varepsilon$ and colloid volume fractions $\eta_{\rm c}$. We first determined the fractions of linkers having neither, one, or both ends bonded using TPT; these fractions are $X_{B_{12}}^{({\rm l})}$, $2(X_{B_1}^{({\rm l})}-X_{B_{12}}^{({\rm l})})$, and $1-2 X_{B_1}^{({\rm l})}+X_{B_{12}}^{({\rm l})}$, respectively, using Eq.~\eqref{eq:rhosigma}. Figure~\ref{fig:bonds} shows a representative result for $\eta_{\rm c} = 0.10$ as a function of $\varepsilon$, which was a composition where the mixture remained a single (homogeneous) phase in our previous simulations \cite{Howard:2019}. The fraction of unbonded linkers decreased monotonically as the attraction strength $\varepsilon$ increased, the fraction of linkers bonded at both ends concomittantly increased monotonically, and the fraction of linkers with only one end bonded had a maximum near $\beta\varepsilon \approx 12.5$. The TPT calculation (dashed blue line) agreed nearly quantitatively with the simulation data (black circles) \cite{Howard:2019}. However, first-order TPT, which completely neglects double bonds between colloids and linkers, yielded comparable results that we omitted here for clarity.

\begin{figure}
    \centering
    \includegraphics{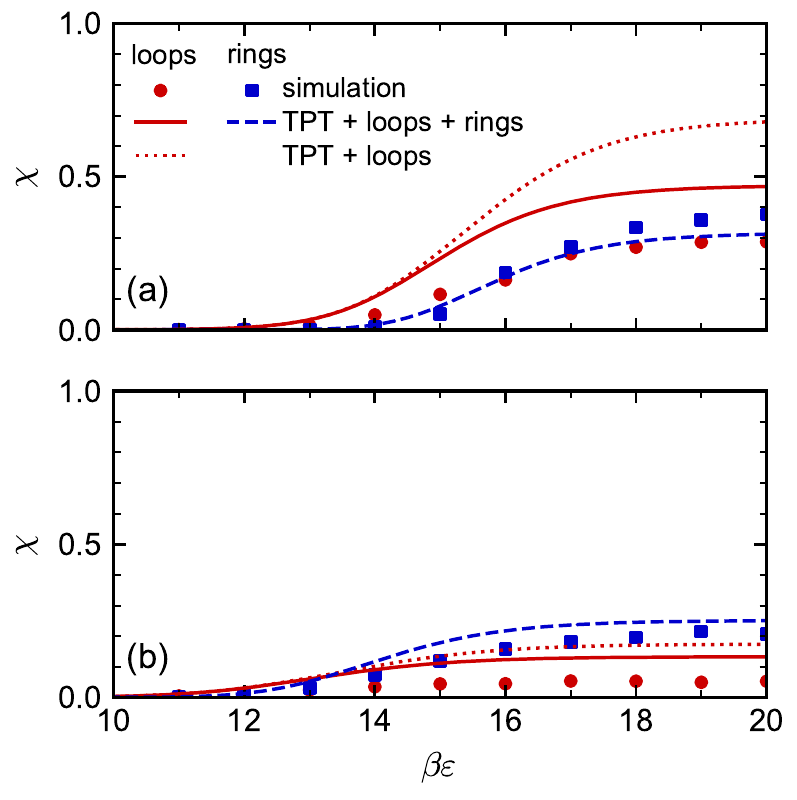}
    \caption{Fraction of linkers $\chi$ in loops (red circles / lines) and two-colloid rings (blue squares / lines) as a function of attraction strength $\varepsilon$ at $\rho_{\rm l}/\rho_{\rm c} = 1.5$ and (a) $\eta_{\rm c} = 0.01$ and (b) $\eta_{\rm c} = 0.10$. Simulation data from Ref.~\citenum{Howard:2019} (symbols) are compared to two TPT calculations: one including only loops ($\chi = \chi_{\rm loop}$ as dotted red lines), and one including both loops ($\chi_{\rm loop}$ as solid red lines) and two-colloid rings ($\chi_{\rm ring}$ as dashed blue lines).}
    \label{fig:loopseps}
\end{figure}

\begin{figure}
    \centering
    \includegraphics{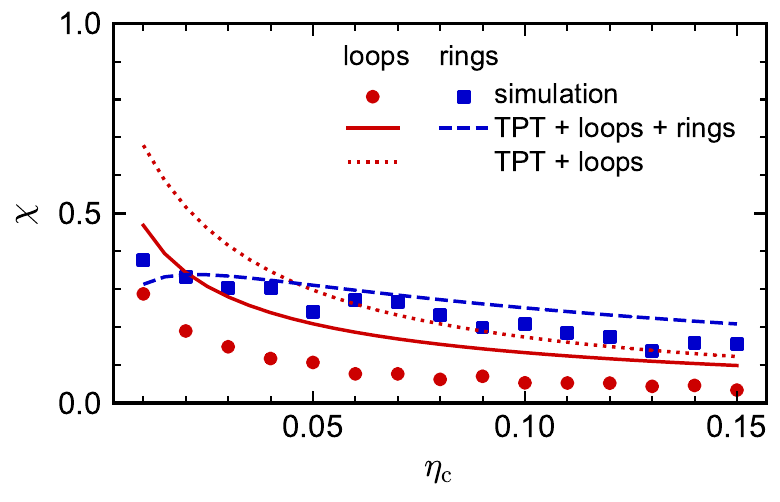}
    \caption{Fraction of linkers $\chi$ in loops (red circles / lines) and two-colloid rings (blue squares / line) at $\beta\varepsilon = 20$ and $\rho_{\rm l}/\rho_{\rm c} = 1.5$ as a function of colloid volume fraction $\eta_{\rm c}$. Simulation data from Ref.~\citenum{Howard:2019} (symbols) are compared to two TPT calculations: one including only loops ($\chi = \chi_{\rm loop}$ as dotted red line), and one including both loops ($\chi_{\rm loop}$ as solid red line) and two-colloid rings ($\chi_{\rm ring}$ as dashed blue line).}
    \label{fig:loopseta}
\end{figure}

The usefulness of the newly developed theory is its ability to predict the fraction of linkers in loops $\chi$, which is absent from TPT1. This fraction can be computed from the double-bond graphs in $\rho_{B_{12}}^{({\rm l})}$, which here is
\begin{equation}
\chi = \frac{\int\dd{\vv{1}}\rho_\varnothing(\vv{1}) \Delta c_{B_{12}}^{({\rm l})}(\vv{1})}{\int \dd{\vv{1}} \rho_{B_{12}}^{({\rm l})}(\vv{1})} = 1 - \frac{(X_{B_1}^{({\rm l})})^2}{X_{B_{12}}^{({\rm l})}}.
\label{eq:chi}
\end{equation}
In TPT1, $\chi = 0$ because $\Delta c_{B_{12}}^{({\rm l})} = 0$ (Section~\ref{sec:tptdb:nodb}), but with double bonding included, $\chi$ should depend on $\varepsilon$ and the mixture composition. We computed $\chi$ as a function of $\varepsilon$ at volume fractions $\eta_{\rm c} = 0.01$ and $\eta_{\rm c} = 0.10$ (dotted red lines in Figure~\ref{fig:loopseps}). At both compositions, more loops formed as $\varepsilon$ increased, and there were significantly more loops at the dilute composition $\eta_{\rm c} = 0.01$ than when $\eta_{\rm c} = 0.10$. The latter is more obvious considering $\chi$ as a function of $\eta_{\rm c}$ in the strong assocation limit, which we take here as $\beta\varepsilon = 20$ (Figure~\ref{fig:loopseta}). Both of these trends in the TPT predictions are qualitatively consistent with our simulations (red circles).

By manipulating the chemical equilibrium equations for the linkers, it can be shown in the strong association limit that
\begin{equation}
\chi \approx \left[1 + K \rho_{\rm c} \bigg(\frac{n_{\rm c}^2}{2\nu_{\rm c}}\bigg)  \bigg(\frac{\Delta_1^2}{\Delta_2}\bigg) \right]^{-1},
\label{eq:chiapprox}
\end{equation}
where $K = (X_{\rm A_{1}}^{({\rm c})})^2 / X_{\rm A_{12}}^{({\rm c})}$ approaches a limiting value as $\beta\varepsilon$ becomes large. Here, we assume that $\Delta_2 \gg \Delta_1$ and $\Delta_2 \sim \Delta_1^2$, which is motivated by Eqs.~\eqref{eq:delta2} and \eqref{eq:delta4} and supported by our calculations. For the range of compositions we studied, $K$ depended weakly on the mixture composition in our numerical calculations but was approximately 1. Consistent with our calculations and simulations (Figure~\ref{fig:loopseta}), smaller colloid number density $\rho_{\rm c}$ favors loop formation (larger $\chi$). $\chi$ also depends on the number of colloid bonding sites $n_{\rm c}$ relative to the number of potential double-bonding pairs $\nu_{\rm c}$ and, significantly, on the ratio of bond volumes $\Delta_1^2/\Delta_2$. This suggests a potential strategy for limiting loop formation. The number of double-bonding pairs $\nu_{\rm c}$ depends on the linker length, and $\Delta_2$ is also roughly proportional to $p(\vv{r}_{A_1 A_2})$, the probabability of the linker having an end-to-end vector commensurate with the distance between colloid bonding sites (Appendix~\ref{sec:bvint}). Reducing the compatibility between the linker and the typical distance between colloid bonding sites could decrease both $\nu_{\rm c}$ and $p(\vv{r}_{A_1A_2})$, and as a result $\chi$. This might be achieved by modifying the linker's length or flexibility and presents an opportunity for engineering colloidal self-assembly.

\subsection{Loop and ring fractions\label{sec:colllink:ring}}
\begin{figure}
    \centering
    \includegraphics{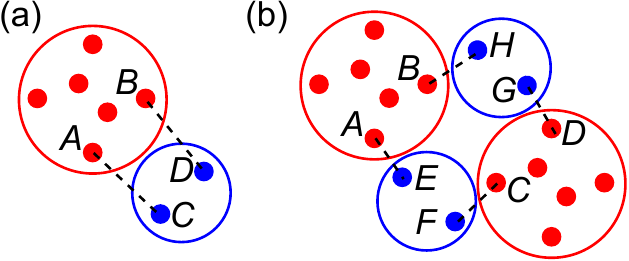}
    \caption{(a) Double-bond (``loop'') graph between one colloid (larger red hyperpoint) and one linker (smaller blue hyperpoint). (b) Ring graph with two colloids and two linkers. Both graphs require simultaneously bonding at two sites on the colloids and linkers. The bonding sites are labeled with general indices, and for clarity, only the $f_{AB}$-bonds (dashed lines) are drawn.}
    \label{fig:ringgraph}
\end{figure}

Despite qualitatively capturing the $\varepsilon$ and $\eta_{\rm c}$ dependences of $\chi$, our TPT calculations consistently overestimated $\chi$ compared to the simulations. We note that the simulations at smaller colloid volume fractions ($\eta_{\rm c} \lesssim 0.06$) phase separated \cite{Howard:2019}, but our TPT calculations assume that the mixture is homogeneous. Some differences might then be expected between TPT and the simulations in this regime; however, similar differences were also obtained at larger volume fractions where the simulated structures were homogeneous. Hence, we suspected there might be additional bonding motifs impacting the thermodynamics that we neglected in our approximation of the irreducible graph sum $\Delta c$. In particular, our simulations suggested that ``rings,'' where two linkers redundantly bridged two colloids (Figure~\ref{fig:snap}b), were also prevalent, but only double-bond graphs (Figure~\ref{fig:ringgraph}a) representing loops were included in Eq.~\eqref{eq:c}.

Rings of two colloids linked by two linkers can be incorporated into our TPT using the approximation of Sear and Jackson \cite{Sear:1994b}, which adds irreducible graphs (Figure~\ref{fig:ringgraph}b) to $\Delta c$,
\begin{align}
\frac{1}{4}&\sum_{\substack{AB \subseteq \Gamma_{\rm c} \\ |AB| = 2}} \sum_{\substack{CD \subseteq \Gamma_{\rm c}\\|CD| = 2}} \int \dd\vv{1} \dd\vv{2} \dd\vv{3} \dd\vv{4} \bigg[\sigma_{\Gamma_{\rm c}\setminus AB}^{({\rm c})}(\vv{1})  \nonumber \\
&\times \sigma_{\Gamma_{\rm c}\setminus CD}^{({\rm c})}(\vv{2}) \sigma_\varnothing^{({\rm l})}(\vv{3}) \sigma_\varnothing^{({\rm l})}(\vv{4}) g_{\rm R}^{(4)}(\vv{1},\vv{2},\vv{3},\vv{4}) \nonumber \\
&\times 16 f_{AE}^{({\rm cl})}(\vv{1},\vv{3}) f_{CF}^{({\rm cl})}(\vv{2},\vv{3}) f_{DG}^{({\rm cl})}(\vv{2},\vv{4}) f_{BH}^{({\rm cl})}(\vv{1},\vv{4}) \bigg],
\end{align}
where $g_{\rm R}^{(4)}$ is the four-body correlation function in the reference mixture approximated by the superposition,
\begin{align}
g_{\rm R}^{(4)}(\vv{1},\vv{2},\vv{3},\vv{4}) &\approx g_{\rm R}^{({\rm cl})}(\vv{1},\vv{3}) g_{\rm R}^{({\rm cl})}(\vv{2},\vv{3}) g_{\rm R}^{({\rm cl})}(\vv{2},\vv{4}) \nonumber \\
&\times g_{\rm R}^{({\rm cl})}(\vv{1},\vv{4}) e_{\rm R}^{({\rm cc})}(\vv{1},\vv{2}).
\end{align}
The sums run over all pairs of bonding sites on the colloids labeled $\vv{1}$ and $\vv{2}$, $\{E,F\}$ and $\{G,H\}$ are the bonding sites on the linkers labeled $\vv{3}$ and $\vv{4}$, respectively, and the factor of 16 accounts for permutations of bonding within these sets of sites. Because this ring graph involves only singlet densities not bonded at two sites, it will contribute only to $\Delta c_{A_{12}}^{({\rm c})}$ and $\Delta c_{B_{12}}^{({\rm l})}$in a similar manner to the double-bond graphs. A similar approximation was also made by Haghmoradi et al. for mixtures of ring-forming colloids with two bonding sites in an article that appeared after our work was completed \cite{Haghmoradi:2020}.

We carried out the functional differentiation of $\Delta c$ including these ring graphs, and we integrated the chemical equilibrium equations for a spatially homogeneous fluid. Two-colloid rings are only likely to form at colloid bonding sites that are nearest neighbors because the linkers would need to stretch significantly ($R \gtrsim d_{\rm c}$) to bond at sites on opposite hemispheres. Hence, the equivalent $\sigma_\alpha^{({\rm c})}$ remain the same as before, and we need only add terms for the ring graph to Eqs.~\eqref{eq:eq2c} and \eqref{eq:eq2p}. The result is
\begin{align}
\frac{X_{\Gamma_{\rm c}\setminus A_{12}}^{({\rm c})}}{X_{\Gamma_{\rm c}}^{({\rm c})}} &- \left(\frac{X_{\Gamma_{\rm c}\setminus A_1}^{({\rm c})}}{X_{\Gamma_{\rm c}}^{({\rm c})}}\right)^2 = 2 \rho_{\rm l} X_{B_{12}}^{({\rm l})} \Delta_2 \nonumber \\
&+16 \bigg(\frac{\nu_{\rm c}}{2}\bigg) \bigg(\rho_{\rm c} X_{A_{12}}^{({\rm c})}\bigg) \bigg(\rho_{\rm l} X_{B_{12}}^{({\rm l})}\bigg)^2 \Delta_4
\label{eq:eq4c} \\
\frac{1}{X_{B_{12}}^{({\rm l})}} &- \left(\frac{X_{B_1}^{({\rm l})}}{X_{B_{12}}^{({\rm l})}}\right)^2 = 2 \nu_{\rm c} \rho_{\rm c} X_{A_{12}}^{({\rm c})} \Delta_2 \nonumber \\
&+ 16 \bigg(\frac{\nu_{\rm c}^2}{2}\bigg) \bigg(\rho_{\rm c} X_{A_{12}}^{({\rm c})}\bigg)^2 \bigg(\rho_{\rm l} X_{B_{12}}^{({\rm l})}\bigg) \Delta_4
\label{eq:eq4p}
\end{align}
with the ring-bond volume
\begin{align}
\Delta_4 = &\Omega_{\rm c}^{-2} \int \dd\vv{r}_{12} \dd\vv{r}_{13} \dd\vv{r}_{14} \dd\vvg{\Omega_1} \dd\vvg{\Omega_2} \dd\vv{R}_3 \dd\vv{R}_4  \nonumber \\
& \bigg[p(\vv{R}_3) p(\vv{R}_4) g_{\rm R}^{(4)}(\vv{r}_{12},\vv{r}_{13},\vv{r}_{14},\vv{R}_3,\vv{R}_4) \nonumber \\
&\times f_{AE}^{({\rm cl})}(\vv{r}_{13},\vvg{\Omega}_1) f_{CF}^{({\rm cl})}(\vv{r}_{12},\vv{r}_{13},\vvg{\Omega}_2,\vv{R}_3) \nonumber \\
&\times f_{DG}^{({\rm cl})}(\vv{r}_{12},\vv{r}_{14},\vvg{\Omega}_2) f_{BH}^{({\rm cl})}(\vv{r}_{14},\vvg{\Omega}_1,\vv{R}_4) \bigg].
\end{align}
Details of how we approximated $\Delta_4$ are given in Appendix~\ref{sec:bvint}.

As in Section~\ref{sec:colllink:loop}, we proceeded to solve the new chemical equilibrium equations. The fractions of linkers with zero, one, or two bonded ends predicted by the TPT with both loops and rings (red line in Figure~\ref{fig:bonds}) was very similar to those predicted by the TPT with only loops, and both agreed well with the simulations. We then determined the fractions of linkers in either loops or rings. With rings included in the TPT, $\chi$ given by Eq.~\eqref{eq:chi} is now the total fraction of linkers in either motif. We accordingly separated the loop-graph contributions to $\Delta c_{B_{12}}^{({\rm l})}$ and $\chi$,
\begin{equation}
\chi_{\rm loop} = 2\nu_{\rm c} \rho_{\rm c} X_{A_{12}}^{({\rm c})} X_{B_{12}}^{({\rm l})} \Delta_2,
\end{equation}
and by subtraction, the ring-graph contribution was $\chi_{\rm ring} = \chi - \chi_{\rm loop}$.

The TPT-predicted fraction of linkers in rings (blue dashed line in Figures~\ref{fig:loopseps} and \ref{fig:loopseta}) is in very good agreement with the simulations (blue squares). The predicted fraction of loops also decreased in the TPT with loops and rings (red line) compared to the TPT with only loops. This improved the agreement between the TPT and simulations, although the fraction of linkers in loops was still slightly overpredicted in the TPT. We believe this overprediction might be due to neglect of other competing bonding motifs such as larger rings or of other irreducible graphs that include effects from steric hindrance \cite{Wertheim:1987}. Nonetheless, the agreement between the simulations and the TPT with loops and two-colloid rings is good, and the TPT with loops and rings is a significant improvement over TPT1, which predicted that $\chi = 0$ for both bonding motifs.

\section{Conclusions\label{sec:conclude}}
We have extended Wertheim's thermodynamic perturbation theory (TPT) for fluids with strong directional attractions to include double bonds between an arbitrary number of pairs of molecular bonding sites in a multicomponent mixture. This extension, which relaxes restrictions in prior TPTs on the number of potential double-bond site pairs, was required to model the assembly of colloidal particles (``colloids'') and difunctional flexible chain molecules (``linkers''). We showed that the fraction of linkers that ``looped'' to make a double bond with both ends attached to the same colloid and/or that formed ``rings'' of two linkers bridging between the same two colloids could be reliably predicted using TPT across a range of compositions and attraction strengths. A large linker loop fraction may inhibit assembly of percolated, self-supporting networks of colloids because looped linkers do not form new bridges between colloids. Our work suggests that the loop fraction can be reduced by making the end-to-end distance of the linker incompatible with the distance between colloid bonding sites, which might be achieved by modifying the linker's molecular weight or flexibility \cite{Howard:2020}. It would be interesting to use the developed theory to compute not only loop fractions but also phase boundaries in order to demonstrate how loops modify the phase behavior of the mixture, including conditions amenable to gelation or equilibrium gels \cite{Bianchi:2006,Zaccarelli2007}.

\begin{acknowledgements}
MPH thanks Ryan Jadrich for introducing him to Wertheim's elegant theory. This research was primarily supported by the National Science Foundation through the Center for Dynamics and Control of Materials: an NSF MRSEC under Cooperative Agreement No.~DMR-1720595, with additional support from an Arnold O. Beckman Postdoctoral Fellowship (ZMS) and the Welch Foundation (Grant Nos.~F-1696 and F-1848). We acknowledge the Texas Advanced Computing Center (TACC) at The University of Texas at Austin for providing HPC resources.
\end{acknowledgements}

\appendix
\section{Site operator algebra\label{sec:alg}}
Useful relationships between $\rho_\alpha^{(i)}$, $\sigma_\alpha^{(i)}$, and $c_\alpha^{(i)}$ can be derived through a formalism of ``site operators'' (or dual numbers) that permits manipulation of equations like Eq.~\eqref{eq:rhoca} \cite{Wertheim:1986a}. For completeness, we will explicitly derive some of Wertheim's key results using this formalism. Let $\varepsilon_A^{(i)}(\vv{1})$ be an operator of a site $A$ on a molecule of component $i$ at a root point labeled $\vv{1}$. All site operators commute and are defined to satisfy
\begin{equation}
\left(\varepsilon_A^{(i)}\right)^2 = 0.
\end{equation}
For some $\alpha \subseteq \Gamma_i$, define
\begin{equation}
\varepsilon_\alpha^{(i)} = \prod_{A \in \alpha} \varepsilon_A^{(i)},
\end{equation}
and as a result,
\begin{equation}
\varepsilon_\alpha^{(i)} \varepsilon_\gamma^{(i)} = \begin{cases}
\varepsilon_{\alpha\cup\gamma}^{(i)},& \alpha\cap\gamma = \varnothing \\
0,& \mathrm{otherwise}
\end{cases}.
\end{equation}
Wertheim considered numbers of the form
\begin{equation}
\mathring{x}^{(i)} = x_\varnothing^{(i)} + \sum_{\substack{\alpha \subseteq \Gamma_i \\ \alpha \ne \varnothing}}
    x_\alpha^{(i)} \varepsilon_\alpha^{(i)}
\label{eq:sitex}
\end{equation}
that follow the arithmetic and algebra for real numbers with the exception that division by $\mathring{x}^{(i)}$ having $x_\varnothing^{(i)} = 0$ is not permitted. Analytical functions of $\mathring{x}^{(i)}$ can be computed by series expansion.

With this algebra, $\mathring{\rho}^{(i)}$ is defined by Eq.~\eqref{eq:sitex},
\begin{equation}
\mathring{\rho}^{(i)} = \rho_\varnothing^{(i)} + \sum_{\substack{\alpha \subseteq \Gamma_i \\ \alpha \ne \varnothing}}
    \rho_\alpha^{(i)} \varepsilon_\alpha^{(i)},
\end{equation}
and we will now connect $\mathring{\sigma}^{(i)}$ to $\mathring{\rho}^{(i)}$. We start with this same definition for $\mathring{\sigma}^{(i)}$ and use Eq.~\eqref{eq:sigmarho} to replace $\sigma_\alpha^{(i)}$,
\begin{align}
\mathring{\sigma}^{(i)} &= \rho_\varnothing^{(i)} + \sum_{\substack{\alpha \subseteq \Gamma_i \\ \alpha \ne \varnothing}}
    \Bigg(\sum_{\gamma \subseteq \alpha} \rho_\gamma^{(i)}\Bigg) \varepsilon_\alpha^{(i)} \nonumber \\
&= \rho_\varnothing^{(i)} + \sum_{\substack{\alpha \subseteq \Gamma_i \\ \alpha \ne \varnothing}}
    \Bigg( \rho_\varnothing^{(i)} + \rho_\alpha^{(i)} + \sum_{\substack{\gamma \subset \alpha \\ \gamma \ne \varnothing}} \rho_\gamma^{(i)}\Bigg) \varepsilon_\alpha^{(i)} \nonumber \\
&= \rho_\varnothing^{(i)} + \sum_{\substack{\alpha \subseteq \Gamma_i \\ \alpha \ne \varnothing}}
    \Bigg( \rho_\varnothing^{(i)} + \rho_\alpha^{(i)} + \sum_{\substack{\gamma \subseteq \Gamma_i \\ \gamma \ne \varnothing}} \rho_\gamma^{(i)}\varepsilon_\gamma^{(i)} \Bigg) \varepsilon_\alpha^{(i)} \nonumber \\
&= \Bigg(\rho_\varnothing^{(i)} + \sum_{\substack{\alpha \subseteq \Gamma_i \\ \alpha \ne \varnothing}} \rho_\alpha^{(i)} \varepsilon_\alpha^{(i)}\Bigg)
    \Bigg(1 + \sum_{\substack{\alpha \subseteq \Gamma_i \\ \alpha \ne \varnothing}} \varepsilon_\alpha^{(i)}\Bigg) \nonumber \\
&= \mathring{\rho}^{(i)} \prod_{A \in \Gamma_i} \left(1+\varepsilon_A^{(i)}\right).
\end{align}
We made use of the identity
\begin{equation}
\sum_{\substack{\alpha \subseteq \Gamma_i \\ \alpha \ne \varnothing}} \varepsilon_\alpha^{(i)} \sum_{\substack{\gamma \subseteq \Gamma_i \\ \gamma \ne \varnothing}} x_\gamma^{(i)} \varepsilon_\gamma^{(i)}
    = \sum_{\substack{\alpha \subseteq \Gamma_i \\ \alpha \ne \varnothing}} \varepsilon_\alpha^{(i)} \sum_{\substack{\gamma \subset \alpha \\ \gamma \ne \varnothing}} x_\gamma^{(i)}
\end{equation}
in the fourth line and
\begin{equation}
\prod_{A \in \Gamma_i} \left(1 \pm \varepsilon_A^{(i)}\right) =
    1 + \sum_{\substack{\alpha \subseteq \Gamma_i \\ \alpha \ne \varnothing}} (\pm 1)^{|\alpha|} \varepsilon_\alpha^{(i)}
\end{equation}
in the last line.

It is now straightforward to invert this relationship,
\begin{equation}
\mathring{\rho}^{(i)} = \mathring{\sigma}^{(i)} \prod_{A \in \Gamma_i}\left(1-\varepsilon_A^{(i)}\right),
\label{eq:rhositesigsite}
\end{equation}
using the series expansion $(1+\varepsilon_A^{(i)})^{-1} = 1-\varepsilon_A^{(i)}$. To find expressions for $\rho_\alpha^{(i)}$ in terms of $\sigma_\alpha^{(i)}$, we expand $\mathring{\sigma}^{(i)}$,
\begin{equation}
\mathring{\rho}^{(i)} = \Bigg(\sigma_\varnothing^{(i)} + \sum_{\substack{\alpha \subseteq \Gamma_i \\ \alpha \ne \varnothing}} \sigma_\alpha^{(i)} \varepsilon_\alpha^{(i)} \Bigg)
\Bigg(1 + \sum_{\substack{\alpha \subseteq \Gamma_i \\ \alpha \ne \varnothing}} (-1)^{|\alpha|} \varepsilon_\alpha^{(i)}\Bigg),
\end{equation}
and collect coefficients of $\varepsilon_\alpha^{(i)}$. This can only include terms in subsets $\gamma\subseteq\alpha$, giving Eq.~\eqref{eq:rhosigma}.

To find expressions for $c_\alpha^{(i)}$ in terms of $\sigma_\alpha^{(i)}$, we start from the definition of $\mathring{c}^{(i)}$ and consider its exponential
\begin{align}
\exp\left(\mathring{c}^{(i)} - c_\varnothing^{(i)}\right)
    &= \exp\Bigg(\sum_{\substack{\alpha \subseteq \Gamma_i \\ \alpha \ne \varnothing}} c_\alpha^{(i)} \varepsilon_\alpha^{(i)} \Bigg) \nonumber \\
    &= \prod_{\substack{\alpha \subseteq \Gamma_i \\ \alpha \ne \varnothing}} \left(1+c_\alpha^{(i)} \varepsilon_\alpha^{(i)}\right),
\label{eq:expsitecprod}
\end{align}
where in the last equality we have separated the exponential and used the series expansion $\exp(\varepsilon_\alpha^{(i)}) = 1+\varepsilon_\alpha^{(i)}$. The product can be reexpressed as a sum over partitions of $\alpha$ because only products of terms in disjoint sets will survive the expansion,
\begin{align}
\exp\left(\mathring{c}^{(i)} - c_\varnothing^{(i)}\right)
&= 1+ \sum_{\substack{\alpha \subseteq \Gamma_i \\ \alpha \ne \varnothing}}
    \Bigg(\sum_{\gamma \in P(\alpha)} \prod_{a \in \gamma} c_a^{(i)}\Bigg) \varepsilon_\alpha^{(i)} \nonumber \\
&= \frac{\mathring{\rho}^{(i)}}{\rho_\varnothing^{(i)}} \nonumber \\
&= \frac{\mathring{\sigma}^{(i)}}{\sigma_\varnothing^{(i)}} \prod_{A \in \Gamma_i}\left(1-\varepsilon_A^{(i)}\right),
\label{eq:expsitec}
\end{align}
using Eq.~\eqref{eq:rhoca} to arrive at the second equality and Eq.~\eqref{eq:rhositesigsite} with the replacement $\rho_\varnothing^{(i)} = \sigma_\varnothing^{(i)}$ to obtain the last equality.

Using Eqs.~\eqref{eq:expsitecprod} and \eqref{eq:expsitec} then inverting, we can immediately show
\begin{align}
\frac{\mathring{\sigma}^{(i)}}{\sigma_\varnothing^{(i)}} &= \prod_{\substack{\alpha \subseteq \Gamma_i \\ \alpha \ne \varnothing}} \left(1+c_\alpha^{(i)} \varepsilon_\alpha^{(i)}\right) \prod_{A \in \Gamma_i}\left(1+\varepsilon_A^{(i)}\right) \nonumber \\
&= \prod_{\substack{\alpha \subseteq \Gamma_i \\ \alpha \ne \varnothing}} \left[1+ \left(\delta_{|\alpha|,1}+c_\alpha^{(i)}\right) \varepsilon_\alpha^{(i)}\right]
\end{align}
Expanding this product and collecting terms as in Eq.~\eqref{eq:expsitec} gives Eq.~\eqref{eq:sigmaca}.

Alternatively, we can invert Eq.~\eqref{eq:expsitec} directly,
\begin{align}
\mathring{c}^{(i)}-c_\varnothing^{(i)}
    &= \ln\left(\frac{\mathring{\sigma}^{(i)}}{\sigma_\varnothing^{(i)}}\right) + \sum_{A \in \Gamma_i} \ln\left(1-\varepsilon_A^{(i)}\right) \nonumber \\
    &= \ln\left(\frac{\mathring{\sigma}^{(i)}}{\sigma_\varnothing^{(i)}}\right) - \sum_{A \in \Gamma_i} \varepsilon_A^{(i)},
\end{align}
where the last equality used the series expansion $\ln(1-\varepsilon_A^{(i)}) = -\varepsilon_A^{(i)}$. In order to determine expressions for $c_\alpha^{(i)}$, we now series expand the logarithm,
\begin{equation}
\ln\left(\frac{\mathring{\sigma}^{(i)}}{\sigma_\varnothing^{(i)}}\right)
    = \ln\Bigg(1+\sum_{\substack{\alpha \subseteq \Gamma_i \\ \alpha \ne \varnothing}} \frac{\sigma_\alpha^{(i)} \varepsilon_\alpha^{(i)}}{\sigma_\varnothing^{(i)}}\Bigg).
\end{equation}
Recall that in general
\begin{equation}
\ln(1+x) = \sum_{n=1}^{\infty} (-1)^{n-1} \frac{x^n}{n},
\end{equation}
so the expansion can be performed in multiple variables by collecting all terms involving disjoint sets whose union is $\alpha$, i.e., these sets must be partitions of $\alpha$,
\begin{align}
\ln\left(\frac{\mathring{\sigma}^{(i)}}{\sigma_\varnothing^{(i)}}\right) =
    \sum_{\substack{\alpha \subseteq \Gamma_i \\ \alpha \ne \varnothing}}
    \Bigg(\sum_{\gamma \in P(\alpha)} (-1)^{|\gamma|-1} \frac{|\gamma|!}{|\gamma|} \prod_{a \in \gamma} \frac{\sigma_a^{(i)}}{\sigma_\varnothing^{(i)}}\Bigg)
    \varepsilon_\alpha^{(i)}.
\end{align}
The factor of $|\gamma|!$ accounts for all permutations of $\gamma$ that occur in the expansion. This gives Eq.~\eqref{eq:casigma} for $c_\alpha^{(i)}$.

We will now show one additional relationship that is useful for computing the thermodynamics. We start by explicitly expanding the sum
\begin{equation}
\sum_{\alpha \subseteq \Gamma_i} \sigma_{\Gamma_i \setminus\alpha}^{(i)} c_\alpha^{(i)}
    = \sigma_{\Gamma_i}^{(i)} c_\varnothing^{(i)} + \sum_{\substack{\alpha\subset\Gamma_i\\\alpha\ne\varnothing}} \sigma_{\Gamma_i\setminus\alpha}^{(i)} c_\alpha^{(i)} + \sigma_\varnothing^{(i)} c_{\Gamma_i}^{(i)}.
\end{equation}
The second term can be reexpressed as a sum over partitions of $\Gamma_i$, excluding the improper partition, by substituting Eq.~\eqref{eq:casigma} and regrouping terms,
\begin{align}
\sum_{\substack{\alpha\subset\Gamma_i\\\alpha\ne\varnothing}} \sigma_{\Gamma_i\setminus\alpha}^{(i)} c_\alpha^{(i)}
&= -\sum_{A\in\Gamma_i} \sigma_{\Gamma_i\setminus A}^{(i)} + \sum_{\substack{\alpha\subset\Gamma_i\\\alpha\ne\varnothing}} \sigma_{\Gamma_i\setminus\alpha}^{(i)} \nonumber \\
    &\times \sum_{\gamma \in P(\alpha)} (-1)^{|\gamma|-1} (|\gamma|-1)! \prod_{a \in \gamma} \frac{\sigma_a^{(i)}}{\sigma_\varnothing^{(i)}} \nonumber \\
&= -\sum_{A\in\Gamma_i} \sigma_{\Gamma_i\setminus A}^{(i)} + \sigma_\varnothing^{(i)} \nonumber \\
    &\times \sum_{\substack{\gamma\in P(\Gamma_i)\\\gamma\ne\{\Gamma_i\}}}
   (-1)^{|\gamma|-2} |\gamma|(|\gamma|-2)! \prod_{a \in \gamma} \frac{\sigma_a^{(i)}}{\sigma_\varnothing^{(i)}}.
\end{align}
Note in the second equality that the sum excludes the improper partition of $\Gamma_i$ and that there is a factor of $|\gamma|$ that accounts for all permutations removing one of the terms in the partition. By direct substitution of Eq.~\eqref{eq:casigma}, the third term can be similarly expressed,
\begin{equation}
\sigma_\varnothing^{(i)} c_{\Gamma_i}^{(i)}
= \sigma_{\Gamma_i}^{(i)} + \sigma_\varnothing^{(i)} \sum_{\substack{\gamma\in P(\Gamma_i) \\\gamma\ne\{\Gamma_i\}}}
    (-1)^{|\gamma|-1} (|\gamma|-1)! \prod_{a \in \gamma} \frac{\sigma_a^{(i)}}{\sigma_\varnothing^{(i)}}.
\end{equation}
Combining the two sums leads to
\begin{equation}
\sum_{\alpha \subseteq \Gamma_i} \sigma_{\Gamma_i\setminus\alpha}^{(i)} c_\alpha^{(i)}
    = \sigma_{\Gamma_i}^{(i)} \left(1+c_\varnothing^{(i)}\right) + Q^{(i)},
\end{equation}
where $Q^{(i)}$ is given by Eq.~\eqref{eq:Q}. Note that $Q^{(i)}$ does not depend on $\sigma_\Gamma^{(i)}$.

\section{Numerics\label{sec:bvint}}
\begin{figure}
    \centering
    \includegraphics{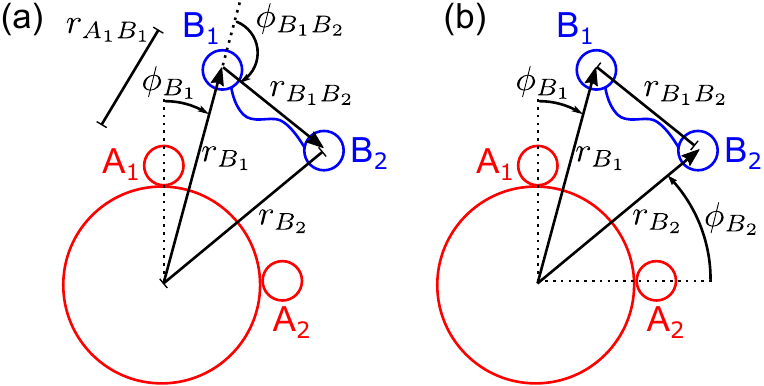}
    \caption{Schematic of linker coordinates (blue) used to evaluate (a) $\Delta_1$ and (b) $\Delta_2$ in the frame that rotates with the colloid (red). Arrows designate integration variables, solid lines with bars designate dependent variables, and dotted lines designate the axis of the spherical coordinate system. Refer to the text for the meaning of each variable.}
    \label{fig:intcoords}
\end{figure}

To evaluate $\Delta_1$, we fixed the colloid at the origin and specified its orientation by a director through $A_1$ having three intrinsic Euler angles $(\theta_{\rm c},\phi_{\rm c},\psi_{\rm c})$---$0 \le \theta_{\rm c} \le 2\pi$ is a right-handed rotation around the $z$-axis, $0 \le \phi_{\rm c} \le \pi$ is a right-handed rotation about the rotated $y$-axis, and $0 \le \psi_{\rm c} \le 2\pi$ is a right-handed rotation about the rotated $z$-axis---so $\Omega_{\rm c} = 8\pi^2$ and $\theta_{\rm c}$ and $\phi_{\rm c}$ have the same meaning as the azimuthal and polar angles in spherical coordinates. We defined the position $\vv{r}_{B_1}$ of the bonded linker site $B_1$ in spherical coordinates $(r_{B_1},\theta_{B_1},\phi_{B_1})$ using the colloid as the origin and $\vv{r}_{A_1}$ as an axis, i.e., in a frame that rotates with the colloid. We will denote the vector between two sites $A$ and $B$ by $\vv{r}_{AB} = \vv{r}_B - \vv{r}_A$ and the distance by $r_{AB} = |\vv{r}_{AB}|$. The end-to-end vector $\vv{R} = \vv{r}_{B_1 B_2}$ was also defined in spherical coordinates $(r_{B_1 B_2},\theta_{B_1 B_2},\phi_{B_1 B_2})$ using $B_1$ as the origin and $\vv{r}_{B_1}$ as the axis. These coordinates are illustrated in Figure~\ref{fig:intcoords}a.

The distance between $A_1$ and $B_1$ was
\begin{equation}
r_{A_1 B_1}^2 = (d_{\rm cl}^*)^2 + r_{B_1}^2 - 2 d_{\rm cl}^* r_{B_1} \cos \phi_{B_1},
\end{equation}
and with the potential given by Eq.~\eqref{eq:ugauss}, $f_{A_1 B_1}$ was nonzero when $d_{\rm cl} \le r_{B_1} \le r^* = d_{\rm cl}^* + 2.5\ell$ and
\begin{equation}
\cos\phi_{B_1} \ge x(r_{B_1}) = \frac{r_{B_1}^2 +(d_{\rm cl}^*)^2 - (2.5\ell)^2}{2 d_{\rm cl}^* r_{B_1}}.
\end{equation}
The end-to-end vector was limited by the segment diameter and the contour length, $d_{\rm l} \le r_{B_1 B_2} \le L$. Conformations where the linker wraps around the colloid are either improbable or forbidden by the hard-sphere exclusions, so we limited $0 \le \phi_{B_1 B_2} \le \pi/2$, effectively approximating the colloid as a flat surface. With these considerations and after integration over all colloid orientations, $\theta_{B_1}$, and $\theta_{B_1 B_2}$, the integral for $\Delta_1$ (Eq.~\eqref{eq:delta2}) was
\begin{align}
\Delta_1 &\approx (2\pi)^2 \int_{d_{\rm cl}}^{r^*} \dd r_{B_1} r_{B_1}^2 \int_{x(r_{B_1})}^1 \dd \cos \phi_{B_1} \int_{d_{\rm l}}^L \dd r_{B_1 B_2} r_{B_1 B_2}^2  \nonumber \\
& \int_0^1 \dd \cos\phi_{B_1 B_2} \bigg[p(r_{B_1 B_2}) g_{\rm R}^{({\rm cl})}(d_{\rm cl}^*,r_{B_2}) f_{A_1 B_1}^{({\rm cl})}(r_{A_1 B_1})  \bigg].
\end{align}
We used a linear fit (Figure~\ref{fig:g}a) for the pair correlation function
\begin{equation}
g_{\rm R}^{({\rm cl})}(d_{\rm cl}^*,r_{B_2}) = [g_{\rm hs}^{({\rm cl})}(d_{\rm cl}^+)](0.137 r_{B_2} - 0.0960)
\label{eq:gcplin}
\end{equation}
with $r_{B_2}$ given by Eq.~\eqref{eq:rB2}. For numerical convenience, we also fit $p(\vv{R})$ for $|\vv{R}| \ge d_{\rm l}$ at $\eta_{\rm c} = 0.01$ (Figure~\ref{fig:endtoend}) with a piecewise function,
\begin{equation}
p(r) = \begin{cases}
p_{\rm max} + a(r-r_{\rm max})^2,& r \le r_{\rm max} \\
p_{\rm max} \exp\left[-b(r-r_{\rm max})^c\right],& r > r_{\rm max}
\end{cases},
\end{equation}
where $r_{\rm max} = 2.76$, $p_{\rm max} = 3.00 \times 10^{-3}$, $a = -5.24 \times 10^{-4}$, $b = 0.276$, and $c = 2.45$ (all units implicit and consistent with $r$ having units of $d_{\rm l}$).

To evaluate $\Delta_2$, we modified the scheme we used for $\Delta_1$ so that $B_2$ was now bonded with a site $A_2$ on the colloid that was a nearest neighbor of $A_1$. The position $\vv{r}_{B_2}$ of $B_2$ was redefined in spherical coordinates $(r_{B_2},\theta_{B_2},\phi_{B_2})$ with the colloid as the origin and $\vv{r}_{A_2}$ as the axis (Figure~\ref{fig:intcoords}b). The distance between $A_1$ and $B_1$ was unchanged, and the distance between $A_2$ and $B_2$ was given by an analogous formula. The distance between $B_1$ and $B_2$ was
\begin{align}
r_{B_1 B_2}^2 = r_{B_1}^2 + r_{B_2}^2 &- 2 r_{B_1} r_{B_2} (\sin\phi_{B_1} \sin\theta_{B_1} \cos\phi_{B_2} \nonumber \\
-& \sin\phi_{B_2} \sin\theta_{B_2} \cos\phi_{B_1} \nonumber \\
+& \sin\phi_{B_1} \cos\theta_{B_1} \sin\phi_{B_2} \cos\theta_{B_2} ).
\end{align}
It was then not possible to trivially integrate over $\theta_{B_1}$ and $\theta_{B_2}$, and we found that this increased dimensionality added a significant computational cost for evaluating $\Delta_2$. However, the distance between the colloid sites $r_{A_1 A_2}$ is much larger than the range of Eq.~\eqref{eq:ugauss}, so to a good approximation $\vv{r}_{B_1 B_2} \approx \vv{r}_{A_1 A_2}$. As a result,
\begin{equation}
\Delta_2 \approx p(\vv{r}_{A_1 A_2}) g_{\rm R}^{({\rm cl})}(d_{\rm cl}^*,\sqrt{2} d_{\rm cl}^*, 3\pi/4) v_{A_1 B_1}^2
\end{equation}
with
\begin{equation}
g_{\rm R}^{({\rm cl})}(d_{\rm cl}^*,\sqrt{2} d_{\rm cl}^*, 3\pi/4)
\approx 0.0485 [g_{\rm hs}^{({\rm cl})}(d_{\rm cl}^+)]^2
\end{equation}
from Figure~\ref{fig:g}b and
\begin{align}
v_{A_1 B_1} = 2\pi \int_{d_{\rm cl}}^{r^*} \dd r_{B_1} r_{B_1}^2 \int_{x(r_{B_1})}^1 \dd\cos\phi_{B_1} f_{A_1 B_1}^{({\rm cl})}(r_{A_1 B_1}).
\end{align}

To evaluate $\Delta_4$, which is an even higher dimensional integral than $\Delta_2$, we adopted a similar approximation for determining $p$ and $g_{\rm R}^{(4)}$ using the positions of the colloid bonding sites rather than the linker ends. We replaced each pairwise $g_{\rm R}^{({\rm cl})}$ in the superposition approximation for $g_{\rm R}^{(4)}$ using Eq.~\eqref{eq:gcplin}. Labeling the additional colloid bonding sites as $C_{12} = \{C_1,C_2\}$ and linker bonding sites as $D_{12} = \{D_1,D_2\}$, we obtained
\begin{align}
\Delta_4 \approx &4\pi v_{A_1 B_1}^4 \int_{d_{\rm c}}^{\sqrt{3}d_{\rm cl}^*+L} \dd{r_{12}} r_{12}^2 \Big\langle p(r_{A_1 C_1}) p(r_{A_2 C_2}) \nonumber \\
&\times g_{\rm R}^{({\rm cl})}(d_{\rm cl}^*,r_{B_1}) g_{\rm R}^{({\rm cl})}(d_{\rm cl}^*,r_{B_2}) g_{\rm R}^{({\rm cl})}(d_{\rm cl}^*,r_{D_1}) \nonumber \\
&\times g_{\rm R}^{({\rm cl})}(d_{\rm cl}^*,r_{D_2}) f_{A_1 B_1}^{({\rm cl})}(r_{A_1 B_1}) f_{C_1 B_2}^{({\rm cl})}(r_{C_1 B_2}) \nonumber \\
&\times f_{C_2 D_1}^{({\rm cl})}(r_{C_2 D_1}) f_{A_2 D_2}^{({\rm cl})}(r_{A_2 D_2})\Big\rangle_{\vv{\Omega}_1,\vvg{\Omega}_2}.
\end{align}
We evaluated the integral over the range of colloid separations $d_{\rm c} \le r_{12} \le \sqrt{3}d_{\rm cl}^*+L$, for which the integrand is nonzero based on $e_{\rm R}^{({\rm cc})}$ and $p(\vv{R})$, using the trapezoidal rule with 21 uniformly spaced points. The angle brackets denote an unweighted average over the orientations of the two colloids, which we evaluated at each separation $r_{12}$ using Monte Carlo sampling. We placed one colloid with sites $A_{12}$ at the origin $(0,0,0)$ and the other with sites $C_{12}$ along the $z$-axis at $(0,0,r_{12})$. The bonding sites on both colloids were initially oriented along the $+z$ and $+y$ axes, and we generated $4 \times 10^6$ configurations by drawing and applying uniformly random axis--angle rotations to each colloid \cite{Miles:1965}. In the averaging, we rejected any configurations in which (1) the distance $r_{A_1 C_1}$ or $r_{A_2 C_2}$ was greater than the contour length $L$ or less than the segment diameter $d_{\rm l}$, (2) either of the site $z$-coordinates $z_{A_1}$ or $z_{A_2}$ was less than zero, or (3) either of the site $z$-coordinates $z_{C_1}$ or $z_{C_2}$ was greater than $r_{12}$. The first condition rejects linker conformations excluded based on $p(\vv{R})$, while the other two conditions reject ring configurations that are unlikely or forbidden because the linkers would need to wrap around or penetrate the colloids.

The chemical equilibrium equations can be written as five nonlinear algebraic equations in five variables: $X_{\Gamma_{\rm c}\setminus A_{12}}^{({\rm c})}$, $X_{\Gamma_{\rm c}\setminus A_1}^{({\rm c})}$, $X_{\Gamma_{\rm c}}^{({\rm c})}$, $X_{B_1}^{({\rm l})}$, and $X_{B_{12}}^{({\rm l})}$. In addition to Eqs.~\eqref{eq:eq1c} and \eqref{eq:eq1p} and Eqs.~\eqref{eq:eq4c} and \eqref{eq:eq4p}, the equations from Eq.~\eqref{eq:Xdb} were
\begin{align}
1 = &\bigg[11 \Big(X_{\Gamma_{\rm c}\setminus A_1}^{({\rm c})}\Big)^6 - 24 X_{\Gamma_{\rm c}\setminus A_{12}}^{({\rm c})} \Big(X_{\Gamma_{\rm c}\setminus A_1}^{({\rm c})}\Big)^4 X_{\Gamma_{\rm c}}^{({\rm c})} \nonumber \\
&+ 6 \Big(X_{\Gamma_{\rm c}\setminus A_{12}}^{({\rm c})}\Big)^2 \Big(X_{\Gamma_{\rm c}\setminus A_1}^{({\rm c})}\Big)^2 \Big(X_{\Gamma_{\rm c}}^{({\rm c})}\Big)^2 \nonumber \\
&+ 8 \Big(X_{\Gamma_{\rm c}\setminus A_{12}}^{({\rm c})}\Big)^3 \Big(X_{\Gamma_{\rm c}}^{({\rm c})}\Big)^3\bigg]\Big/\Big(X_{\Gamma_{\rm c}}^{({\rm c})}\Big)^5 \\
X_{A_1}^{({\rm c})} = &\bigg[3 \Big(X_{\Gamma_{\rm c}\setminus A_1}^{({\rm c})}\Big)^5 - 12 X_{\Gamma_{\rm c}\setminus A_{12}}^{({\rm c})} \Big(X_{\Gamma_{\rm c}\setminus A_1}^{({\rm c})}\Big)^3 X_{\Gamma_{\rm c}}^{({\rm c})} \nonumber \\
&+ 10 \Big(X_{\Gamma_{\rm c}\setminus A_{12}}^{({\rm c})}\Big)^2 X_{\Gamma_{\rm c}\setminus A_1}^{({\rm c})} \Big(X_{\Gamma_{\rm c}}^{({\rm c})}\Big)^2\bigg]\Big/\Big(X_{\Gamma_{\rm c}}^{({\rm c})}\Big)^4 \\
X_{A_{12}}^{({\rm c})} = &\bigg[-2 \Big(X_{\Gamma_{\rm c}\setminus A_1}^{({\rm c})}\Big)^4 + X_{\Gamma_{\rm c}\setminus A_{12}}^{({\rm c})} \Big(X_{\Gamma_{\rm c}\setminus A_1}^{({\rm c})}\Big)^2 X_{\Gamma_{\rm c}}^{({\rm c})} \nonumber \\
&+ 2 \Big(X_{\Gamma_{\rm c}\setminus A_{12}}^{({\rm c})}\Big)^2 \Big(X_{\Gamma_{\rm c}}^{({\rm c})}\Big)^2\bigg]\Big/\Big(X_{\Gamma_{\rm c}}^{({\rm c})}\Big)^3.
\end{align}
The expressions for $X_{A_1}^{({\rm c})}$ and $X_{A_{12}}^{({\rm c})}$ were directly substituted before solving.

To obtain numerical solutions, we first evaluated the bond volume integrals for $0 \le \beta\varepsilon \le 20$ in increments of 0.5 using the multivariable quadrature method in SciPy (version 1.3.1) with numba (version 0.50.1) \cite{numpy,scipy,numba}. We then iteratively solved the equations for a given composition $\eta_{\rm c}$ and $\rho_{\rm l}/\rho_{\rm c}$ as a function of $\varepsilon$, starting from $\beta\varepsilon = 0$ where all $X_\alpha^{(i)} = 1$ and increasing in steps of 0.01. The bond volumes at intermediate values of $\varepsilon$ were interpolated; for $\beta\varepsilon < 5$, we used a cubic spline interpolation of the bond volumes, while for larger $\varepsilon$, we used a cubic spline interpolation of the natural logarithms of the bond volumes. The equations were solved using the SciPy implementation of the Levenberg--Marquadt algorithm with an explicitly specified Jacobian matrix, relative error tolerance of $10^{-12}$ in the sum of squares, relative error tolerance of $10^{-12}$ in the solution, and a maximum of 5000 iterations. We used the solution from the previous value of $\varepsilon$ as an initial guess, and we checked that $0 \le X_\alpha^{(i)} \le 1$ for both components and all $\alpha$ (including dependent values of $X_\alpha^{({\rm c})}$) to ensure a physically meaningful converged solution.

\section*{Data Availability}
The data that support the findings of this study are available from the authors upon reasonable request.

\section*{References}
\bibliography{tpt}

\begin{thebibliography}{64}%
\makeatletter
\providecommand \@ifxundefined [1]{%
 \@ifx{#1\undefined}
}%
\providecommand \@ifnum [1]{%
 \ifnum #1\expandafter \@firstoftwo
 \else \expandafter \@secondoftwo
 \fi
}%
\providecommand \@ifx [1]{%
 \ifx #1\expandafter \@firstoftwo
 \else \expandafter \@secondoftwo
 \fi
}%
\providecommand \natexlab [1]{#1}%
\providecommand \enquote  [1]{``#1''}%
\providecommand \bibnamefont  [1]{#1}%
\providecommand \bibfnamefont [1]{#1}%
\providecommand \citenamefont [1]{#1}%
\providecommand \href@noop [0]{\@secondoftwo}%
\providecommand \href [0]{\begingroup \@sanitize@url \@href}%
\providecommand \@href[1]{\@@startlink{#1}\@@href}%
\providecommand \@@href[1]{\endgroup#1\@@endlink}%
\providecommand \@sanitize@url [0]{\catcode `\\12\catcode `\$12\catcode
  `\&12\catcode `\#12\catcode `\^12\catcode `\_12\catcode `\%12\relax}%
\providecommand \@@startlink[1]{}%
\providecommand \@@endlink[0]{}%
\providecommand \url  [0]{\begingroup\@sanitize@url \@url }%
\providecommand \@url [1]{\endgroup\@href {#1}{\urlprefix }}%
\providecommand \urlprefix  [0]{URL }%
\providecommand \Eprint [0]{\href }%
\providecommand \doibase [0]{https://doi.org/}%
\providecommand \selectlanguage [0]{\@gobble}%
\providecommand \bibinfo  [0]{\@secondoftwo}%
\providecommand \bibfield  [0]{\@secondoftwo}%
\providecommand \translation [1]{[#1]}%
\providecommand \BibitemOpen [0]{}%
\providecommand \bibitemStop [0]{}%
\providecommand \bibitemNoStop [0]{.\EOS\space}%
\providecommand \EOS [0]{\spacefactor3000\relax}%
\providecommand \BibitemShut  [1]{\csname bibitem#1\endcsname}%
\let\auto@bib@innerbib\@empty
\bibitem [{\citenamefont {Wertheim}(1984{\natexlab{a}})}]{Wertheim:1984a}%
  \BibitemOpen
  \bibfield  {author} {\bibinfo {author} {\bibfnamefont {M.~S.}\ \bibnamefont
  {Wertheim}},\ }\bibfield  {title} {\enquote {\bibinfo {title} {{Fluids with
  Highly Directional Attractive Forces. I. Statistical Thermodynamics}},}\
  }\href@noop {} {\bibfield  {journal} {\bibinfo  {journal} {J. Stat. Phys.}\
  }\textbf {\bibinfo {volume} {35}},\ \bibinfo {pages} {19--34} (\bibinfo
  {year} {1984}{\natexlab{a}})}\BibitemShut {NoStop}%
\bibitem [{\citenamefont {Wertheim}(1984{\natexlab{b}})}]{Wertheim:1984b}%
  \BibitemOpen
  \bibfield  {author} {\bibinfo {author} {\bibfnamefont {M.~S.}\ \bibnamefont
  {Wertheim}},\ }\bibfield  {title} {\enquote {\bibinfo {title} {{Fluids with
  Highly Directional Attractive Forces. II. Thermodynamic Perturbation Theory
  and Integral Equations}},}\ }\href@noop {} {\bibfield  {journal} {\bibinfo
  {journal} {J. Stat. Phys.}\ }\textbf {\bibinfo {volume} {35}},\ \bibinfo
  {pages} {35--47} (\bibinfo {year} {1984}{\natexlab{b}})}\BibitemShut
  {NoStop}%
\bibitem [{\citenamefont {Wertheim}(1986{\natexlab{a}})}]{Wertheim:1986a}%
  \BibitemOpen
  \bibfield  {author} {\bibinfo {author} {\bibfnamefont {M.~S.}\ \bibnamefont
  {Wertheim}},\ }\bibfield  {title} {\enquote {\bibinfo {title} {{Fluids with
  Highly Directional Attractive Forces. III. Multiple Attraction Sites}},}\
  }\href@noop {} {\bibfield  {journal} {\bibinfo  {journal} {J. Stat. Phys.}\
  }\textbf {\bibinfo {volume} {42}},\ \bibinfo {pages} {459--476} (\bibinfo
  {year} {1986}{\natexlab{a}})}\BibitemShut {NoStop}%
\bibitem [{\citenamefont {Wertheim}(1986{\natexlab{b}})}]{Wertheim:1986b}%
  \BibitemOpen
  \bibfield  {author} {\bibinfo {author} {\bibfnamefont {M.~S.}\ \bibnamefont
  {Wertheim}},\ }\bibfield  {title} {\enquote {\bibinfo {title} {{Fluids with
  Highly Directional Attractive Forces. IV. Equilibrium Polymerization}},}\
  }\href@noop {} {\bibfield  {journal} {\bibinfo  {journal} {J. Stat. Phys.}\
  }\textbf {\bibinfo {volume} {42}},\ \bibinfo {pages} {477--492} (\bibinfo
  {year} {1986}{\natexlab{b}})}\BibitemShut {NoStop}%
\bibitem [{\citenamefont {Rovigatti}, \citenamefont {Bomboi},\ and\
  \citenamefont {Sciortino}(2014)}]{Rovigatti:2014}%
  \BibitemOpen
  \bibfield  {author} {\bibinfo {author} {\bibfnamefont {L.}~\bibnamefont
  {Rovigatti}}, \bibinfo {author} {\bibfnamefont {F.}~\bibnamefont {Bomboi}},\
  and\ \bibinfo {author} {\bibfnamefont {F.}~\bibnamefont {Sciortino}},\
  }\bibfield  {title} {\enquote {\bibinfo {title} {{Accurate phase diagram of
  tetravalent DNA nanostars}},}\ }\href@noop {} {\bibfield  {journal} {\bibinfo
   {journal} {J. Chem. Phys.}\ }\textbf {\bibinfo {volume} {140}},\ \bibinfo
  {pages} {154903} (\bibinfo {year} {2014})}\BibitemShut {NoStop}%
\bibitem [{\citenamefont {Locatelli}\ \emph {et~al.}(2017)\citenamefont
  {Locatelli}, \citenamefont {Handle}, \citenamefont {Likos}, \citenamefont
  {Sciortino},\ and\ \citenamefont {Rovigatti}}]{Locatelli:2017}%
  \BibitemOpen
  \bibfield  {author} {\bibinfo {author} {\bibfnamefont {E.}~\bibnamefont
  {Locatelli}}, \bibinfo {author} {\bibfnamefont {P.~H.}\ \bibnamefont
  {Handle}}, \bibinfo {author} {\bibfnamefont {C.~N.}\ \bibnamefont {Likos}},
  \bibinfo {author} {\bibfnamefont {F.}~\bibnamefont {Sciortino}},\ and\
  \bibinfo {author} {\bibfnamefont {L.}~\bibnamefont {Rovigatti}},\ }\bibfield
  {title} {\enquote {\bibinfo {title} {{Condensation and Demixing in Solutions
  of DNA Nanostars and Their Mixtures}},}\ }\href@noop {} {\bibfield  {journal}
  {\bibinfo  {journal} {ACS Nano}\ }\textbf {\bibinfo {volume} {11}},\ \bibinfo
  {pages} {2094--2102} (\bibinfo {year} {2017})}\BibitemShut {NoStop}%
\bibitem [{\citenamefont {Bianchi}\ \emph
  {et~al.}(2006{\natexlab{a}})\citenamefont {Bianchi}, \citenamefont {Largo},
  \citenamefont {Tartaglia}, \citenamefont {Zaccarelli},\ and\ \citenamefont
  {Sciortino}}]{Bianchi2006}%
  \BibitemOpen
  \bibfield  {author} {\bibinfo {author} {\bibfnamefont {E.}~\bibnamefont
  {Bianchi}}, \bibinfo {author} {\bibfnamefont {J.}~\bibnamefont {Largo}},
  \bibinfo {author} {\bibfnamefont {P.}~\bibnamefont {Tartaglia}}, \bibinfo
  {author} {\bibfnamefont {E.}~\bibnamefont {Zaccarelli}},\ and\ \bibinfo
  {author} {\bibfnamefont {F.}~\bibnamefont {Sciortino}},\ }\bibfield  {title}
  {\enquote {\bibinfo {title} {{Phase Diagram of Patchy Colloids: Towards Empty
  Liquids}},}\ }\href@noop {} {\bibfield  {journal} {\bibinfo  {journal} {Phys.
  Rev. Lett.}\ }\textbf {\bibinfo {volume} {97}},\ \bibinfo {pages} {168301}
  (\bibinfo {year} {2006}{\natexlab{a}})}\BibitemShut {NoStop}%
\bibitem [{\citenamefont {Bianchi}\ \emph {et~al.}(2007)\citenamefont
  {Bianchi}, \citenamefont {Tartaglia}, \citenamefont {La~Nave},\ and\
  \citenamefont {Sciortino}}]{Bianchi2007}%
  \BibitemOpen
  \bibfield  {author} {\bibinfo {author} {\bibfnamefont {E.}~\bibnamefont
  {Bianchi}}, \bibinfo {author} {\bibfnamefont {P.}~\bibnamefont {Tartaglia}},
  \bibinfo {author} {\bibfnamefont {E.}~\bibnamefont {La~Nave}},\ and\ \bibinfo
  {author} {\bibfnamefont {F.}~\bibnamefont {Sciortino}},\ }\bibfield  {title}
  {\enquote {\bibinfo {title} {{Fully Solvable Equilibrium Self-Assembly
  Process: Fine-Tuning the Clusters Size and the Connectivity in Patchy
  Particle Systems}},}\ }\href@noop {} {\bibfield  {journal} {\bibinfo
  {journal} {J. Phys. Chem. B}\ }\textbf {\bibinfo {volume} {111}},\ \bibinfo
  {pages} {11765--11769} (\bibinfo {year} {2007})}\BibitemShut {NoStop}%
\bibitem [{\citenamefont {Russo}, \citenamefont {Tartaglia},\ and\
  \citenamefont {Sciortino}(2009)}]{Russo:2009}%
  \BibitemOpen
  \bibfield  {author} {\bibinfo {author} {\bibfnamefont {J.}~\bibnamefont
  {Russo}}, \bibinfo {author} {\bibfnamefont {P.}~\bibnamefont {Tartaglia}},\
  and\ \bibinfo {author} {\bibfnamefont {F.}~\bibnamefont {Sciortino}},\
  }\bibfield  {title} {\enquote {\bibinfo {title} {{Reversible gels of patchy
  particles: Role of the valence}},}\ }\href@noop {} {\bibfield  {journal}
  {\bibinfo  {journal} {J. Chem. Phys.}\ }\textbf {\bibinfo {volume} {131}},\
  \bibinfo {pages} {014504} (\bibinfo {year} {2009})}\BibitemShut {NoStop}%
\bibitem [{\citenamefont {Russo}\ \emph {et~al.}(2011)\citenamefont {Russo},
  \citenamefont {Tavares}, \citenamefont {Teixeira}, \citenamefont {Telo~da
  Gama},\ and\ \citenamefont {Sciortino}}]{Russo:2011}%
  \BibitemOpen
  \bibfield  {author} {\bibinfo {author} {\bibfnamefont {J.}~\bibnamefont
  {Russo}}, \bibinfo {author} {\bibfnamefont {J.~M.}\ \bibnamefont {Tavares}},
  \bibinfo {author} {\bibfnamefont {P.~I.~C.}\ \bibnamefont {Teixeira}},
  \bibinfo {author} {\bibfnamefont {M.~M.}\ \bibnamefont {Telo~da Gama}},\ and\
  \bibinfo {author} {\bibfnamefont {F.}~\bibnamefont {Sciortino}},\ }\bibfield
  {title} {\enquote {\bibinfo {title} {{Reentrant Phase Diagram of Network
  Fluids}},}\ }\href@noop {} {\bibfield  {journal} {\bibinfo  {journal} {Phys.
  Rev. Lett.}\ }\textbf {\bibinfo {volume} {106}},\ \bibinfo {pages} {085703}
  (\bibinfo {year} {2011})}\BibitemShut {NoStop}%
\bibitem [{\citenamefont {Chapman}\ \emph {et~al.}(1986)\citenamefont
  {Chapman}, \citenamefont {Gubbins}, \citenamefont {Joslin},\ and\
  \citenamefont {Gray}}]{Chapman:1986}%
  \BibitemOpen
  \bibfield  {author} {\bibinfo {author} {\bibfnamefont {W.~G.}\ \bibnamefont
  {Chapman}}, \bibinfo {author} {\bibfnamefont {K.~E.}\ \bibnamefont
  {Gubbins}}, \bibinfo {author} {\bibfnamefont {C.~G.}\ \bibnamefont
  {Joslin}},\ and\ \bibinfo {author} {\bibfnamefont {C.~G.}\ \bibnamefont
  {Gray}},\ }\bibfield  {title} {\enquote {\bibinfo {title} {{Theory and
  Simulation of Associating Liquid Mixtures}},}\ }\href@noop {} {\bibfield
  {journal} {\bibinfo  {journal} {Fluid Phase Equilib.}\ }\textbf {\bibinfo
  {volume} {29}},\ \bibinfo {pages} {337--346} (\bibinfo {year}
  {1986})}\BibitemShut {NoStop}%
\bibitem [{\citenamefont {Joslin}\ \emph {et~al.}(1987)\citenamefont {Joslin},
  \citenamefont {Gray}, \citenamefont {Chapman},\ and\ \citenamefont
  {Gubbins}}]{Joslin:1987}%
  \BibitemOpen
  \bibfield  {author} {\bibinfo {author} {\bibfnamefont {C.~G.}\ \bibnamefont
  {Joslin}}, \bibinfo {author} {\bibfnamefont {C.~G.}\ \bibnamefont {Gray}},
  \bibinfo {author} {\bibfnamefont {W.~G.}\ \bibnamefont {Chapman}},\ and\
  \bibinfo {author} {\bibfnamefont {K.~E.}\ \bibnamefont {Gubbins}},\
  }\bibfield  {title} {\enquote {\bibinfo {title} {{Theory and Simulation of
  Associating Liquid Mixtures. II}},}\ }\href@noop {} {\bibfield  {journal}
  {\bibinfo  {journal} {Mol. Phys.}\ }\textbf {\bibinfo {volume} {62}},\
  \bibinfo {pages} {843--860} (\bibinfo {year} {1987})}\BibitemShut {NoStop}%
\bibitem [{\citenamefont {Jackson}, \citenamefont {Chapman},\ and\
  \citenamefont {Gubbins}(1988)}]{Jackson:1988}%
  \BibitemOpen
  \bibfield  {author} {\bibinfo {author} {\bibfnamefont {G.}~\bibnamefont
  {Jackson}}, \bibinfo {author} {\bibfnamefont {W.~G.}\ \bibnamefont
  {Chapman}},\ and\ \bibinfo {author} {\bibfnamefont {K.~E.}\ \bibnamefont
  {Gubbins}},\ }\bibfield  {title} {\enquote {\bibinfo {title} {{Phase
  equilibria of associating fluids: Spherical molecules with multiple bonding
  sites}},}\ }\href@noop {} {\bibfield  {journal} {\bibinfo  {journal} {Mol.
  Phys.}\ }\textbf {\bibinfo {volume} {65}},\ \bibinfo {pages} {1--31}
  (\bibinfo {year} {1988})}\BibitemShut {NoStop}%
\bibitem [{\citenamefont {Chapman}, \citenamefont {Jackson},\ and\
  \citenamefont {Gubbins}(1988)}]{Chapman:1988}%
  \BibitemOpen
  \bibfield  {author} {\bibinfo {author} {\bibfnamefont {W.~G.}\ \bibnamefont
  {Chapman}}, \bibinfo {author} {\bibfnamefont {G.}~\bibnamefont {Jackson}},\
  and\ \bibinfo {author} {\bibfnamefont {K.~E.}\ \bibnamefont {Gubbins}},\
  }\bibfield  {title} {\enquote {\bibinfo {title} {{Phase equilibria of
  associating fluids: Chain molecules with multiple bonding sites}},}\
  }\href@noop {} {\bibfield  {journal} {\bibinfo  {journal} {Mol. Phys.}\
  }\textbf {\bibinfo {volume} {65}},\ \bibinfo {pages} {1057--1079} (\bibinfo
  {year} {1988})}\BibitemShut {NoStop}%
\bibitem [{\citenamefont {M\"uller}\ and\ \citenamefont
  {Gubbins}(2001)}]{Muller:2001}%
  \BibitemOpen
  \bibfield  {author} {\bibinfo {author} {\bibfnamefont {E.~A.}\ \bibnamefont
  {M\"uller}}\ and\ \bibinfo {author} {\bibfnamefont {K.~E.}\ \bibnamefont
  {Gubbins}},\ }\bibfield  {title} {\enquote {\bibinfo {title}
  {{Molecular-Based Equations of State for Associating Fluids: A Review of SAFT
  and Related Approaches}},}\ }\href@noop {} {\bibfield  {journal} {\bibinfo
  {journal} {Ind. Eng. Chem. Res.}\ }\textbf {\bibinfo {volume} {40}},\
  \bibinfo {pages} {2193--2211} (\bibinfo {year} {2001})}\BibitemShut {NoStop}%
\bibitem [{\citenamefont {Howard}\ \emph {et~al.}(2019)\citenamefont {Howard},
  \citenamefont {Jadrich}, \citenamefont {Lindquist}, \citenamefont {Khabaz},
  \citenamefont {Bonnecaze}, \citenamefont {Milliron},\ and\ \citenamefont
  {Truskett}}]{Howard:2019}%
  \BibitemOpen
  \bibfield  {author} {\bibinfo {author} {\bibfnamefont {M.~P.}\ \bibnamefont
  {Howard}}, \bibinfo {author} {\bibfnamefont {R.~B.}\ \bibnamefont {Jadrich}},
  \bibinfo {author} {\bibfnamefont {B.~A.}\ \bibnamefont {Lindquist}}, \bibinfo
  {author} {\bibfnamefont {F.}~\bibnamefont {Khabaz}}, \bibinfo {author}
  {\bibfnamefont {R.~T.}\ \bibnamefont {Bonnecaze}}, \bibinfo {author}
  {\bibfnamefont {D.~J.}\ \bibnamefont {Milliron}},\ and\ \bibinfo {author}
  {\bibfnamefont {T.~M.}\ \bibnamefont {Truskett}},\ }\bibfield  {title}
  {\enquote {\bibinfo {title} {{Structure and phase behavior of polymer-linked
  colloidal gels}},}\ }\href@noop {} {\bibfield  {journal} {\bibinfo  {journal}
  {J. Chem. Phys.}\ }\textbf {\bibinfo {volume} {151}},\ \bibinfo {pages}
  {124901} (\bibinfo {year} {2019})}\BibitemShut {NoStop}%
\bibitem [{\citenamefont {Lindquist}\ \emph {et~al.}(2016)\citenamefont
  {Lindquist}, \citenamefont {Jadrich}, \citenamefont {Milliron},\ and\
  \citenamefont {Truskett}}]{Lindquist:2016}%
  \BibitemOpen
  \bibfield  {author} {\bibinfo {author} {\bibfnamefont {B.~A.}\ \bibnamefont
  {Lindquist}}, \bibinfo {author} {\bibfnamefont {R.~B.}\ \bibnamefont
  {Jadrich}}, \bibinfo {author} {\bibfnamefont {D.~J.}\ \bibnamefont
  {Milliron}},\ and\ \bibinfo {author} {\bibfnamefont {T.~M.}\ \bibnamefont
  {Truskett}},\ }\bibfield  {title} {\enquote {\bibinfo {title} {{On the
  formation of equilibrium gels via a macroscopic bond limitation}},}\
  }\href@noop {} {\bibfield  {journal} {\bibinfo  {journal} {J. Chem. Phys.}\
  }\textbf {\bibinfo {volume} {145}},\ \bibinfo {pages} {074906} (\bibinfo
  {year} {2016})}\BibitemShut {NoStop}%
\bibitem [{\citenamefont {Saez~Cabezas}\ \emph {et~al.}(2018)\citenamefont
  {Saez~Cabezas}, \citenamefont {Ong}, \citenamefont {Jadrich}, \citenamefont
  {Lindquist}, \citenamefont {Agrawal}, \citenamefont {Truskett},\ and\
  \citenamefont {Milliron}}]{SaezCabezas:2018}%
  \BibitemOpen
  \bibfield  {author} {\bibinfo {author} {\bibfnamefont {C.~A.}\ \bibnamefont
  {Saez~Cabezas}}, \bibinfo {author} {\bibfnamefont {G.~K.}\ \bibnamefont
  {Ong}}, \bibinfo {author} {\bibfnamefont {R.~B.}\ \bibnamefont {Jadrich}},
  \bibinfo {author} {\bibfnamefont {B.~A.}\ \bibnamefont {Lindquist}}, \bibinfo
  {author} {\bibfnamefont {A.}~\bibnamefont {Agrawal}}, \bibinfo {author}
  {\bibfnamefont {T.~M.}\ \bibnamefont {Truskett}},\ and\ \bibinfo {author}
  {\bibfnamefont {D.~J.}\ \bibnamefont {Milliron}},\ }\bibfield  {title}
  {\enquote {\bibinfo {title} {{Gelation of plasmonic metal oxide nanocrystals
  by polymer-induced depletion attractions}},}\ }\href@noop {} {\bibfield
  {journal} {\bibinfo  {journal} {Proc. Natl. Acad. Sci. U.S.A.}\ }\textbf
  {\bibinfo {volume} {115}},\ \bibinfo {pages} {8925--8930} (\bibinfo {year}
  {2018})}\BibitemShut {NoStop}%
\bibitem [{\citenamefont {Dominguez}\ \emph {et~al.}(2020)\citenamefont
  {Dominguez}, \citenamefont {Howard}, \citenamefont {Maier}, \citenamefont
  {Valenzuela}, \citenamefont {Sherman}, \citenamefont {Reimnitz},
  \citenamefont {Kang}, \citenamefont {Cho}, \citenamefont {Gibbs},
  \citenamefont {Menta}, \citenamefont {Zhuang}, \citenamefont {van~der Stok},
  \citenamefont {Kline}, \citenamefont {Anslyn}, \citenamefont {Truskett},\
  and\ \citenamefont {Milliron}}]{Dominguez:2020}%
  \BibitemOpen
  \bibfield  {author} {\bibinfo {author} {\bibfnamefont {M.~N.}\ \bibnamefont
  {Dominguez}}, \bibinfo {author} {\bibfnamefont {M.~P.}\ \bibnamefont
  {Howard}}, \bibinfo {author} {\bibfnamefont {J.~M.}\ \bibnamefont {Maier}},
  \bibinfo {author} {\bibfnamefont {S.}~\bibnamefont {Valenzuela}}, \bibinfo
  {author} {\bibfnamefont {Z.~M.}\ \bibnamefont {Sherman}}, \bibinfo {author}
  {\bibfnamefont {L.~C.}\ \bibnamefont {Reimnitz}}, \bibinfo {author}
  {\bibfnamefont {J.}~\bibnamefont {Kang}}, \bibinfo {author} {\bibfnamefont
  {S.~H.}\ \bibnamefont {Cho}}, \bibinfo {author} {\bibfnamefont {S.~L.}\
  \bibnamefont {Gibbs}}, \bibinfo {author} {\bibfnamefont {A.~K.}\ \bibnamefont
  {Menta}}, \bibinfo {author} {\bibfnamefont {D.~L.}\ \bibnamefont {Zhuang}},
  \bibinfo {author} {\bibfnamefont {A.}~\bibnamefont {van~der Stok}}, \bibinfo
  {author} {\bibfnamefont {S.~J.}\ \bibnamefont {Kline}}, \bibinfo {author}
  {\bibfnamefont {E.~V.}\ \bibnamefont {Anslyn}}, \bibinfo {author}
  {\bibfnamefont {T.~M.}\ \bibnamefont {Truskett}},\ and\ \bibinfo {author}
  {\bibfnamefont {D.~J.}\ \bibnamefont {Milliron}},\ }\bibfield  {title}
  {\enquote {\bibinfo {title} {{Assembly of Linked Nanocrystal Colloids by
  Reversible Covalent Bonds}},}\ }\href@noop {} {\bibfield  {journal} {\bibinfo
   {journal} {Chem. Mater.}\ }\textbf {\bibinfo {volume} {32}},\ \bibinfo
  {pages} {10235--10245} (\bibinfo {year} {2020})}\BibitemShut {NoStop}%
\bibitem [{\citenamefont {Stukowski}(2010)}]{Stukowski:2010}%
  \BibitemOpen
  \bibfield  {author} {\bibinfo {author} {\bibfnamefont {A.}~\bibnamefont
  {Stukowski}},\ }\bibfield  {title} {\enquote {\bibinfo {title}
  {{Visualization and analysis of atomistic simulation data with OVITO--the
  Open Visualization Tool}},}\ }\href@noop {} {\bibfield  {journal} {\bibinfo
  {journal} {Modell. Simul. Mater. Sci. Eng.}\ }\textbf {\bibinfo {volume}
  {18}},\ \bibinfo {pages} {015012} (\bibinfo {year} {2010})}\BibitemShut
  {NoStop}%
\bibitem [{\citenamefont {Sear}\ and\ \citenamefont
  {Jackson}(1994{\natexlab{a}})}]{Sear:1994a}%
  \BibitemOpen
  \bibfield  {author} {\bibinfo {author} {\bibfnamefont {R.~P.}\ \bibnamefont
  {Sear}}\ and\ \bibinfo {author} {\bibfnamefont {G.}~\bibnamefont {Jackson}},\
  }\bibfield  {title} {\enquote {\bibinfo {title} {{Thermodynamic perturbation
  theory for association into doubly bonded dimers}},}\ }\href@noop {}
  {\bibfield  {journal} {\bibinfo  {journal} {Mol. Phys.}\ }\textbf {\bibinfo
  {volume} {82}},\ \bibinfo {pages} {1033--1048} (\bibinfo {year}
  {1994}{\natexlab{a}})}\BibitemShut {NoStop}%
\bibitem [{\citenamefont {Sear}\ and\ \citenamefont
  {Jackson}(1994{\natexlab{b}})}]{Sear:1994b}%
  \BibitemOpen
  \bibfield  {author} {\bibinfo {author} {\bibfnamefont {R.~P.}\ \bibnamefont
  {Sear}}\ and\ \bibinfo {author} {\bibfnamefont {G.}~\bibnamefont {Jackson}},\
  }\bibfield  {title} {\enquote {\bibinfo {title} {{Thermodynamic perturbation
  theory for association into chains and rings}},}\ }\href@noop {} {\bibfield
  {journal} {\bibinfo  {journal} {Phys. Rev. E}\ }\textbf {\bibinfo {volume}
  {50}},\ \bibinfo {pages} {386--394} (\bibinfo {year}
  {1994}{\natexlab{b}})}\BibitemShut {NoStop}%
\bibitem [{\citenamefont {Galindo}\ \emph {et~al.}(2002)\citenamefont
  {Galindo}, \citenamefont {Burton}, \citenamefont {Jackson}, \citenamefont
  {Visco~Jr.},\ and\ \citenamefont {Kofke}}]{Galindo:2002}%
  \BibitemOpen
  \bibfield  {author} {\bibinfo {author} {\bibfnamefont {A.}~\bibnamefont
  {Galindo}}, \bibinfo {author} {\bibfnamefont {S.~J.}\ \bibnamefont {Burton}},
  \bibinfo {author} {\bibfnamefont {G.}~\bibnamefont {Jackson}}, \bibinfo
  {author} {\bibfnamefont {D.~P.}\ \bibnamefont {Visco~Jr.}},\ and\ \bibinfo
  {author} {\bibfnamefont {D.~A.}\ \bibnamefont {Kofke}},\ }\bibfield  {title}
  {\enquote {\bibinfo {title} {{Improved models for the phase behaviour of
  hydrogen fluoride: chain and ring aggregates in the SAFT approach and the
  AEOS model}},}\ }\href@noop {} {\bibfield  {journal} {\bibinfo  {journal}
  {Mol. Phys.}\ }\textbf {\bibinfo {volume} {100}},\ \bibinfo {pages}
  {2241--2259} (\bibinfo {year} {2002})}\BibitemShut {NoStop}%
\bibitem [{\citenamefont {Avlund}, \citenamefont {Kontogeorgis},\ and\
  \citenamefont {Chapman}(2011)}]{Avlund:2011}%
  \BibitemOpen
  \bibfield  {author} {\bibinfo {author} {\bibfnamefont {A.~S.}\ \bibnamefont
  {Avlund}}, \bibinfo {author} {\bibfnamefont {G.~M.}\ \bibnamefont
  {Kontogeorgis}},\ and\ \bibinfo {author} {\bibfnamefont {W.~G.}\ \bibnamefont
  {Chapman}},\ }\bibfield  {title} {\enquote {\bibinfo {title} {{Intramolecular
  association within the SAFT framework}},}\ }\href@noop {} {\bibfield
  {journal} {\bibinfo  {journal} {Mol. Phys.}\ }\textbf {\bibinfo {volume}
  {109}},\ \bibinfo {pages} {1759--1769} (\bibinfo {year} {2011})}\BibitemShut
  {NoStop}%
\bibitem [{\citenamefont {Tavares}, \citenamefont {Rovigatti},\ and\
  \citenamefont {Sciortino}(2012)}]{Tavares:2012}%
  \BibitemOpen
  \bibfield  {author} {\bibinfo {author} {\bibfnamefont {J.~M.}\ \bibnamefont
  {Tavares}}, \bibinfo {author} {\bibfnamefont {L.}~\bibnamefont {Rovigatti}},\
  and\ \bibinfo {author} {\bibfnamefont {F.}~\bibnamefont {Sciortino}},\
  }\bibfield  {title} {\enquote {\bibinfo {title} {{Quantitative description of
  the self-assembly of patchy particles into chains and rings}},}\ }\href@noop
  {} {\bibfield  {journal} {\bibinfo  {journal} {J. Chem. Phys.}\ }\textbf
  {\bibinfo {volume} {137}},\ \bibinfo {pages} {044901} (\bibinfo {year}
  {2012})}\BibitemShut {NoStop}%
\bibitem [{\citenamefont {Rovigatti}, \citenamefont {Tavares},\ and\
  \citenamefont {Sciortino}(2013)}]{Rovigatti:2013}%
  \BibitemOpen
  \bibfield  {author} {\bibinfo {author} {\bibfnamefont {L.}~\bibnamefont
  {Rovigatti}}, \bibinfo {author} {\bibfnamefont {J.~M.}\ \bibnamefont
  {Tavares}},\ and\ \bibinfo {author} {\bibfnamefont {F.}~\bibnamefont
  {Sciortino}},\ }\bibfield  {title} {\enquote {\bibinfo {title}
  {{Self-Assembly in Chains, Rings, and Branches: A Single Component System
  with Two Critical Points}},}\ }\href@noop {} {\bibfield  {journal} {\bibinfo
  {journal} {Phys. Rev. Lett.}\ }\textbf {\bibinfo {volume} {111}},\ \bibinfo
  {pages} {168302} (\bibinfo {year} {2013})}\BibitemShut {NoStop}%
\bibitem [{\citenamefont {Marshall}\ and\ \citenamefont
  {Chapman}(2013{\natexlab{a}})}]{Marshall:2013a}%
  \BibitemOpen
  \bibfield  {author} {\bibinfo {author} {\bibfnamefont {B.~D.}\ \bibnamefont
  {Marshall}}\ and\ \bibinfo {author} {\bibfnamefont {W.~G.}\ \bibnamefont
  {Chapman}},\ }\bibfield  {title} {\enquote {\bibinfo {title} {{Thermodynamic
  perturbation theory for associating fluids with small bond angles: Effects of
  steric hindrance, ring formation, and double bonding}},}\ }\href@noop {}
  {\bibfield  {journal} {\bibinfo  {journal} {Phys. Rev. E}\ }\textbf {\bibinfo
  {volume} {87}},\ \bibinfo {pages} {052307} (\bibinfo {year}
  {2013}{\natexlab{a}})}\BibitemShut {NoStop}%
\bibitem [{\citenamefont {Marshall}\ and\ \citenamefont
  {Chapman}(2013{\natexlab{b}})}]{Marshall:2013b}%
  \BibitemOpen
  \bibfield  {author} {\bibinfo {author} {\bibfnamefont {B.~D.}\ \bibnamefont
  {Marshall}}\ and\ \bibinfo {author} {\bibfnamefont {W.~G.}\ \bibnamefont
  {Chapman}},\ }\bibfield  {title} {\enquote {\bibinfo {title} {{Molecular
  theory for the phase equilibria and cluster distribution of associating
  fluids with small bond angles}},}\ }\href@noop {} {\bibfield  {journal}
  {\bibinfo  {journal} {J. Chem. Phys.}\ }\textbf {\bibinfo {volume} {139}},\
  \bibinfo {pages} {054902} (\bibinfo {year} {2013}{\natexlab{b}})}\BibitemShut
  {NoStop}%
\bibitem [{\citenamefont {Marshall}(2018)}]{Marshall:2018}%
  \BibitemOpen
  \bibfield  {author} {\bibinfo {author} {\bibfnamefont {B.~D.}\ \bibnamefont
  {Marshall}},\ }\bibfield  {title} {\enquote {\bibinfo {title} {{A general
  mixture equation of state for double bonding carboxylic acids with $\ge$2
  association sites}},}\ }\href@noop {} {\bibfield  {journal} {\bibinfo
  {journal} {J. Chem. Phys.}\ }\textbf {\bibinfo {volume} {148}},\ \bibinfo
  {pages} {174103} (\bibinfo {year} {2018})}\BibitemShut {NoStop}%
\bibitem [{\citenamefont {Hoppe}\ \emph {et~al.}(2002)\citenamefont {Hoppe},
  \citenamefont {Geidel}, \citenamefont {Weller},\ and\ \citenamefont
  {Eychm\"uller}}]{Hoppe:2002}%
  \BibitemOpen
  \bibfield  {author} {\bibinfo {author} {\bibfnamefont {K.}~\bibnamefont
  {Hoppe}}, \bibinfo {author} {\bibfnamefont {E.}~\bibnamefont {Geidel}},
  \bibinfo {author} {\bibfnamefont {H.}~\bibnamefont {Weller}},\ and\ \bibinfo
  {author} {\bibfnamefont {A.}~\bibnamefont {Eychm\"uller}},\ }\bibfield
  {title} {\enquote {\bibinfo {title} {{Covalently bound CdTe nanocrystals}},}\
  }\href@noop {} {\bibfield  {journal} {\bibinfo  {journal} {Phys. Chem. Chem.
  Phys.}\ }\textbf {\bibinfo {volume} {4}},\ \bibinfo {pages} {1704--1706}
  (\bibinfo {year} {2002})}\BibitemShut {NoStop}%
\bibitem [{\citenamefont {Maneeprakorn}, \citenamefont {Malik},\ and\
  \citenamefont {O'Brien}(2010)}]{Maneeprakorn:2010}%
  \BibitemOpen
  \bibfield  {author} {\bibinfo {author} {\bibfnamefont {W.}~\bibnamefont
  {Maneeprakorn}}, \bibinfo {author} {\bibfnamefont {M.~A.}\ \bibnamefont
  {Malik}},\ and\ \bibinfo {author} {\bibfnamefont {P.}~\bibnamefont
  {O'Brien}},\ }\bibfield  {title} {\enquote {\bibinfo {title} {{Developing
  Chemical Strategies for the Assembly of Nanoparticles into Mesoscopic
  Objects}},}\ }\href@noop {} {\bibfield  {journal} {\bibinfo  {journal} {J.
  Am. Chem. Soc.}\ }\textbf {\bibinfo {volume} {132}},\ \bibinfo {pages}
  {1780--1781} (\bibinfo {year} {2010})}\BibitemShut {NoStop}%
\bibitem [{\citenamefont {Macfarlane}\ \emph {et~al.}(2011)\citenamefont
  {Macfarlane}, \citenamefont {Lee}, \citenamefont {Jones}, \citenamefont
  {Harris}, \citenamefont {Schatz},\ and\ \citenamefont
  {Mirkin}}]{Macfarlane:2011}%
  \BibitemOpen
  \bibfield  {author} {\bibinfo {author} {\bibfnamefont {R.~J.}\ \bibnamefont
  {Macfarlane}}, \bibinfo {author} {\bibfnamefont {B.}~\bibnamefont {Lee}},
  \bibinfo {author} {\bibfnamefont {M.}~\bibnamefont {Jones}}, \bibinfo
  {author} {\bibfnamefont {N.}~\bibnamefont {Harris}}, \bibinfo {author}
  {\bibfnamefont {G.~C.}\ \bibnamefont {Schatz}},\ and\ \bibinfo {author}
  {\bibfnamefont {C.~A.}\ \bibnamefont {Mirkin}},\ }\bibfield  {title}
  {\enquote {\bibinfo {title} {{Nanoparticle Superlattice Engineering with
  DNA}},}\ }\href@noop {} {\bibfield  {journal} {\bibinfo  {journal} {Science}\
  }\textbf {\bibinfo {volume} {334}},\ \bibinfo {pages} {204--208} (\bibinfo
  {year} {2011})}\BibitemShut {NoStop}%
\bibitem [{\citenamefont {Borsley}\ and\ \citenamefont
  {Kay}(2016)}]{Borsley:2016}%
  \BibitemOpen
  \bibfield  {author} {\bibinfo {author} {\bibfnamefont {S.}~\bibnamefont
  {Borsley}}\ and\ \bibinfo {author} {\bibfnamefont {E.~R.}\ \bibnamefont
  {Kay}},\ }\bibfield  {title} {\enquote {\bibinfo {title} {{Dynamic Covalent
  Assembly and Disassembly of Nanoparticle Aggregates}},}\ }\href@noop {}
  {\bibfield  {journal} {\bibinfo  {journal} {Chem. Commun.}\ }\textbf
  {\bibinfo {volume} {52}},\ \bibinfo {pages} {9117--9120} (\bibinfo {year}
  {2016})}\BibitemShut {NoStop}%
\bibitem [{\citenamefont {Wang}\ \emph {et~al.}(2019)\citenamefont {Wang},
  \citenamefont {Santos}, \citenamefont {Kubiak}, \citenamefont {Guo},
  \citenamefont {Lee},\ and\ \citenamefont {Macfarlane}}]{Wang:2019}%
  \BibitemOpen
  \bibfield  {author} {\bibinfo {author} {\bibfnamefont {Y.}~\bibnamefont
  {Wang}}, \bibinfo {author} {\bibfnamefont {P.~J.}\ \bibnamefont {Santos}},
  \bibinfo {author} {\bibfnamefont {J.~M.}\ \bibnamefont {Kubiak}}, \bibinfo
  {author} {\bibfnamefont {X.}~\bibnamefont {Guo}}, \bibinfo {author}
  {\bibfnamefont {M.~S.}\ \bibnamefont {Lee}},\ and\ \bibinfo {author}
  {\bibfnamefont {R.~J.}\ \bibnamefont {Macfarlane}},\ }\bibfield  {title}
  {\enquote {\bibinfo {title} {{Multistimuli Responsive Nanocomposite Tectons
  for Pathway Dependent Self-Assembly and Acceleration of Covalent Bond
  Formation}},}\ }\href@noop {} {\bibfield  {journal} {\bibinfo  {journal} {J.
  Am. Chem. Soc.}\ }\textbf {\bibinfo {volume} {141}},\ \bibinfo {pages}
  {13234--13243} (\bibinfo {year} {2019})}\BibitemShut {NoStop}%
\bibitem [{\citenamefont {Marro}, \citenamefont {della Sala},\ and\
  \citenamefont {Kay}(2020)}]{Marro:2020}%
  \BibitemOpen
  \bibfield  {author} {\bibinfo {author} {\bibfnamefont {N.}~\bibnamefont
  {Marro}}, \bibinfo {author} {\bibfnamefont {F.}~\bibnamefont {della Sala}},\
  and\ \bibinfo {author} {\bibfnamefont {E.~R.}\ \bibnamefont {Kay}},\
  }\bibfield  {title} {\enquote {\bibinfo {title} {{Programmable Dynamic
  Covalent Nanoparticle Building Blocks with Complementary Reactivity}},}\
  }\href@noop {} {\bibfield  {journal} {\bibinfo  {journal} {Chem. Sci.}\
  }\textbf {\bibinfo {volume} {11}},\ \bibinfo {pages} {372--383} (\bibinfo
  {year} {2020})}\BibitemShut {NoStop}%
\bibitem [{\citenamefont {Mirkin}\ \emph {et~al.}(1996)\citenamefont {Mirkin},
  \citenamefont {Letsinger}, \citenamefont {Mucic},\ and\ \citenamefont
  {Storhoff}}]{Mirkin:1996}%
  \BibitemOpen
  \bibfield  {author} {\bibinfo {author} {\bibfnamefont {C.~A.}\ \bibnamefont
  {Mirkin}}, \bibinfo {author} {\bibfnamefont {R.~L.}\ \bibnamefont
  {Letsinger}}, \bibinfo {author} {\bibfnamefont {R.~C.}\ \bibnamefont
  {Mucic}},\ and\ \bibinfo {author} {\bibfnamefont {J.~J.}\ \bibnamefont
  {Storhoff}},\ }\bibfield  {title} {\enquote {\bibinfo {title} {{A DNA-based
  method for rationally assembling nanoparticles into macroscopic
  materials}},}\ }\href@noop {} {\bibfield  {journal} {\bibinfo  {journal}
  {Nature}\ }\textbf {\bibinfo {volume} {382}},\ \bibinfo {pages} {607--609}
  (\bibinfo {year} {1996})}\BibitemShut {NoStop}%
\bibitem [{\citenamefont {Alivisatos}\ \emph {et~al.}(1996)\citenamefont
  {Alivisatos}, \citenamefont {Johnsson}, \citenamefont {Peng}, \citenamefont
  {Wilson}, \citenamefont {Loweth}, \citenamefont {Bruchez},\ and\
  \citenamefont {Schultz}}]{Alivisatos:1996}%
  \BibitemOpen
  \bibfield  {author} {\bibinfo {author} {\bibfnamefont {A.~P.}\ \bibnamefont
  {Alivisatos}}, \bibinfo {author} {\bibfnamefont {K.~P.}\ \bibnamefont
  {Johnsson}}, \bibinfo {author} {\bibfnamefont {X.~G.}\ \bibnamefont {Peng}},
  \bibinfo {author} {\bibfnamefont {T.~E.}\ \bibnamefont {Wilson}}, \bibinfo
  {author} {\bibfnamefont {C.~J.}\ \bibnamefont {Loweth}}, \bibinfo {author}
  {\bibfnamefont {M.~P.}\ \bibnamefont {Bruchez}},\ and\ \bibinfo {author}
  {\bibfnamefont {P.~G.}\ \bibnamefont {Schultz}},\ }\bibfield  {title}
  {\enquote {\bibinfo {title} {{Organization of 'nanocrystal molecules' using
  DNA}},}\ }\href@noop {} {\bibfield  {journal} {\bibinfo  {journal} {Nature}\
  }\textbf {\bibinfo {volume} {382}},\ \bibinfo {pages} {609--611} (\bibinfo
  {year} {1996})}\BibitemShut {NoStop}%
\bibitem [{\citenamefont {Xiong}, \citenamefont {van~der Lelie},\ and\
  \citenamefont {Gang}(2009)}]{Xiong:2009}%
  \BibitemOpen
  \bibfield  {author} {\bibinfo {author} {\bibfnamefont {H.}~\bibnamefont
  {Xiong}}, \bibinfo {author} {\bibfnamefont {D.}~\bibnamefont {van~der
  Lelie}},\ and\ \bibinfo {author} {\bibfnamefont {O.}~\bibnamefont {Gang}},\
  }\bibfield  {title} {\enquote {\bibinfo {title} {{Phase Behavior of
  Nanoparticles Assembled by DNA Linkers}},}\ }\href@noop {} {\bibfield
  {journal} {\bibinfo  {journal} {Phys. Rev. Lett.}\ }\textbf {\bibinfo
  {volume} {102}},\ \bibinfo {pages} {015504} (\bibinfo {year}
  {2009})}\BibitemShut {NoStop}%
\bibitem [{\citenamefont {Zanchet}\ \emph {et~al.}(2001)\citenamefont
  {Zanchet}, \citenamefont {Micheel}, \citenamefont {Parak}, \citenamefont
  {Gerion},\ and\ \citenamefont {Alivisatos}}]{Zanchet:2001}%
  \BibitemOpen
  \bibfield  {author} {\bibinfo {author} {\bibfnamefont {D.}~\bibnamefont
  {Zanchet}}, \bibinfo {author} {\bibfnamefont {C.~M.}\ \bibnamefont
  {Micheel}}, \bibinfo {author} {\bibfnamefont {W.~J.}\ \bibnamefont {Parak}},
  \bibinfo {author} {\bibfnamefont {D.}~\bibnamefont {Gerion}},\ and\ \bibinfo
  {author} {\bibfnamefont {A.~P.}\ \bibnamefont {Alivisatos}},\ }\bibfield
  {title} {\enquote {\bibinfo {title} {{Electrophoretic Isolation of Discrete
  Au Nanocrystal/DNA Conjugates}},}\ }\href@noop {} {\bibfield  {journal}
  {\bibinfo  {journal} {Nano Lett.}\ }\textbf {\bibinfo {volume} {1}},\
  \bibinfo {pages} {32--35} (\bibinfo {year} {2001})}\BibitemShut {NoStop}%
\bibitem [{\citenamefont {Marshall}\ and\ \citenamefont
  {Chapman}(2016)}]{Marshall:2016}%
  \BibitemOpen
  \bibfield  {author} {\bibinfo {author} {\bibfnamefont {B.~D.}\ \bibnamefont
  {Marshall}}\ and\ \bibinfo {author} {\bibfnamefont {W.~G.}\ \bibnamefont
  {Chapman}},\ }\bibfield  {title} {\enquote {\bibinfo {title} {{Thermodynamic
  Perturbation Theory for Associating Molecules}},}\ }in\ \href@noop {} {\emph
  {\bibinfo {booktitle} {{Adv. Chem. Phys.}}}},\ Vol.\ \bibinfo {volume}
  {160},\ \bibinfo {editor} {edited by\ \bibinfo {editor} {\bibfnamefont
  {S.~A.}\ \bibnamefont {Rice}}\ and\ \bibinfo {editor} {\bibfnamefont {A.~R.}\
  \bibnamefont {Dinner}}}\ (\bibinfo  {publisher} {John Wiley \& Sons, Inc.},\
  \bibinfo {year} {2016})\ pp.\ \bibinfo {pages} {1--47}\BibitemShut {NoStop}%
\bibitem [{\citenamefont {Zmpitas}\ and\ \citenamefont
  {Gross}(2016)}]{Zmpitas:2016}%
  \BibitemOpen
  \bibfield  {author} {\bibinfo {author} {\bibfnamefont {W.}~\bibnamefont
  {Zmpitas}}\ and\ \bibinfo {author} {\bibfnamefont {J.}~\bibnamefont
  {Gross}},\ }\bibfield  {title} {\enquote {\bibinfo {title} {{Detailed
  pedagogical review and analysis of Wertheim's thermodynamic perturbation
  theory}},}\ }\href@noop {} {\bibfield  {journal} {\bibinfo  {journal} {Fluid
  Phase Equilib.}\ }\textbf {\bibinfo {volume} {428}},\ \bibinfo {pages}
  {121--152} (\bibinfo {year} {2016})}\BibitemShut {NoStop}%
\bibitem [{\citenamefont {Andersen}(1977)}]{Andersen:1977}%
  \BibitemOpen
  \bibfield  {author} {\bibinfo {author} {\bibfnamefont {H.~C.}\ \bibnamefont
  {Andersen}},\ }\bibfield  {title} {\enquote {\bibinfo {title} {{Cluster
  Methods in Equilibrium Statistical Mechanics}},}\ }in\ \href@noop {} {\emph
  {\bibinfo {booktitle} {{Statistical Mechanics. Modern Theoretical
  Chemistry.}}}},\ Vol.~\bibinfo {volume} {5},\ \bibinfo {editor} {edited by\
  \bibinfo {editor} {\bibfnamefont {B.~J.}\ \bibnamefont {Berne}}}\ (\bibinfo
  {publisher} {Springer},\ \bibinfo {year} {1977})\ pp.\ \bibinfo {pages}
  {1--45}\BibitemShut {NoStop}%
\bibitem [{Note1()}]{Note1}%
  \BibitemOpen
  \bibinfo {note} {An articulation point of a connected graph is a point that,
  if removed, will disconnect the graph into two or more graphs \cite
  {Andersen:1977}. An irreducible graph is free of articulation points. For
  example, a pair of points or any closed cycle of points is irreducible, but
  three points bonded colinearly are not irreducible because the middle point
  is an articulation point.}\BibitemShut {Stop}%
\bibitem [{Note2()}]{Note2}%
  \BibitemOpen
  \bibinfo {note} {$A \subseteq B$ denotes that $A$ is a subset of $B$,
  including the improper subset $A = B$.}\BibitemShut {Stop}%
\bibitem [{Note3()}]{Note3}%
  \BibitemOpen
  \bibinfo {note} {A partition of a set $A$ is a grouping of the elements of
  $A$ into one or more non-empty sets using every element exactly once. For
  example, if $A = \protect \{a,b,c\protect \}$, then $\protect \{\protect
  \{a\protect \},\protect \{b\protect \},\protect \{c\protect \}\protect \}$,
  $\protect \{\protect \{a\protect \},\protect \{b,c\protect \}\protect \}$,
  $\protect \{\protect \{a,b\protect \},\protect \{c\protect \}\protect \}$,
  $\protect \{\protect \{a,c\protect \},\protect \{b\protect \}\protect \}$,
  and $\protect \{\protect \{a,b,c\protect \}\protect \}$ are all partitions of
  $A$. $P(A)$ denotes the set of all possible partitions of $A$. The last
  partition, into only a single subset $\protect \{A\protect \}$, is called an
  improper partition.}\BibitemShut {Stop}%
\bibitem [{Note4()}]{Note4}%
  \BibitemOpen
  \bibinfo {note} {$A \setminus B = \protect \{a \in A | a \protect \notin B
  \protect \}$ denotes the set difference, i.e., all elements that are in $A$
  but not in $B$.}\BibitemShut {Stop}%
\bibitem [{Note5()}]{Note5}%
  \BibitemOpen
  \bibinfo {note} {Forgiving some abuse of notation, the label of a single site
  $A$ should be replaced by a set $\protect \{A\protect \}$ when it represents
  a set of bonded sites.}\BibitemShut {Stop}%
\bibitem [{Note6()}]{Note6}%
  \BibitemOpen
  \bibinfo {note} {$|A|$ denotes the number of elements in a set
  $A$.}\BibitemShut {Stop}%
\bibitem [{Note7()}]{Note7}%
  \BibitemOpen
  \bibinfo {note} {$A \subset B$ denotes $A$ is a proper subset of $B$, i.e.,
  there is at least one element of $B$ that is not in $A$ so $A \not
  =B$.}\BibitemShut {Stop}%
\bibitem [{\citenamefont {Boubl{\'i}k}(1970)}]{Boublik:1970}%
  \BibitemOpen
  \bibfield  {author} {\bibinfo {author} {\bibfnamefont {T.}~\bibnamefont
  {Boubl{\'i}k}},\ }\bibfield  {title} {\enquote {\bibinfo {title}
  {{Hard-Sphere Equation of State}},}\ }\href@noop {} {\bibfield  {journal}
  {\bibinfo  {journal} {J. Chem. Phys.}\ }\textbf {\bibinfo {volume} {53}},\
  \bibinfo {pages} {471--472} (\bibinfo {year} {1970})}\BibitemShut {NoStop}%
\bibitem [{\citenamefont {Sear}\ and\ \citenamefont
  {Jackson}(1996)}]{Sear:1996}%
  \BibitemOpen
  \bibfield  {author} {\bibinfo {author} {\bibfnamefont {R.~P.}\ \bibnamefont
  {Sear}}\ and\ \bibinfo {author} {\bibfnamefont {G.}~\bibnamefont {Jackson}},\
  }\bibfield  {title} {\enquote {\bibinfo {title} {{The ring integral in a
  thermodynamic perturbation theory for association}},}\ }\href@noop {}
  {\bibfield  {journal} {\bibinfo  {journal} {Mol. Phys.}\ }\textbf {\bibinfo
  {volume} {87}},\ \bibinfo {pages} {517--521} (\bibinfo {year}
  {1996})}\BibitemShut {NoStop}%
\bibitem [{\citenamefont {Wertheim}(1987)}]{Wertheim:1987}%
  \BibitemOpen
  \bibfield  {author} {\bibinfo {author} {\bibfnamefont {M.~S.}\ \bibnamefont
  {Wertheim}},\ }\bibfield  {title} {\enquote {\bibinfo {title} {{Thermodynamic
  perturbation theory of polymerization}},}\ }\href@noop {} {\bibfield
  {journal} {\bibinfo  {journal} {J. Chem. Phys.}\ }\textbf {\bibinfo {volume}
  {87}},\ \bibinfo {pages} {7323--7331} (\bibinfo {year} {1987})}\BibitemShut
  {NoStop}%
\bibitem [{\citenamefont {Weeks}, \citenamefont {Chandler},\ and\ \citenamefont
  {Andersen}(1971)}]{Weeks:1971}%
  \BibitemOpen
  \bibfield  {author} {\bibinfo {author} {\bibfnamefont {J.~D.}\ \bibnamefont
  {Weeks}}, \bibinfo {author} {\bibfnamefont {D.}~\bibnamefont {Chandler}},\
  and\ \bibinfo {author} {\bibfnamefont {H.~C.}\ \bibnamefont {Andersen}},\
  }\bibfield  {title} {\enquote {\bibinfo {title} {{Role of Repulsive Forces in
  Determining Equilibrium Structure of Simple Liquids}},}\ }\href@noop {}
  {\bibfield  {journal} {\bibinfo  {journal} {J. Chem. Phys.}\ }\textbf
  {\bibinfo {volume} {54}},\ \bibinfo {pages} {5237--5247} (\bibinfo {year}
  {1971})}\BibitemShut {NoStop}%
\bibitem [{\citenamefont {Fuchs}\ and\ \citenamefont
  {Schweizer}(2002)}]{Fuchs:2002}%
  \BibitemOpen
  \bibfield  {author} {\bibinfo {author} {\bibfnamefont {M.}~\bibnamefont
  {Fuchs}}\ and\ \bibinfo {author} {\bibfnamefont {K.~S.}\ \bibnamefont
  {Schweizer}},\ }\bibfield  {title} {\enquote {\bibinfo {title} {{Structure of
  colloid--polymer suspensions}},}\ }\href@noop {} {\bibfield  {journal}
  {\bibinfo  {journal} {J. Phys.: Condens. Matter}\ }\textbf {\bibinfo {volume}
  {14}},\ \bibinfo {pages} {R239--R269} (\bibinfo {year} {2002})}\BibitemShut
  {NoStop}%
\bibitem [{\citenamefont {Plimpton}(1995)}]{Plimpton:1995}%
  \BibitemOpen
  \bibfield  {author} {\bibinfo {author} {\bibfnamefont {S.}~\bibnamefont
  {Plimpton}},\ }\bibfield  {title} {\enquote {\bibinfo {title} {{Fast Parallel
  Algorithms for Short-Range Molecular Dynamics}},}\ }\href@noop {} {\bibfield
  {journal} {\bibinfo  {journal} {J. Comput. Phys.}\ }\textbf {\bibinfo
  {volume} {117}},\ \bibinfo {pages} {1--19} (\bibinfo {year}
  {1995})}\BibitemShut {NoStop}%
\bibitem [{\citenamefont {Grest}\ and\ \citenamefont
  {Kremer}(1986)}]{Grest:1986}%
  \BibitemOpen
  \bibfield  {author} {\bibinfo {author} {\bibfnamefont {G.~S.}\ \bibnamefont
  {Grest}}\ and\ \bibinfo {author} {\bibfnamefont {K.}~\bibnamefont {Kremer}},\
  }\bibfield  {title} {\enquote {\bibinfo {title} {{Molecular dynamics
  simulation for polymers in the presence of a heat bath}},}\ }\href@noop {}
  {\bibfield  {journal} {\bibinfo  {journal} {Phys. Rev. A}\ }\textbf {\bibinfo
  {volume} {33}},\ \bibinfo {pages} {3628--3631} (\bibinfo {year}
  {1986})}\BibitemShut {NoStop}%
\bibitem [{\citenamefont {Haghmoradi}, \citenamefont {Marshall},\ and\
  \citenamefont {Chapman}(2020)}]{Haghmoradi:2020}%
  \BibitemOpen
  \bibfield  {author} {\bibinfo {author} {\bibfnamefont {A.}~\bibnamefont
  {Haghmoradi}}, \bibinfo {author} {\bibfnamefont {B.~D.}\ \bibnamefont
  {Marshall}},\ and\ \bibinfo {author} {\bibfnamefont {W.~G.}\ \bibnamefont
  {Chapman}},\ }\bibfield  {title} {\enquote {\bibinfo {title} {{Beyond
  Wertheim’s Multi-density Theory: Steric Hindrance and Associated Rings in a
  Two-Density Formalism for Binary Mixtures of Molecules with Two Associating
  Sites}},}\ }\href {https://doi.org/10.1021/acs.jced.0c00695} {\bibfield
  {journal} {\bibinfo  {journal} {J. Chem. Eng. Data}\ } (\bibinfo {year}
  {2020}),\ 10.1021/acs.jced.0c00695}\BibitemShut {NoStop}%
\bibitem [{\citenamefont {Howard}\ \emph {et~al.}(2020)\citenamefont {Howard},
  \citenamefont {Sherman}, \citenamefont {Sreenivasan}, \citenamefont
  {Valenzuela}, \citenamefont {Anslyn}, \citenamefont {Milliron},\ and\
  \citenamefont {Truskett}}]{Howard:2020}%
  \BibitemOpen
  \bibfield  {author} {\bibinfo {author} {\bibfnamefont {M.~P.}\ \bibnamefont
  {Howard}}, \bibinfo {author} {\bibfnamefont {Z.~M.}\ \bibnamefont {Sherman}},
  \bibinfo {author} {\bibfnamefont {A.~N.}\ \bibnamefont {Sreenivasan}},
  \bibinfo {author} {\bibfnamefont {S.~A.}\ \bibnamefont {Valenzuela}},
  \bibinfo {author} {\bibfnamefont {E.~V.}\ \bibnamefont {Anslyn}}, \bibinfo
  {author} {\bibfnamefont {D.~J.}\ \bibnamefont {Milliron}},\ and\ \bibinfo
  {author} {\bibfnamefont {T.~M.}\ \bibnamefont {Truskett}},\ }\bibfield
  {title} {\enquote {\bibinfo {title} {{Effects of linker flexibility on phase
  behavior and structure of linked colloidal gels}},}\ }\href@noop {} {\ ,\
  \bibinfo {pages} {arXiv:2011.12512} (\bibinfo {year} {2020})}\BibitemShut
  {NoStop}%
\bibitem [{\citenamefont {Bianchi}\ \emph
  {et~al.}(2006{\natexlab{b}})\citenamefont {Bianchi}, \citenamefont {Largo},
  \citenamefont {Tartaglia}, \citenamefont {Zaccarelli},\ and\ \citenamefont
  {Sciortino}}]{Bianchi:2006}%
  \BibitemOpen
  \bibfield  {author} {\bibinfo {author} {\bibfnamefont {E.}~\bibnamefont
  {Bianchi}}, \bibinfo {author} {\bibfnamefont {J.}~\bibnamefont {Largo}},
  \bibinfo {author} {\bibfnamefont {P.}~\bibnamefont {Tartaglia}}, \bibinfo
  {author} {\bibfnamefont {E.}~\bibnamefont {Zaccarelli}},\ and\ \bibinfo
  {author} {\bibfnamefont {F.}~\bibnamefont {Sciortino}},\ }\bibfield  {title}
  {\enquote {\bibinfo {title} {{Phase Diagram of Patchy Colloids: Towards Empty
  Liquids}},}\ }\href {https://doi.org/10.1103/PhysRevLett.97.168301}
  {\bibfield  {journal} {\bibinfo  {journal} {Phys. Rev. Lett.}\ }\textbf
  {\bibinfo {volume} {97}},\ \bibinfo {pages} {168301} (\bibinfo {year}
  {2006}{\natexlab{b}})}\BibitemShut {NoStop}%
\bibitem [{\citenamefont {Zaccarelli}(2007)}]{Zaccarelli2007}%
  \BibitemOpen
  \bibfield  {author} {\bibinfo {author} {\bibfnamefont {E.}~\bibnamefont
  {Zaccarelli}},\ }\bibfield  {title} {\enquote {\bibinfo {title} {{Colloidal
  Gels: Equilibrium and Nonequilibrium Routes}},}\ }\href@noop {} {\bibfield
  {journal} {\bibinfo  {journal} {J. Phys.: Condens. Matter}\ }\textbf
  {\bibinfo {volume} {19}},\ \bibinfo {pages} {323101} (\bibinfo {year}
  {2007})}\BibitemShut {NoStop}%
\bibitem [{\citenamefont {Miles}(1965)}]{Miles:1965}%
  \BibitemOpen
  \bibfield  {author} {\bibinfo {author} {\bibfnamefont {R.~E.}\ \bibnamefont
  {Miles}},\ }\bibfield  {title} {\enquote {\bibinfo {title} {{On random
  rotations in $R^3$}},}\ }\href@noop {} {\bibfield  {journal} {\bibinfo
  {journal} {Biometrika}\ }\textbf {\bibinfo {volume} {52}},\ \bibinfo {pages}
  {636--639} (\bibinfo {year} {1965})}\BibitemShut {NoStop}%
\bibitem [{\citenamefont {Harris}\ \emph {et~al.}(2020)\citenamefont {Harris},
  \citenamefont {Millman}, \citenamefont {van~der Walt}, \citenamefont
  {Gommers}, \citenamefont {Virtanen}, \citenamefont {Cournapeau},
  \citenamefont {Wieser}, \citenamefont {Taylor}, \citenamefont {Berg},
  \citenamefont {Smith}, \citenamefont {Kern}, \citenamefont {Picus},
  \citenamefont {Hoyer}, \citenamefont {van Kerkwijk}, \citenamefont {Brett},
  \citenamefont {Haldane}, \citenamefont {Fernández~del Río}, \citenamefont
  {Wiebe}, \citenamefont {Peterson}, \citenamefont {Gérard-Marchant},
  \citenamefont {Sheppard}, \citenamefont {Reddy}, \citenamefont {Weckesser},
  \citenamefont {Abbasi}, \citenamefont {Gohlke},\ and\ \citenamefont
  {Oliphant}}]{numpy}%
  \BibitemOpen
  \bibfield  {author} {\bibinfo {author} {\bibfnamefont {C.~R.}\ \bibnamefont
  {Harris}}, \bibinfo {author} {\bibfnamefont {K.~J.}\ \bibnamefont {Millman}},
  \bibinfo {author} {\bibfnamefont {S.~J.}\ \bibnamefont {van~der Walt}},
  \bibinfo {author} {\bibfnamefont {R.}~\bibnamefont {Gommers}}, \bibinfo
  {author} {\bibfnamefont {P.}~\bibnamefont {Virtanen}}, \bibinfo {author}
  {\bibfnamefont {D.}~\bibnamefont {Cournapeau}}, \bibinfo {author}
  {\bibfnamefont {E.}~\bibnamefont {Wieser}}, \bibinfo {author} {\bibfnamefont
  {J.}~\bibnamefont {Taylor}}, \bibinfo {author} {\bibfnamefont
  {S.}~\bibnamefont {Berg}}, \bibinfo {author} {\bibfnamefont {N.~J.}\
  \bibnamefont {Smith}}, \bibinfo {author} {\bibfnamefont {R.}~\bibnamefont
  {Kern}}, \bibinfo {author} {\bibfnamefont {M.}~\bibnamefont {Picus}},
  \bibinfo {author} {\bibfnamefont {S.}~\bibnamefont {Hoyer}}, \bibinfo
  {author} {\bibfnamefont {M.~H.}\ \bibnamefont {van Kerkwijk}}, \bibinfo
  {author} {\bibfnamefont {M.}~\bibnamefont {Brett}}, \bibinfo {author}
  {\bibfnamefont {A.}~\bibnamefont {Haldane}}, \bibinfo {author} {\bibfnamefont
  {J.}~\bibnamefont {Fernández~del Río}}, \bibinfo {author} {\bibfnamefont
  {M.}~\bibnamefont {Wiebe}}, \bibinfo {author} {\bibfnamefont
  {P.}~\bibnamefont {Peterson}}, \bibinfo {author} {\bibfnamefont
  {P.}~\bibnamefont {Gérard-Marchant}}, \bibinfo {author} {\bibfnamefont
  {K.}~\bibnamefont {Sheppard}}, \bibinfo {author} {\bibfnamefont
  {T.}~\bibnamefont {Reddy}}, \bibinfo {author} {\bibfnamefont
  {W.}~\bibnamefont {Weckesser}}, \bibinfo {author} {\bibfnamefont
  {H.}~\bibnamefont {Abbasi}}, \bibinfo {author} {\bibfnamefont
  {C.}~\bibnamefont {Gohlke}},\ and\ \bibinfo {author} {\bibfnamefont {T.~E.}\
  \bibnamefont {Oliphant}},\ }\bibfield  {title} {\enquote {\bibinfo {title}
  {Array programming with {NumPy}},}\ }\href@noop {} {\bibfield  {journal}
  {\bibinfo  {journal} {Nature}\ }\textbf {\bibinfo {volume} {585}},\ \bibinfo
  {pages} {357–362} (\bibinfo {year} {2020})}\BibitemShut {NoStop}%
\bibitem [{\citenamefont {Virtanen}\ \emph {et~al.}(2020)\citenamefont
  {Virtanen}, \citenamefont {Gommers}, \citenamefont {Oliphant}, \citenamefont
  {Haberland}, \citenamefont {Reddy}, \citenamefont {Cournapeau}, \citenamefont
  {Burovski}, \citenamefont {Peterson}, \citenamefont {Weckesser},
  \citenamefont {Bright}, \citenamefont {{van der Walt}}, \citenamefont
  {Brett}, \citenamefont {Wilson}, \citenamefont {Millman}, \citenamefont
  {Mayorov}, \citenamefont {Nelson}, \citenamefont {Jones}, \citenamefont
  {Kern}, \citenamefont {Larson}, \citenamefont {Carey}, \citenamefont {Polat},
  \citenamefont {Feng}, \citenamefont {Moore}, \citenamefont {{VanderPlas}},
  \citenamefont {Laxalde}, \citenamefont {Perktold}, \citenamefont {Cimrman},
  \citenamefont {Henriksen}, \citenamefont {Quintero}, \citenamefont {Harris},
  \citenamefont {Archibald}, \citenamefont {Ribeiro}, \citenamefont
  {Pedregosa}, \citenamefont {{van Mulbregt}},\ and\ \citenamefont {{SciPy 1.0
  Contributors}}}]{scipy}%
  \BibitemOpen
  \bibfield  {author} {\bibinfo {author} {\bibfnamefont {P.}~\bibnamefont
  {Virtanen}}, \bibinfo {author} {\bibfnamefont {R.}~\bibnamefont {Gommers}},
  \bibinfo {author} {\bibfnamefont {T.~E.}\ \bibnamefont {Oliphant}}, \bibinfo
  {author} {\bibfnamefont {M.}~\bibnamefont {Haberland}}, \bibinfo {author}
  {\bibfnamefont {T.}~\bibnamefont {Reddy}}, \bibinfo {author} {\bibfnamefont
  {D.}~\bibnamefont {Cournapeau}}, \bibinfo {author} {\bibfnamefont
  {E.}~\bibnamefont {Burovski}}, \bibinfo {author} {\bibfnamefont
  {P.}~\bibnamefont {Peterson}}, \bibinfo {author} {\bibfnamefont
  {W.}~\bibnamefont {Weckesser}}, \bibinfo {author} {\bibfnamefont
  {J.}~\bibnamefont {Bright}}, \bibinfo {author} {\bibfnamefont {S.~J.}\
  \bibnamefont {{van der Walt}}}, \bibinfo {author} {\bibfnamefont
  {M.}~\bibnamefont {Brett}}, \bibinfo {author} {\bibfnamefont
  {J.}~\bibnamefont {Wilson}}, \bibinfo {author} {\bibfnamefont {K.~J.}\
  \bibnamefont {Millman}}, \bibinfo {author} {\bibfnamefont {N.}~\bibnamefont
  {Mayorov}}, \bibinfo {author} {\bibfnamefont {A.~R.~J.}\ \bibnamefont
  {Nelson}}, \bibinfo {author} {\bibfnamefont {E.}~\bibnamefont {Jones}},
  \bibinfo {author} {\bibfnamefont {R.}~\bibnamefont {Kern}}, \bibinfo {author}
  {\bibfnamefont {E.}~\bibnamefont {Larson}}, \bibinfo {author} {\bibfnamefont
  {C.~J.}\ \bibnamefont {Carey}}, \bibinfo {author} {\bibfnamefont
  {{\.I}.}~\bibnamefont {Polat}}, \bibinfo {author} {\bibfnamefont
  {Y.}~\bibnamefont {Feng}}, \bibinfo {author} {\bibfnamefont {E.~W.}\
  \bibnamefont {Moore}}, \bibinfo {author} {\bibfnamefont {J.}~\bibnamefont
  {{VanderPlas}}}, \bibinfo {author} {\bibfnamefont {D.}~\bibnamefont
  {Laxalde}}, \bibinfo {author} {\bibfnamefont {J.}~\bibnamefont {Perktold}},
  \bibinfo {author} {\bibfnamefont {R.}~\bibnamefont {Cimrman}}, \bibinfo
  {author} {\bibfnamefont {I.}~\bibnamefont {Henriksen}}, \bibinfo {author}
  {\bibfnamefont {E.~A.}\ \bibnamefont {Quintero}}, \bibinfo {author}
  {\bibfnamefont {C.~R.}\ \bibnamefont {Harris}}, \bibinfo {author}
  {\bibfnamefont {A.~M.}\ \bibnamefont {Archibald}}, \bibinfo {author}
  {\bibfnamefont {A.~H.}\ \bibnamefont {Ribeiro}}, \bibinfo {author}
  {\bibfnamefont {F.}~\bibnamefont {Pedregosa}}, \bibinfo {author}
  {\bibfnamefont {P.}~\bibnamefont {{van Mulbregt}}},\ and\ \bibinfo {author}
  {\bibnamefont {{SciPy 1.0 Contributors}}},\ }\bibfield  {title} {\enquote
  {\bibinfo {title} {{{SciPy} 1.0: Fundamental Algorithms for Scientific
  Computing in Python}},}\ }\href@noop {} {\bibfield  {journal} {\bibinfo
  {journal} {Nature Methods}\ }\textbf {\bibinfo {volume} {17}},\ \bibinfo
  {pages} {261--272} (\bibinfo {year} {2020})}\BibitemShut {NoStop}%
\bibitem [{\citenamefont {Lam}, \citenamefont {Pitrou},\ and\ \citenamefont
  {Seibert}(2015)}]{numba}%
  \BibitemOpen
  \bibfield  {author} {\bibinfo {author} {\bibfnamefont {S.~K.}\ \bibnamefont
  {Lam}}, \bibinfo {author} {\bibfnamefont {A.}~\bibnamefont {Pitrou}},\ and\
  \bibinfo {author} {\bibfnamefont {S.}~\bibnamefont {Seibert}},\ }\bibfield
  {title} {\enquote {\bibinfo {title} {{Numba: A LLVM-Based Python JIT
  Compiler}},}\ }in\ \href {https://doi.org/10.1145/2833157.2833162} {\emph
  {\bibinfo {booktitle} {{Proceedings of the Second Workshop on the LLVM
  Compiler Infrastructure in HPC}}}},\ \bibinfo {series and number} {LLVM '15}\
  (\bibinfo {year} {2015})\BibitemShut {NoStop}%
\end{thebibliography}%

\end{document}